\documentclass{svjour3}     
\usepackage{graphicx} 
\usepackage[hidelinks]{hyperref}
\usepackage[utf8]{inputenc}
\usepackage{csquotes}
\usepackage{xcolor}
\usepackage{amsmath}
\usepackage{amsfonts}
\usepackage{amssymb}
\usepackage{mathrsfs}
\usepackage{dsfont}

\newcommand{\pH}{port-Hamiltonian }
\newcommand{\cf}{\textit{cf.} }
\renewcommand{\qed}{$\hfill\blacksquare$}

\newcommand{\half}{\frac{1}{2}}

\newcommand{\B}[1]{\boldsymbol{#1}}
\newcommand{\bb}[1]{\mathbb{#1}}
\newcommand{\cl}[1]{\mathcal{#1}}
\newcommand{\scr}[1]{\mathscr{#1}}

\newcommand{\tr}{\mathrm{tr}}

\newcommand{\TwoTwoMat}[4]{
	\begin{pmatrix}
		#1 & #2 \\
		#3 & #4
\end{pmatrix}}
\newcommand{\TwoVec}[2]{
	\begin{pmatrix}
		#1\\
		#2 
\end{pmatrix}}

\newcommand{\ThrVec}[3]{
	\begin{pmatrix}
		#1\\
		#2\\
		#3
\end{pmatrix}}

\newcommand{\map}[3]{{#1}:{#2}\rightarrow {#3}}
\newcommand{\fullmap}[5]{
	\begin{split}
		#1 : {#2} &\rightarrow {#3}\\
		{#4} &\mapsto {#5}
	\end{split}
}

\newcommand{\subTxt}[2]{{#1}_{\mathrm{#2}}}

\newcommand{\dds}{\left.\frac{d}{ds}\right |_{s=0}}
\newcommand{\sym}{\mathrm{sym}}
\newcommand{\asym}{\mathrm{skew}}

\newcommand{\spFn}[1]{C^\infty(#1)}
\newcommand{\spVec}[1]{\Gamma(T#1)}
\newcommand{\spFrm}[2]{\Omega^{#2}(#1)}
\newcommand{\spFrmB}[1]{\spFrm{\cl{B}}{#1}}
\newcommand{\spFrmS}[1]{\spFrm{\cl{S}}{#1}}

\newcommand{\spVecS}{\Gamma(T\cl{S})}
\newcommand{\spVecB}{\Gamma(T\cl{B})}

\newcommand{\spvecFrmB}[1]{\spFrm{\cl{B};T\cl{B}}{#1}}
\newcommand{\spcovFrmB}[1]{\spFrm{\cl{B};T^*\cl{B}}{#1}}
\newcommand{\spvecFrmS}[1]{\spFrm{\cl{S};T\cl{S}}{#1}}
\newcommand{\spcovFrmS}[1]{\spFrm{\cl{S};T^*\cl{S}}{#1}}
\newcommand{\spvecFrmPhi}[1]{\Omega_\varphi^{#1}(\cl{B};T\cl{S})}
\newcommand{\spcovFrmPhi}[1]{\Omega_\varphi^{#1}(\cl{B};T^*\cl{S})}

\newcommand{\spvecFrmBbnd}[1]{\spFrm{\partial\cl{B};T\cl{B}}{#1}}
\newcommand{\spcovFrmBbnd}[1]{\spFrm{\partial\cl{B};T^*\cl{B}}{#1}}
\newcommand{\spvecFrmSbnd}[1]{\spFrm{\partial\cl{S};T\cl{S}}{#1}}
\newcommand{\spcovFrmSbnd}[1]{\spFrm{\partial\cl{S};T^*\cl{S}}{#1}}
\newcommand{\spvecFrmPhibnd}[1]{\Omega_\varphi^{#1}(\partial\cl{B};T\cl{S})}
\newcommand{\spcovFrmPhibnd}[1]{\Omega_\varphi^{#1}(\partial\cl{B};T^*\cl{S})}

\newcommand{\spFrmM}[2]{\spFrm{M;#2}{#1}}

\newcommand{\intB}{\int_{\cl{B}}}
\newcommand{\intS}{\int_{\cl{S}}}

\newcommand{\spC}{\mathscr{C}}
\newcommand{\spMetB}{\cl{M}(\cl{B})}

\newcommand{\bndB}{\partial\cl{B}}
\newcommand{\cfgSp}{\mathscr{C}}
\newcommand{\cfgT}{\varphi_t}
\newcommand{\cfgInvT}{\varphi{\scriptstyle^{-1}_t}}
\newcommand{\cfgP}{\varphi_t^*}

\newcommand{\mFM}{\tilde{\mu}}
\newcommand{\mFC}{\hat{\mu}}
\newcommand{\mFS}{\mu}

\newcommand{\mDS}{\rho}
\newcommand{\mDC}{\hat{\rho}}
\newcommand{\mDM}{\tilde{\rho}}

\newcommand{\vFS}{\omega_{\gS}}
\newcommand{\vFC}{\hat{\omega}_{\gC}}

\newcommand{\vS}{v}
\newcommand{\vM}{\tilde{v}}
\newcommand{\vC}{{\hat{v}}}

\newcommand{\vfS}{v^\flat}
\newcommand{\vfM}{\tilde{v}^\flat}
\newcommand{\vfC}{\hat{v}^\flat}

\newcommand{\gS}{g}

\newcommand{\gC}{\hat{g}}

\newcommand{\nabS}{\nabla}
\newcommand{\nabM}{\tilde{\nabla}}
\newcommand{\nabC}{\hat{\nabla}}
\newcommand{\Lie}[2]{\cl{L}_{#1}{#2}}
\newcommand{\divrS}{\mathrm{div}}
\newcommand{\divrC}{\widehat{\mathrm{div}}}
\newcommand{\divrM}{\widetilde{\mathrm{div}}}

\newcommand{\epS}{\varepsilon}
\newcommand{\epC}{\hat{\varepsilon}}

\newcommand{\extd}{\mathrm{d}}
\newcommand{\extD}{\mathrm{D}}

\newcommand{\cfgPleg}[1]{\varphi_{\mathrm{#1}}^*}
\newcommand{\cfgFleg}[1]{\varphi_{\mathrm{#1},*}}

\newcommand{\wedgedot}{\ \dot{\wedge}\ }
\newcommand{\extcdS}{\extd_{\nabS}}
\newcommand{\extcdC}{\hat{\extd}_{\nabC}}
\newcommand{\extcdM}{\tilde{\extd}_{\nabM}}

\newcommand{\duPair}[3]{\langle  #1 | #2 \rangle_{#3}}
\newcommand{\duPairB}[2]{\duPair{#1}{#2}{\cl{B}}}

\newcommand{\hodgeS}{\star_c}
\newcommand{\invhodgeS}{\star_c^{-1}}
\newcommand{\hodgeC}{\hat{\star}_c}
\newcommand{\invhodgeC}{\hat{\star}_c^{-1}}
\newcommand{\hodgeM}{\tilde{\star}_c}
\newcommand{\invhodgeM}{\tilde{\star}_c^{-1}}

\newcommand{\Sbound}{{\partial\cl{S}}}
\newcommand{\Bbound}{{\partial\cl{B}}}
\newcommand{\stS}{\cl{T}}
\newcommand{\stC}{\hat{\cl{T}}}
\newcommand{\stM}{\tilde{\cl{T}}}
\newcommand{\ptr}{i_{\mathrm{f}}^*}

\newcommand{\momM}{\tilde{\cl{M}}}
\newcommand{\momC}{\hat{\cl{M}}}
\newcommand{\momS}{\cl{M}}

\newcommand{\stTauS}{\tau}
\newcommand{\stTauC}{\hat{\tau}}

\newcommand{\stSigS}{\sigma}
\newcommand{\stSigC}{\hat{\sigma}}
\newcommand{\stSigM}{\tilde{\sigma}}

\newcommand{\parD}[2]{\frac{\partial #1}{\partial #2}}

\newcommand{\varD}[2]{\delta_{#2}{#1}}
\newcommand{\varDbnd}[2]{\delta^\cup_{#2}{#1}}

\newcommand{\LkinM}{\subTxt{\tilde{L}}{k}}
\newcommand{\HkinM}{\subTxt{\tilde{H}}{k}}
\newcommand{\HkinMdot}{\subTxt{\dot{\tilde{H}}}{k}}
\newcommand{\spPkinM}{\subTxt{\tilde{\cl{P}}}{k}}
\newcommand{\spXkinM}{\subTxt{\tilde{\cl{X}}}{k}}

\newcommand{\chiM}{\tilde{\chi}}
\newcommand{\forceM}{\tilde{\cl{F}}}
\newcommand{\delphiM}{\delta\tilde{\varphi}}

\newcommand{\diracKinM}{\subTxt{\tilde{\cl{D}}}{k}}
\newcommand{\intKinM}{\subTxt{\tilde{\cl{I}}}{k}}
\newcommand{\FKinM}{\subTxt{\tilde{\mathfrak{F}}}{k}}
\newcommand{\FdlKinM}{\subTxt{\tilde{\mathfrak{F}}^*}{k}}

\newcommand{\scrTld}[1]{\tilde{\mathscr{#1}}}

\newcommand{\fM}[1]{\tilde{f}_{#1}}
\newcommand{\eM}[1]{\tilde{e}_{#1}}

\newcommand{\diracStrM}{\subTxt{\tilde{\cl{D}}}{s}}

\newcommand{\HkinC}{\subTxt{\hat{H}}{k}}
\newcommand{\HkinCdot}{\subTxt{\dot{\hat{H}}}{k}}
\newcommand{\spPkinC}{\subTxt{\hat{\cl{P}}}{k}}
\newcommand{\spXkinC}{\subTxt{\hat{\cl{X}}}{k}}
\newcommand{\chiC}{\hat{\chi}}
\newcommand{\forceC}{\hat{\cl{F}}}
\newcommand{\delgC}{{\delta\gC}}

\newcommand{\delphiC}{\delta\hat{\varphi}}

\newcommand{\diracKinC}{\subTxt{\hat{\cl{D}}}{k}}

\newcommand{\FKinC}{\subTxt{\hat{\mathfrak{F}}}{k}}
\newcommand{\FdlKinC}{\subTxt{\hat{\mathfrak{F}}^*}{k}}
\newcommand{\diracStrC}{\subTxt{\hat{\cl{D}}}{s}}
\newcommand{\FStrC}{\subTxt{\hat{\mathfrak{F}}}{s}}
\newcommand{\FdlStrC}{\subTxt{\hat{\mathfrak{F}}^*}{s}}

\newcommand{\scrHat}[1]{\hat{\mathscr{#1}}}
\newcommand{\fC}[1]{\hat{f}_{#1}}
\newcommand{\eC}[1]{\hat{e}_{#1}}

\newcommand{\duPairS}[2]{\duPair{#1}{#2}{\cl{S}}}

\newcommand{\HkinS}{\subTxt{{H}}{k}}
\newcommand{\HkinSdot}{\subTxt{\dot{{H}}}{k}}
\newcommand{\spPkinS}{\subTxt{{\cl{P}}}{k}}
\newcommand{\spXkinS}{\subTxt{{\cl{X}}}{k}}
\newcommand{\chiS}{{\chi}}
\newcommand{\forceS}{{\cl{F}}}

\newcommand{\diracKinS}{\subTxt{{\cl{D}}}{k}}

\newcommand{\FKinS}{\subTxt{{\mathfrak{F}}}{k}}
\newcommand{\FdlKinS}{\subTxt{{\mathfrak{F}}^*}{k}}

\newcommand{\fS}[1]{{f}_{#1}}
\newcommand{\eS}[1]{{e}_{#1}}

\newcommand{\diracStrS}{\subTxt{{\cl{D}}}{s}}

\newcommand{\spvecFrmBbndi}[2]{\spFrm{\partial\cl{B}_{#2};T\cl{B}}{#1}}
\newcommand{\spcovFrmBbndi}[2]{\spFrm{\partial\cl{B}_{#2};T^*\cl{B}}{#1}}

\newcommand{\EpsC}{\hat{\varepsilon}}

\newcommand{\piCsym}{\subTxt{\hat{\pi}}{sym}}
\newcommand{\piCasy}{\subTxt{\hat{\pi}}{asy}}

\newcommand{\spvecFrmBsub}[2]{\Omega_{\mathrm{#2}}^{#1}(\cl{B};T\cl{B})}
\newcommand{\spcovFrmBsub}[2]{\Omega_{\mathrm{#2}}^{#1}(\cl{B};T^*\cl{B})}

\newcommand{\spvecFrmSsub}[2]{\Omega_{\mathrm{#2}}^{#1}(\cl{S};T\cl{S})}

\newcommand{\piCvol}{\subTxt{\hat{\pi}}{vol}}
\newcommand{\piCdev}{\subTxt{\hat{\pi}}{dev}}

\newcommand{\StrC}{{\hat{\zeta}}}

\newcommand{\cfgref}{\subTxt{\varphi}{ref}}
\newcommand{\cfgPref}{\subTxt{\varphi}{ref}^*}
\newcommand{\Exp}{\mathrm{exp}}
\newcommand{\Ln}{\mathrm{ln}}

\newcommand{\epCvol}{\subTxt{\epC}{vol}}
\newcommand{\epCdev}{\subTxt{\epC}{dev}}
\newcommand{\StrCvol}{\subTxt{\StrC}{vol}}
\newcommand{\StrCdev}{\subTxt{\StrC}{dev}}

\newcommand{\psiCvol}{\subTxt{\hat{\psi}}{vol}}
\newcommand{\psiCdev}{\subTxt{\hat{\psi}}{dev}}

\newcommand{\stTauCvol}{\subTxt{\stTauC}{vol}}
\newcommand{\stTauCdev}{\subTxt{\stTauC}{dev}}

\newcommand{\epSvol}{\subTxt{\varepsilon}{vol}}
\newcommand{\epSdev}{\subTxt{\varepsilon}{dev}}
\newcommand{\stTauSvol}{\subTxt{\stTauS}{vol}}
\newcommand{\stTauSdev}{\subTxt{\stTauS}{dev}}

\newcommand{\stefano}[1]{#1}

\graphicspath{{figures/}}
\title{The port-Hamiltonian structure of continuum mechanics}
\author{Ramy Rashad and Stefano Stramigioli}
\institute{R. Rashad and S. Stramigioli \at
	Robotics and Mechatronics Department, University of Twente, The Netherlands.\\
	\email{\{r.a.m.rashadhashem, s.stramigioli\}@utwente.nl}           
}
\date{\today}

\begin{document}

\maketitle

\begin{abstract}
In this paper we present a novel approach to the geometric formulation of solid and fluid mechanics within the \pH framework, which extends the standard Hamiltonian formulation to non-conservative and open dynamical systems. Leveraging Dirac structures, instead of symplectic or Poisson structures, this formalism allows the incorporation of energy exchange within the spatial domain or through its boundary, which allows for a more  comprehensive description of continuum mechanics. Building upon our recent work in describing nonlinear elasticity using exterior calculus and bundle-valued differential forms, this paper focuses on the systematic derivation of \pH models for solid and fluid mechanics in the material, spatial, and convective representations using Hamiltonian reduction theory. This paper also discusses constitutive relations for stress within this framework including hyper-elasticity, for both finite- and infinite-strains, as well as viscous fluid flow governed by the Navier-Stokes equations.

\keywords{port-Hamiltonian \and Dirac structures \and bundle-valued forms \and exterior calculus}
\end{abstract}

\section{Introduction}

\stefano{Due to the invariance of physical laws and Noether's theorem, energy is at}
the heart of many physical theories
\stefano{and}
is fundamental to the Lagrangian and Hamiltonian approaches to continuum mechanics.
In the Hamiltonian formulation, the governing equations of continuum mechanics are expressed using symplectic or Poisson structures which are geometric objects that \stefano{embody the invariance of physical laws in the state space and} 
\stefano{describe}
the \textit{energetic structure} of the equations of motion.
This 
\stefano{holds
in all classical fields and clearly also for}
both fluid and solid mechanics in the spatial, convective, or material representation \cite{arnold1965topologie,marsden1974reduction,Simo1988ThePlates}.

Due to the inherent skew-symmetric nature of symplectic and Poisson structures, the standard Hamiltonian formalism is limited to conservative closed systems that do not exchange energy with their surrounding.
In fluid mechanics this is usually imposed by assuming the velocity field to be tangent to the spatial domain's boundary \cite{arnold1965topologie}, whereas in solid mechanics one is limited to fixed boundary conditions \cite{Simo1988ThePlates}.

On the contrary, the port-Hamiltonian formalism \cite{van2002hamiltonian} is applicable to non-conservative and open systems 
\stefano{and}
relies on Dirac structures, instead of symplectic or Poisson structures, for characterizing the energetic structure of the equations of motion.
This framework is suitable for describing fluid and solid mechanics with energy exchange within the spatial domain or through its boundary:
it can incorporate state constraints (e.g. incompressibility) as well as viscous effects unlike the 
\stefano{canonical}
Hamiltonian formalism.
Being based on energy, which is the \enquote{lingua franca} of physics, the port-Hamiltonian formalism has been widely applied to numerous multi-physics domains such as fluid mechanics \cite{califano2021geometric}, solid mechanics \cite{brugnoli2021port}, fluid structure interaction \cite{califano2022energetic}, thermodynamics \cite{califano2022differential}, and magneto-hydrodynamics \cite{vu2016structured}.

This paper is a sequel to our recent work \cite{rashad2023intrinsic} dealing with the geometric modeling of nonlinear elasticity using exterior calculus. In the current paper we focus on the formulation of nonlinear elasticity in the port-Hamiltonian framework using bundle-valued differential forms.
The main goal is to express the governing equations of elasticity as a network of energetic modules interconnected together using power-ports and Dirac structures, graphically represented in Fig. \ref{fig:pH_model_conv}.
Such identification of the energetic structure underlying the equations has numerous benefits.
On one hand, it provides conceptual insights to better understand the theory of continuum mechanics in a coordinate-free manner.
On the other hand, it can \stefano{be} exploited for structure-preserving discretization, model order-reduction, energy-based control, and analysis. The interested reader may refer to \cite{rashad2020twenty} for an extensive survey of the potential of the \pH framework.\\

The contributions of this paper are as follows:
\begin{enumerate}
	\item We formulate the governing equations of motion of solid and fluid mechanics in the \pH framework.
	\item In contrast to other efforts in the literature that usually rewrite the equations in a \pH form, we use Hamiltonian reduction theory to derive the governing equations from first principles. Demonstrating this \pH modeling process in itself is a contribution because it can be applied to other physical systems.
	\item We treat the material, convective, and spatial representations of continuum mechanics and highlight the similarities between them explicated by our exterior calculus formulation.
	\item We also provide a number of constitutive relations for stress showing how they fit within the \pH framework and highlighting the importance of the convective representation for constitutive modeling.
\end{enumerate}
The outline of the paper is as follows: In Sec. 2, we provide a summary of the geometric formulation of continuum mechanics in tensor calculus and exterior calculus detailed in \cite{rashad2023intrinsic}. In Sec. 3 and 4, we present the \pH model for nonlinear elasticity and fluid mechanics, separately to highlight the similarities and differences between them. We present the constitutive relations for hyper-elasticity and viscous fluid flow in Sec. 5, and conclude the paper in Sec. 6.

\section{Geometric formulation of nonlinear elasticity}
In this section we recall the geometric approach to continuum mechanics both in tensor calculus and exterior calculus.
We refer the reader to \cite{Marsden1994} for an introduction to the tensor calculus formulation and to \cite{rashad2023intrinsic} for an intrinsic treatment of the topic in exterior calculus using bundle-valued forms.\\

\subsubsection*{Notation}

Throughout this paper, we will distinguish material quantities by a $\sim$ on top, convective quantities by a $\wedge$ on top, and spatial quantities with none.
For any manifold $M$, we shall denote the space of (real-valued) smooth functions on $M$ by $\spFn{M}$ and sections of any vector-bundle $\bb{E}$ over $M$ by $\Gamma(\bb{E})$. For instance, the space of $p$-contravariant and $q$-covariant smooth tensor fields is denoted by $\spVec{^p_q M}$.
The space of (scalar-valued) differential $k$-forms, i.e. totally asymmetric $k$-covariant tensor fields, will be denoted by $\spFrm{M}{k}$. The space of generic two-point tensor-fields on the manifolds $M$ and $N$ related by $\map{\varphi}{M}{N}$ will be denoted by $\spVec{^p_q M \otimes \varphi^* T^r_s N}$. If $M$ is endowed with a Riemannian metric $g$, we shall interchangeably denote the index lowering operation of a vector field $u\in\spVec{M}$ to a one-form in $\spFrm{M}{1}$ either by $u^\flat$ or $g\cdot u$.
The latter explicit notation will be useful since we will use different metrics for the spatial, convective, and material representations.

\subsection{Tensor calculus}\label{sec:geom_mech_tens_calc}
The configuration of an elastic body is described by a smooth embedding $\map{\varphi}{\cl{B}}{\mathscr{A}}$ of the body manifold $\cl{B}$ into the ambient space $\mathscr{A}$.
A motion of the elastic body is represented by a smooth curve $\map{c_\varphi}{\bb{R}}{\cfgSp}$ with $\cfgSp$ denoting the configuration space of smooth embeddings of $\cl{B}$ in $\mathscr{A}$, i.e. $\cfgSp:= \text{Emb}^\infty(\cl{B},\mathscr{A})$.
We denote the image of the whole body by $\cl{S}:= \varphi(\cl{B}) \subset \mathscr{A}$ and consider only the case $\dim(\cl{B}) = \dim(\cl{S}) = 3$.
We assume $\mathscr{A}$ has a Riemannian structure with $\gS$ denoting its metric, while $\cl{B}$ does not have an intrinsic Riemannian structure and only inherits a configuration-dependent metric $\gC:=\cfgP(\gS)$ through the embedding $\varphi$ via pullback. Instead, $\cl{B}$ is equipped intrinsically with the mass form $\mFC\in\spFrmB{n}$.

The material (Lagrangian) velocity of the body is denoted by $\map{\vM_t}{\cl{B}}{T\cl{S}}$ while the spatial (Eulerian) and convective velocities are denoted by $\vS_t\in \spVecS$ and $\vC_t\in \spVecB$, respectively.
The three representations of the velocity are related by
\begin{equation}\label{eq:vel_relations}
	\vS_t := \vM_t \circ \cfgInvT, \qquad\qquad \vC_t := T\cfgInvT \circ \vM_t = T\cfgInvT \circ \vS_t \circ \cfgT,
\end{equation}
where $\map{T\cfgInvT}{T\cl{S}}{T\cl{B}}$ denotes the tangent map of $\cfgInvT$.
The convective, material, and spatial mass density functions will be denoted by $\mDC, \mDM \in \spFn{\cl{B}},\mDS \in \spFn{\cl{S}}$ which are related by
$$\mDM = J_{\cfgT} \mDC_t = J_{\cfgT}(\mDS_t \circ \cfgT),$$
with $J_{\cfgT}\in \spFn{\cl{B}}$ denoting the Jacobian of $\cfgT$.

The material equations of motion governing the evolution of $(\varphi,\vM)\in T\cfgSp$ on the tangent bundle of $\cfgSp$ are given by
\begin{align}
	\partial_t \varphi =& \vM\\
	D_t \vM =& \frac{1}{\mDM} \divrM(\stSigM)\label{eq:EoM_M_vM}\\
	D_t F =& \nabM \vM\\
	\stSigM =& \mDM\ \left(\parD{\tilde{e}}{F}\right)^\sharp(F), \label{eq:EoM_M_sig}
\end{align}
where $\stSigM\in \spVec{\cl{B}\otimes\varphi^*T\cl{S}}$ denotes the 1\textsuperscript{st} Piola-Kirchhoff stress, $F:= T\varphi$ denotes the deformation gradient, $D_t$ denotes the material derivative, and $\tilde{e}(F) \in \spFn{\cl{B}}$ denotes the material internal energy function. 
Furthermore, $\nabM$ denotes the material covariant derivative, $\divrM:=\tr\circ \nabM$ denotes its corresponding divergence operator, and the index raising (i.e. $\sharp$ map) in (\ref{eq:EoM_M_sig}) is with respect to $ \gS^{ij} \circ \cfgT \in \spFn{\cl{B}}$.

The spatial equations of motion governing the evolution of $(\mDS,\vS) \in \spFn{\cl{S}}\times\spVecS$ are given by
\begin{align}
	\partial_t \mDS =& -\divrS(\mDS\vS) \label{eq:EoM_S_mDS}\\
	\partial_t \vS =& - \nabS_{\vS}\vS + \frac{1}{\mDS} \divrS(\stSigS)\label{eq:EoM_S_vS}\\
	\stSigS =&  2 \mDS \parD{e}{\gS}(F,\gS). \label{eq:EoM_S_Sig}
\end{align}
where $\stSigS\in \spVec{_0^2\cl{S}}$ denotes the symmetric Cauchy stress tensor field and $e(F,\gS) \in \spFn{\cl{S}}$ denotes the spatial internal energy function. We denote by $\nabS$ the spatial covariant derivative and by $\divrS:= \tr \circ \nabS$ its corresponding divergence operator.

Finally, the convective counterparts of (\ref{eq:EoM_S_mDS}-\ref{eq:EoM_S_Sig}) that govern the evolution of $(\mDC,\vC)\in \spFn{\cl{B}}\times\spVecB$ are given by
\begin{align}
	\partial_t \mDC =& - \mDC \divrC(\vC) \label{eq:mDC_dot}\\
	\partial_t \vC =& - \nabC_{\vC}\vC + \frac{1}{\mDC} \divrC(\stSigC)\label{eq:EoM_C_vC}\\
	\partial_t \gC =& \Lie{\vC}{\gC} \label{eq:gC_dot}\\
		\stSigC =& 2 \mDC \parD{\hat{e}}{\gC} (\gC) \label{eq:DE_C_std}
\end{align}
where $\stSigC\in \spVec{_0^2\cl{B}}$ denotes the symmetric convective stress tensor field, $\hat{e}(\gC)\in\spFn{\cl{B}}$ denotes the convective internal energy function, and $\gC := \cfgP(\gS) \in \spMetB$ denotes the convective metric.
Furthermore, $\nabC$ denotes the convective covariant derivative, and $\divrC:=\tr\circ \nabC$ denotes its corresponding divergence operator.
We denote by $\spMetB$ the space of Riemannian metrics on $\cl{B}$, which represents the space of deformations.
As described in \cite{rashad2023intrinsic,Fiala2016,Kolev2021ObjectiveMetrics} and discussed later, strain is
geometrically represented as a geodesic on this space. Thus, $\spMetB$ will play an important role for constitutive modeling in Sec. 5.

%
%
%

\subsection{Exterior calculus} \label{sec:geom_mech_ext_calc}

The geometric formulation presented above relies on the representation of the physical variables appearing in the equations of motion as thermodynamically-intensive (i.e. volume independent) variables.
The mathematical objects used in such formulation consist of scalar fields, vector-fields, and second-rank tensor-fields.
In the spatial, convective, and material representations these respectively are $(\mDS,\vS,\stSigS)$, $(\mDC,\vC,\stSigC)$, and $(\varphi,\vM,\stSigM)$, in addition to the variables that characterize elastic deformation.

A key limitation associated with the tensor calculus operations used in such geometric formulation is that the underlying geometric and topological structure of the equations of motion are entangled together.
Exterior calculus on the other hand provides an elegant machinery for distinguishing between topology and geometry that can highlight the rich structures hidden in the partial differential equations above.
The physical variables of continuum mechanics are represented in this exterior calculus formulation using differential forms, both scalar-valued and bundle-valued.

The space of scalar-valued differential $k$-forms on an $n$-dimensional manifold $M$ is denoted by $\spFrm{M}{k}$. A k-form $\phi^k\in \spFrm{M}{k}$ is locally a totally asymmetric $(0,k)$ tensor with values in $\bb{R}$ at the point $p\in M$.
One has that 0-forms are isomorphic to scalar fields and 1-forms isomorphic to covector-fields, i.e. $\spFrm{M}{0}\cong \spFn{M}$ and $\spFrm{M}{1}\cong \spVec{^*M}$.
We denote the \textit{wedge} product, \textit{exterior} derivative, and \textit{Hodge star} operators, respectively by
$$\map{\wedge}{\spFrm{M}{k}\times\spFrm{M}{l}}{\spFrm{M}{k+l}}$$
$$\map{\extd}{\spFrm{M}{k}}{\spFrm{M}{k+1}} \qquad \qquad \map{\star}{\spFrm{M}{k}}{\spFrm{M}{n-k}}.$$
Other operations will be introduced later in the sequel.

The space of bundle-valued differential $k$-forms on $M$ will be denoted by $\spFrm{M;\bb{E}}{k}$ where $\bb{E}$ denotes any vector bundle over $M$.
A $k$-form $\Phi^k\in \spFrm{M;\bb{E}}{k}$ is locally a totally asymmetric $(0,k)$ tensor that takes values in the fiber $\bb{E}$ at the point $p\in M$.
One has that $\spFrm{M;\bb{E}}{0} \cong \Gamma(\bb{E})$, thus $\spFrm{M;TM}{0}$ and $\spFrm{M;T^*M}{0}$ are isomorphic to vector-fields and covector-fields, respectively.
Throughout this paper, we shall refer to elements of $\spFrm{M;TM}{0}$ and $\spFrm{M;T^*M}{0}$ as vector-valued and covector-valued forms, respectively.
A trivial vector-(or covector-) valued $k$-form is one which is equivalent to the tensor product of a vector-field (or covector-field) and an ordinary $k$-form.
For example, we say that $\zeta \in \spFrmM{k}{TM}$ is trivial if it is composed of a  vector-field $u \in \Gamma(TM)$ and a $k$-form $\alpha^k \in \spFrm{M}{k}$ such that $\zeta = \alpha^k\otimes u.$

Bundle-valued forms can also elegantly incorporate two-point tensor fields that appear in the material representation of continuum mechanics.
We shall denote the space of such forms by $\Omega^k_\varphi(M;\bb{F})$ where $\bb{F}$ is any vector bundle over the manifold $N$ related to $M$ via the map $\map{\varphi}{M}{N}$.
We denote the \textit{wedge-dot} product, \textit{exterior-covariant} derivative, and \textit{complementary Hodge star} operators, respectively by
$$\map{\wedgedot}{\spFrm{M;\bb{E}}{k}\times\spFrm{M;\bb{E}^*}{l}}{\spFrm{M}{k+l}}$$
$$\map{\extcdS}{\spFrm{M;\bb{E}}{k}}{\spFrm{M;\bb{E}}{k+1}} \qquad \qquad \map{\hodgeS}{\spFrm{M;\bb{E}}{k}}{\spFrm{M;\bb{E}^*}{n-k}}.$$
The exterior covariant derivative $\extcdS$ generalizes the standard covariant derivative $\nabS$ and for the case $k=0$ one has that $\extcdS = \nabS$.
The exact definitions and coordinate-based expressions of the above operators\footnote{In \cite{rashad2023intrinsic} we used the notation $\star^\flat$ to denote the complementary Hodge-star operator instead of $\hodgeS$, but we opt for the latter in this work because it is more expressive for general bundle-valued forms.} can be found in \cite{rashad2023intrinsic}.

In our exterior calculus formulation of continuum mechanics we shall represent physical variables as thermodynamically-extensive variables using a combination of scalar-valued and bundle-valued forms.
In particular, we shall represent the spatial mass, momentum, and stress by 
$$(\mFS,\momS,\stS) \in \spFrmS{n}\times\spcovFrmS{n}\times\spcovFrmS{n-1}.$$
Furthermore, we shall represent the convective metric, momentum, and stress variables by
$$(\gC,\momC,\stC) \in \spMetB\times\spcovFrmB{n}\times\spcovFrmB{n-1},$$
with the tangent and cotangent bundles of $\spMetB$ identified with $\spcovFrmBsub{1}{sym}\subset\spcovFrmB{1}$ and $\spvecFrmBsub{n-1}{sym}\subset \spvecFrmB{n-1}$, respectively. Working with symmetric subsets of the above spaces follows from the geometric structure of $\spMetB$ that comprises symmetric positive-definite tensor-fields \cite{rashad2023intrinsic}.
Finally, we shall represent the configuration, material momentum and stress variables by
$$(\varphi,\momM,\stM) \in \spC\times\spcovFrmPhi{n}\times\spcovFrmPhi{n-1},$$
with the tangent and cotangent bundles of $\spC$ identified with $\spvecFrmPhi{0}$ and $\spcovFrmPhi{n}$, respectively.

\subsection{Related works}

Our geometric formulation based on exterior calculus relies on the work of \cite{Frankel2019,Kanso2007OnMechanics,Gilbert2023} where stress and momentum are treated as bundle-valued differential forms, which is the natural mathematical representation of these physical variables from a topological perspective.

The majority of geometric formulations of continuum mechanics are based on tensor calculus.  
In the Hamiltonian mechanics literature, solid and fluid mechanics have been studied extensively using Poisson reduction in \cite{Simo1988ThePlates,marsden1982hamiltonian,marsden1983coadjoint,marsden1984semidirect,holm1986hamiltonian,Lewis1986TheProblems,Mazer1989HamiltonianFlows}.
On the Lagrangian side, Euler-Poincare reduction has been also applied e.g. in \cite{holm1998euler,Gay-Balmaz2012}.
In contrast to these works, our approach to deriving the governing equations of motion in the \pH framework is based on Dirac structures and uses the bottom-up philosophy of port-based modeling to represent the dynamics decomposed into a number of energetic subsystems connected to each other using so called \textit{power ports}, or ports for short.

In previous works of ours, we developed \pH models for Euler equations (i.e. ideal fluid flow) using scalar-valued forms only in \cite{Rashad2021Port-HamiltonianEnergy,Rashad2021Port-HamiltonianFlow} which was later extended to Navier-Stokes equations in \cite{califano2021geometric,Rashad2021ExteriorModels}, to Fourier-Navier-Stokes in \cite{califano2022differential}, and fluid-structure interaction in \cite{califano2022energetic}.
In these extensions, the momentum balance was usually written in terms of scalar-valued forms with only the stress treated as a bundle-valued form.
On the other hand, the \pH model we derive in this paper will represent both momentum and momentum flux as bundle-valued forms which will highlight more their geometric characteristics.

Other works in the \pH literature that treat fluid mechanics and simplified elastic models for beams and plates can be found in \cite{cheng2023port,Trivedi2016,Brugnoli2019Port-HamiltonianPlates,Brugnoli2021MixedBeams,Macchelli2004b}
In contrast to these works that usually mathematically manipulate the governing equations of motion into a \pH form, we shall use in this work Hamiltonian reduction theory to derive the \pH models of continuum mechanics from first principles in a geometric coordinate-free manner.

%
%
%
%
%

\section{Port-Hamiltonian model of nonlinear elasticity}\label{sec:pH_modeling_solid}

\begin{figure}
	\centering
	\includegraphics[width=\columnwidth]{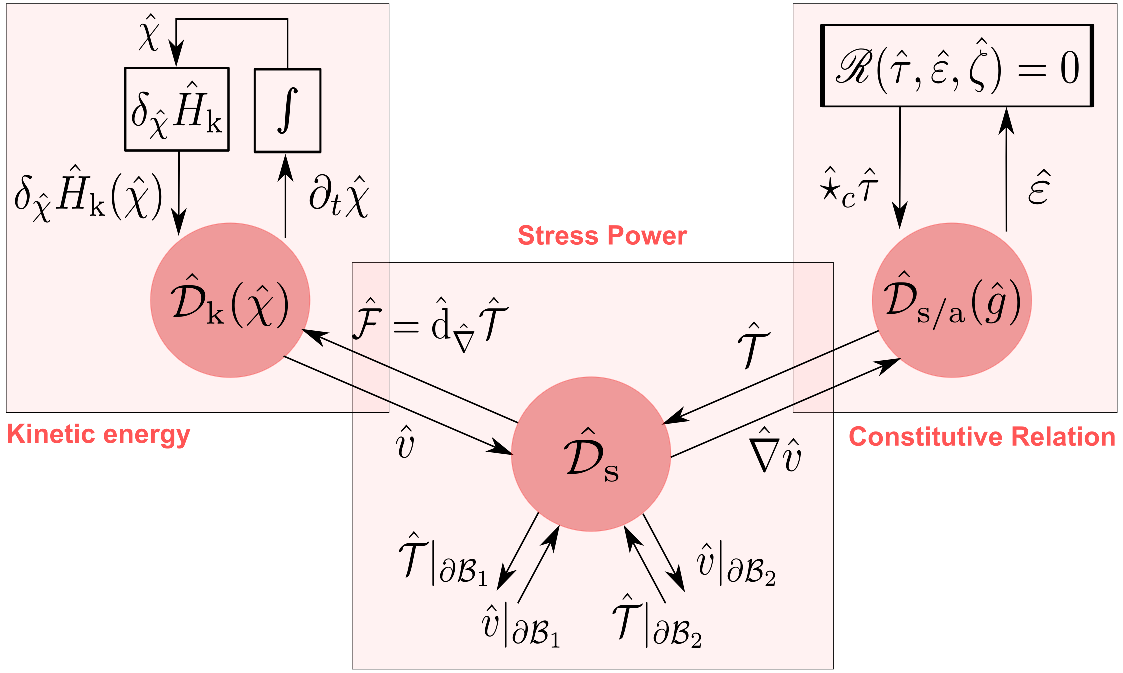}
	\caption{Convective \pH model of nonlinear elasticity showing the kinetic energy, stress power, and constitutive relations subsystems.}
	\label{fig:pH_model_conv}
\end{figure}

In this section we present the \pH dynamic model of nonlinear elasticity in the material, spatial, and convective representations. The unique characteristic of the \pH paradigm that distinguishes it from the standard Hamiltonian approach in \cite{Simo1988ThePlates,Lewis1986TheProblems,Mazer1989HamiltonianFlows} is that the physical system is modeled as a network of energetic units interconnected by ports.

Each port $(f,e) \in V\times V^*$ consists of a pair of power-conjugate variables $e\in V^*$ and $f\in V$, referred to, respectively, as effort and flow variables in port-based modeling.
In our work, the space of flow variables $V$ will be given by vector-valued forms while its dual space $V^*$ will be given by covector-valued pseudo-forms of complementary degree.
The duality pairing of the port variables $e$ and $f$ is denoted by $\duPair{e}{f}{M}:= \int_M f\wedgedot e$ and characterizes the power flowing in the port $(f,e)$. One has that $M=\cl{B}$ in case $(f,e)$ are represented in the material or convective representation, while $M=\cl{S}$ in the case of the spatial representation. On some occasions, we will also represent the port variables using scalar valued forms. With a slight abuse of notation, we will also denote their corresponding duality pairing by $\duPair{e}{f}{M}:= \int_M f\wedge e$, as it will be clear from the context.

In the standard Hamiltonian formalism, one follows a top-down approach by starting from a Hamiltonian functional, that characterizes the elastic body's kinetic energy, strain energy, and boundary conditions. The governing equations are represented using a Poisson structure such that the Hamiltonian is conserved \cite{Simo1988ThePlates}.
On the other hand, in the \pH formalism, we follow a bottom-up approach by representing each energetic subsystem separate from the others using a Dirac structure that characterizes the subsystem's unique power balance.
For nonlinear elasticity, its \pH model will consist of 1) a kinetic energy subsystem, 2) a stress power subsystem, and 3) a stress constitutive relation subsystem, as depicted in Fig. \ref{fig:pH_model_conv}.
By interconnecting these three subsystems to each other using power ports, the result is a decomposed model of the overall system that explicates the underlying energetic structure of the theory and emphasizes the natural duality of the port-variables.

We shall present the three representations of the aforementioned energetic units using the exterior calculus formulation introduced earlier.
We also show how one recovers the mass, momentum and energy balance laws from the presented \pH models. In fact the \pH procedure we will present will allow us to derive the equations of motion from first principles using Hamiltonian reduction techniques.
The reader is assumed to be familiar with the standard Hamiltonian framework of mechanics e.g. in \cite{marsden2013introduction,holm2009geometric}.

\subsection*{Frechet and variational derivatives}

Before proceeding, we introduce the Frechet derivative and variational derivatives of functionals of differential forms which will be used extensively in this paper.
Consider any functional $\map{\mathscr{F}}{\spFrmM{k}{\bb{E}}}{\bb{R}}$ with $\bb{E}$ denoting a vector bundle over the manifold $M$.
The Frechet derivative $\extD_\alpha\mathscr{F}$ of $\mathscr{F}$ with respect to $\alpha \in \spFrmM{k}{\bb{E}}$ is defined for any $\delta\alpha\in \spFrmM{k}{\bb{E}}$ as the functional
\begin{equation}
	\extD_\alpha\mathscr{F}[\alpha,\delta\alpha] := \dds \mathscr{F}[\alpha + s \delta\alpha],
\end{equation}
while the variational derivative $\map{\varD{\mathscr{F}}{\alpha}}{\spFrmM{k}{\bb{E}}}{\spFrmM{n-k}{\bb{E}^*}}$ is defined by
\begin{equation}
	 \int_M \delta\alpha \wedgedot \varD{\mathscr{F}}{\alpha}(\alpha) = \extD_\alpha\mathscr{F}[\alpha,\delta\alpha],
\end{equation}
for any $\delta\alpha\in \spFrmM{k}{\bb{E}}$.

The above definitions can be trivially extended to define 1) partial Frechet derivatives, 2) partial variational derivatives, and 3) functionals of scalar-valued forms. Furthermore, we shall deal later with the case that $\alpha$ belongs to an infinite dimensional manifold or a bundle, which will be essential for the \pH modeling procedure.
For notational simplicity in these more complex cases, we shall opt for denoting $\extD_\alpha\mathscr{F}[\alpha,\delta\alpha]$ as $\extD_\alpha\mathscr{F}(\alpha) \cdot\delta\alpha$ following \cite{Marsden1994} and usually denote $\varD{\mathscr{F}}{\alpha}(\alpha)$ simply by $\varD{\mathscr{F}}{\alpha}$.

\subsection{Kinetic energy subsystem}

For an elastic body undergoing a motion described by $\map{\cfgT}{\cl{B}}{\cl{S}}$, its kinetic energy as a function of time is expressed using spatial, material, and convective variables, respectively as
\begin{equation}\label{eq:kinetic_energy}
		\subTxt{E}{kin}(t) = \int_\cl{S} \half \gS(\vS_t,\vS_t) \mFS_t = \int_\cl{B} \half (\gS\circ \cfgT)(\vM_t,\vM_t) \mFM = \int_\cl{B} \half \gC_t(\vC_t,\vC_t) \mFC,
\end{equation}
where $\mFM = \mFC \in \spFrmB{n}$ denote the (time-independent) intrinsic mass form associated o the body manifold.
The second equality in (\ref{eq:kinetic_energy}) follows from the change of variables formula and the third one can be seen in local components:
$$(\gS\circ \cfgT)(\vM,\vM) = (\gS_{ij}\circ\cfgT) \vM^i\vM^j = (\gS_{ij}\circ\cfgT)F^i_I F^j_J \vC^I\vC^J = \gC_{IJ}\vC^I\vC^J = \gC(\vC,\vC).$$

In the spatial representation, the kinetic energy has state dependency on the spatial mass form and velocity field $(\mFS_t,\vS_t)\in \spFrmS{n}\times \spVecS$ and parametric dependency on the Riemannian metric $\gS\in \cl{M}(\cl{S})$.
In the material representation, it has state dependency on the configuration and material velocity field $(\cfgT,\vM_t) \in T\mathscr{C}$ and parametric dependency on $(\gS,\mFM) \in \cl{M}(\cl{S})\times \spFrmB{n}$.
In the convective representation, it has state dependency on the convective metric and velocity field $(\gC_t,\vC_t)\in\spMetB\times \spVecB$ and parametric dependency on $\mFC\in \spFrmB{n}$.

The kinetic energy gives rise to three Lagrangian functionals on the state spaces detailed above.
In what follows we shall derive their Hamiltonian counterparts and the associated energetic structures that characterize conservation of energy.
We start with the material representation exploiting the canonical Poisson structure on the cotangent bundle $T^*\mathscr{C}$.
Then we present the Hamiltonian reduction procedure to the spatial representation, followed by the convective one.
We perform this reduction by direct calculation similar to \cite{Simo1988ThePlates,Mazer1989HamiltonianFlows} with two distinctions.
First, instead of using tensor calculus, our formulation is based fully on exterior calculus which is (arguably) more mathematically elegant because it explicates the intrinsic relation between differential forms and integration.
Second, instead of applying Hamiltonian reduction to the total energy of the system, we follow the bottom-up approach described earlier which simplifies the reduction process greatly as we deal with the kinetic energy separately.

\subsubsection*{Material Poisson structure}

Let $\spPkinM:=\cl{M}(\cl{S})\times \spFrmB{n}$ denote the space of parameters in the material representation and its state space given by the tangent bundle $T\mathscr{C}$.
The material Lagrangian functional $\map{\LkinM}{T\mathscr{C}\times \spPkinM}{\bb{R}}$ is defined by
\begin{equation}\label{eq:LkinM}
	\LkinM[\varphi,\vM;\gS,\mFM] := \intB \half (g\circ\varphi)(\vM,\vM) \mFM,
\end{equation}
which depends parametrically on $(\gS,\mFM) \in \spPkinM$ while its time dependency is implicit through its state variables $(\varphi,\vM) \in T\mathscr{C}$.
In what follows, we will omit the time dependency and usually suppress the parametric dependency unless needed.
It is important to note that $T\mathscr{C}$ is not a product space and thus caution should be made when defining variations \cite{Simo1988ThePlates}.

In terms of the exterior calculus construction in Sec. \ref{sec:geom_mech_ext_calc}, we can rewrite (\ref{eq:LkinM}) as
\begin{equation}\label{eq:LkinM_ext}
	\LkinM[\varphi,\vM] := \intB \half \vM \wedgedot \hodgeM \vM,
\end{equation}
where we consider $\vM\in \spvecFrmPhi{0}$ while the state dependency on $\varphi$ and parametric dependency on $\gS$ and $\mFM$ have been absorbed in the material Hodge star operator $\map{\hodgeM}{\spvecFrmPhi{k}}{\spcovFrmPhi{n-k}}$.


By applying a partial Legendre transformation on $\vM$, we define on the cotangent bundle $T^*\mathscr{C} =: \spXkinM$ the material Hamiltonian functional $\map{\HkinM}{\spXkinM\times \spPkinM}{\bb{R}}$ as
\begin{equation}\label{eq:HkinM}
	\HkinM[\varphi,\momM] := \intB \half \invhodgeM \momM \wedgedot \momM,
\end{equation}
where $\momM:= \hodgeM \vM \in \spcovFrmPhi{n} $ denotes the material momentum of the body and  $\map{\invhodgeM}{\spcovFrmPhi{k}}{\spvecFrmPhi{n-k}}$ is the inverse map of $\hodgeM$.

\begin{remark}\label{remark:momM}
	As shown in\cite{rashad2023intrinsic}, $\momM$ is locally expressed as $\momM = \vM_i \mFM\otimes e^i$ where $\vM_i := (\gS_{ij}\circ \varphi) \vM^j \in \spFrmB{0}$ denote the components of the covector velocity field $\vfM$. Note that $\momM$ is not a trivial covector-valued top-form, i.e. $\momM \neq \mFM\otimes\vfM$. Indeed both expressions are equivalent when treated as a two-point tensor, \cf \cite[eq. 16]{Kanso2007OnMechanics}. However, as a bundle-valued form its form part is given by $\vM_i \mFM \in \spFrmB{n}$.
\end{remark}

In the theory of Hamiltonian mechanics, the canonical equations of motion on $\spXkinM$ are given implicitly by $\dot{\scrTld{F}} = {}^m\{\scrTld{F},\HkinM\}$, where $\map{\scrTld{F}}{\spXkinM}{\bb{R}}$ is an arbitrary functional and $\map{{}^m\{\cdot,\cdot\}}{\spFn{\spXkinM}\times \spFn{\spXkinM}}{\bb{R}}$ denotes the canonical Poisson bracket in the material representation, to be introduced shortly.
To write the explicit expressions of the above canonical equations and bracket, it is essential to introduce the partial Frechet and variational derivatives of functionals on the cotangent bundle $\spXkinM$.

In our work we identify tangent and cotangent spaces of $\spC$ by $T_{\varphi}\spC \cong \spvecFrmPhi{0}$ and $T_{\varphi}^*\spC \cong \spcovFrmPhi{n}$ \cite{rashad2023intrinsic}. Consequently, we have that
\begin{align*}
	T_{\chiM}\spXkinM \cong&\ \spvecFrmPhi{0} \times \spcovFrmPhi{n} \times \spvecFrmPhibnd{0} \\
	T^*_{\chiM}\spXkinM \cong&\ \spcovFrmPhi{n} \times \spvecFrmPhi{0} \times \spcovFrmPhibnd{n-1}.
\end{align*}
A functional $\map{\scrTld{F}}{\spXkinM}{\bb{R}}$ is said to have partial variational derivatives, if for any $(\varphi,\momM) \in \spXkinM$ there exists the bundle-valued forms
$$\varD{\scrTld{F}}{\varphi} \in \spcovFrmPhi{n}, \qquad \varD{\scrTld{F}}{\momM} \in \spvecFrmPhi{0}, \qquad \varDbnd{\scrTld{F}}{\varphi} \in \spcovFrmPhibnd{n-1}, $$
that satisfy for any $(\delta\tilde{\varphi},\delta\momM) \in \spvecFrmPhi{0}\times\spcovFrmPhi{n}$
\begin{align}
	\extD_\varphi\scrTld{F}(\varphi,\momM) \cdot \delta\tilde{\varphi} =& \intB \delta\tilde{\varphi} \wedgedot \varD{\scrTld{F}}{\varphi} + \int_{\bndB} \delta\tilde{\varphi}|_{\Bbound} \wedgedot \varDbnd{\scrTld{F}}{\varphi} \label{eq:FrechD_1_M}\\
		\extD_{\momM}\scrTld{F}(\varphi,\momM) \cdot \delta\momM =& \intB \delta\momM \wedgedot \varD{\scrTld{F}}{\momM}, \label{eq:FrechD_2_M}
\end{align}
where for any $\tilde{\alpha} \in \spvecFrmPhi{k}$, we denote by $\tilde{\alpha}|_{\Bbound} := i_\text{f}^*(\tilde{\alpha}) \in \spvecFrmPhibnd{k}$ the trace of the form part of $\tilde{\alpha}$ to the boundary where $\map{i}{\Bbound}{\cl{B}}$ denotes the inclusion map and $i_\text{f}^*$ denotes the pullback of the form part of a bundle-valued form.

The partial Frechet derivative $\extD_{\momM}\scrTld{F}$ is defined such that
\begin{equation}\label{eq:Frech_F_M}
	\extD_{\momM}\scrTld{F}(\varphi,\momM) \cdot \delta\momM := \dds \scrTld{F}[\varphi,\momM + s \delta\momM].
\end{equation}
Intuitively, $\extD_{\momM}\scrTld{F}$ accounts to varying $\momM$ as a covector in $T^*_\varphi\spC$ while keeping $\varphi$ fixed, i.e. a fiber derivative.
On the other hand, to define the partial Frechet derivative $\extD_\varphi\scrTld{F}$, one must intuitively fix the covector $\momM$ while allowing the base point $\varphi$ to vary.
This must be done with caution since $\spXkinM = T^*\spC$ is not a product space similar to $T\spC$.	
Let the variation $\delphiM \in T_\varphi\spC \cong \spvecFrmPhi{0}$ be the tangent vector to the smooth curve $s \mapsto \varphi_s \in \spC$ at $s=0$, i.e. $\delphiM := \dds \varphi_s$ with $\varphi_s|_{s=0} = \varphi$.
The partial Frechet derivative $\extD_{\varphi}\scrTld{F}$ is defined such that
\begin{equation}\label{eq:Frech_F_phi}
	\extD_{\varphi}\scrTld{F}(\varphi,\momM) \cdot \delphiM := \dds \scrTld{F}[\varphi_s,\momM_s],
\end{equation}
where $\momM_s$ would be the induced variation on the covector in $T_{\varphi}^*\spC$ due to varying $\varphi$.
The details of the above construction can be found in \cite{Lewis1986TheProblems,Mazer1989HamiltonianFlows} which we shall refer to later.
Consequently, the rate of change of any $\map{\scrTld{F}}{\spXkinM}{\bb{R}}$ with respect to time is expressed as
\begin{equation}\label{eq:Fdot_M}
	\dot{\scrTld{F}} = \intB \partial_t \varphi \wedgedot \varD{\scrTld{F}}{\varphi} + \extD_t\momM \wedgedot \varD{\scrTld{F}}{\momM} +  \int_{\bndB} \partial_t\varphi|_{\Bbound} \wedgedot \varDbnd{\scrTld{F}}{\varphi},
\end{equation}
with $\partial_t \varphi \in \spvecFrmPhi{0}$ and $\extD_t \momM \in \spcovFrmPhi{n}$.


The canonical Poisson bracket on the cotangent bundle $\spXkinM$ then takes the form\cite{Lewis1986TheProblems}
\begin{equation}\label{eq:can_poisson}
	\begin{split}
		{}^m\{\scrTld{F},\scrTld{G}\} :=& 
		\intB \varD{\scrTld{G}}{\momM} \wedgedot \varD{\scrTld{F}}{\varphi} - \varD{\scrTld{F}}{\momM} \wedgedot \varD{\scrTld{G}}{\varphi} \\
		&+ \int_{\bndB} \varD{\scrTld{G}}{\momM}|_{\bndB} \wedgedot \varDbnd{\scrTld{F}}{\varphi} - \varD{\scrTld{F}}{\momM}|_{\bndB} \wedgedot \varDbnd{\scrTld{G}}{\varphi}.
	\end{split}
\end{equation}
Using the above construction, the canonical equations of motion are given explicitly by the following result.

\begin{proposition}\label{prop:Ham_eqns_M}
	For the kinetic energy as a Hamiltonian functional with state variables $ \chiM := (\varphi,\momM) \in \spXkinM$, the equations of motion in the material representation are given by:
	\begin{equation}\label{eq:kin_dynamics_closed_M}
		\TwoVec{\partial_t \varphi}{\extD_t \momM} = {\TwoTwoMat{0}{1}{-1}{0}} \TwoVec{\varD{\HkinM}{\varphi}}{\varD{\HkinM}{\momM}}, \qquad \partial_t \varphi|_{\Bbound} = \varD{\HkinM}{\momM}|_{\Bbound} , \qquad 0 = \varDbnd{\HkinM}{\varphi}.
	\end{equation}
	The Hamiltonian functional (\ref{eq:HkinM}) admits its rate of change such that along trajectories $(\varphi(t),\momM(t))$ of (\ref{eq:kin_dynamics_closed_M}), it holds that
	\begin{equation}\label{eq:HkinM_closed}
		\HkinMdot = 0.
	\end{equation}
\end{proposition}
\begin{proof}
	First, By comparing (\ref{eq:Fdot_M}) and (\ref{eq:can_poisson}), one can easily deduce the equations of motion (\ref{eq:kin_dynamics_closed_M}).
	As for the energy balance (\ref{eq:HkinM_closed}), it follows immediately  from the skew-symmetry of the bracket (\ref{eq:can_poisson}) that $\HkinMdot = {}^m\{\HkinM,\HkinM\} = 0$.
	\qed
\end{proof}

\begin{corollary}\label{cor:varD_H_M}
	The variational derivatives of the functional $\HkinM$ in (\ref{eq:HkinM}) with respect to $\varphi \in \cfgSp$ and $\momM \in \spcovFrmPhi{n}$ are given respectively by
	$$\varD{\HkinM}{\varphi} = 0 \in \spcovFrmPhi{n}, \qquad \varD{\HkinM}{\momM} = \vM \in \spvecFrmPhi{0}, \qquad \varDbnd{\HkinM}{\varphi} = 0 \in \spcovFrmPhibnd{n-1}.$$
	Consequently, one can rewrite the Hamilton equations (\ref{eq:kin_dynamics_closed_M}) as
	$$\partial_t \varphi = \vM, \qquad \qquad D_t \momM = 0, \qquad\qquad \partial_t \varphi|_{\Bbound} = \vM|_{\Bbound},$$
	where the first and third equations represent the kinematic definition of the material velocity while the second represents the conservation of momentum.
\end{corollary}
\begin{proof}
	Second, By construction from the Legendre transformation, we have that $\varD{\HkinM}{\momM} = \vM \in \spvecFrmPhi{0}.$
	In \cite{Simo1988ThePlates}, the procedure for computing $\varD{\HkinM}{\varphi}$ and $\varDbnd{\HkinM}{\varphi}$ have been detailed. The construction is involved due to addressing the whole dynamic system at once. However, in our case where the Hamiltonian functional consists only of kinetic energy, and one can see from \cite[Prop. 3.1]{Simo1988ThePlates} that both $\varD{\HkinM}{\varphi}$ and $\varDbnd{\HkinM}{\varphi}$ are equal to $0$. 
	\qed
\end{proof}

%

In summary, the \textit{closed} Hamiltonian system, defined by the tuple $(\spXkinM,\HkinM,{}^m\{\cdot,\cdot\})$, describes the conservation of energy and the corresponding evolution of the state $\chiM\in \spXkinM$ as a conservative system isolated from any energy exchange with its surroundings.

\begin{remark}
	Note that the Hamilton equations of motion (\ref{eq:kin_dynamics_closed_M}) are valid for any continuum (whether solid or fluid) and for any Hamiltonian $\map{\tilde{H}}{T^*\spC}{\bb{R}}$. 
	However, the condition $\varDbnd{\tilde{H}}{\varphi} = 0$ immediately restricts the class of admissible functions such that the Hamiltonian is conserved. 
	It will be shown later in the spatial representation that this condition amounts to having vanishing momentum flux at the image of the boundary in the ambient space.
	While this condition is naturally satisfied for solid mechanics in Corollary \ref{cor:varD_H_M}, it will have interesting implications for the case of fluid mechanics that we shall discuss later in Sec. \ref{sec:pH_fluid}.
\end{remark}

\subsubsection*{Material Dirac structure}

Dirac structures are geometric objects that generalize symplectic and Poisson structures and play a central role in the theory of \pH systems.
Let $T\spXkinM\oplus T^*\spXkinM$ denote the Whitney sum bundle over $\spXkinM = T^*\spC$, i.e. the bundle over the base $\spXkinM$ with fiber over $\chiM := (\varphi,\momM)$ equal to $T_{\chiM}\spXkinM\times T_{\chiM}^*\spXkinM$. 
As per the construction in (\ref{eq:FrechD_1_M}-\ref{eq:Fdot_M}), the duality pairing between any $\fM{}:=(\fM{\varphi},\fM{\momM},\fM{\varphi}^\cup) \in T_{\chiM}\spXkinM$ and $\eM{}:=(\eM{\varphi},\eM{\momM},\eM{\varphi}^\cup) \in T^*_{\chiM}\spXkinM$ is expressed as
$$\duPair{\eM{}}{\fM{}}{T\spXkinM} := \duPairB{\eM{\varphi}}{\fM{\varphi}} + \duPairB{\eM{\momM}}{\fM{\momM}} + \duPair{\eM{\varphi}^\cup}{\fM{\varphi}^\cup}{\bndB}.$$
A Dirac structure on $\spXkinM$ is defined as the sub-bundle $\tilde{\cl{D}} \subset T\spXkinM\oplus T^*\spXkinM$ such that $\tilde{\cl{D}} = \tilde{\cl{D}}^\perp$, with $\tilde{\cl{D}}^\perp$ denoting the annihilator with respect to the symmetric bilinear form:
$$\langle\langle  (\fM{}^1,\eM{}^1) | (\fM{}^2,\eM{}^2) \rangle\rangle := \langle{\eM{}^1}|{\fM{}^2}\rangle_{T\spXkinM} + \langle{\eM{}^2}|{\fM{}^1}\rangle_{T\spXkinM},$$ 
for all $(\fM{}^1,\eM{}^1) , (\fM{}^2,\eM{}^2) \in T_{\chiM}\spXkinM\times T_{\chiM}^*\spXkinM.$

The canonical Dirac structure on the cotangent bundle $\spXkinM$ corresponding to the Poisson bracket (\ref{eq:can_poisson}) will be denoted by $\tilde{J}_\text{k} \subset T\spXkinM\oplus T^*\spXkinM$ whose fiber at any $\chiM\in\spXkinM$ is given by:
\begin{equation}\label{eq:can_Dirac_kin_M}
	\begin{split}
		\tilde{J}_\text{k}(\chiM) := &\left \{ (\fM{\text{k}} ,\eM{\text{k}}) \in T_{\chiM}\spXkinM \times T_{\chiM}^*\spXkinM | \right.\\
		& \TwoVec{\fM{\varphi}}{\fM{\momM}} = \TwoTwoMat{0}{1}{-1}{0} \TwoVec{\eM{\varphi}}{\eM{\momM}} ,\\
		& \left. \fM{\varphi}^\cup = \eM{\momM}|_{\Bbound}, \qquad \qquad 0 = \eM{\varphi}^\cup \right\},
	\end{split}
\end{equation}
which characterizes the power balance $\duPair{\eM{\text{k}}}{\fM{\text{k}}}{T\spXkinM} = 0$.
The Hamilton equations in Prop. \ref{prop:Ham_eqns_M} immediately follow by setting $\eM{\text{k}} = (\varD{\HkinM}{\varphi},\varD{\HkinM}{\momM},\varDbnd{\HkinM}{\varphi})$ as inputs and $\fM{\text{k}} = (\partial_t\varphi,\partial_t \momM, \partial_t\varphi|_{\Bbound})$ as outputs yielding the conservation of the Hamiltonian (\ref{eq:HkinM}).

\begin{figure}
	\centering
	\includegraphics[width=1\columnwidth]{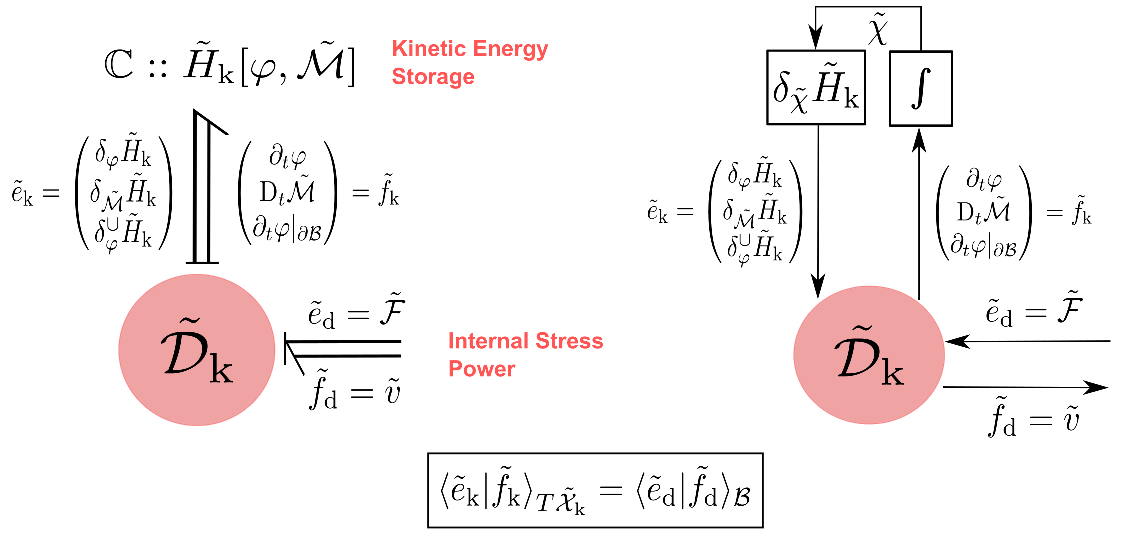}
	\caption{Material \pH model of the kinetic energy subsystem expressed in terms of block diagrams (left) and bond graphs (right)}
	\label{fig:pH_kin_M}
\end{figure}

Unlike Poisson structures that characterize conservative dynamical systems, Dirac structures can be used to characterize open dynamical systems that allow for non-zero energy exchange either through the boundary or within the spatial domain.
Let $\forceM \in \spcovFrmPhi{n}$ be an external body force field and $\intKinM := \spvecFrmPhi{0} \times \spcovFrmPhi{n}$ denote the interaction space such that $(\vM,\forceM) \in \intKinM$ define an open external port used to model the interaction with other subsystems within the spatial domain\footnote{We can also allow interaction through the boundary $\Bbound$, which will be done later for fluids.}. 
This interaction is characterized by the power given by the duality pairing  $\duPairB{\forceM}{\vM}$ and will be used later to model internal stress forces. In principle, one could add any body force, e.g. electrostatics or gravity in the same manner.

Now we can extend the canonical Dirac structure (\ref{eq:can_Dirac_kin_M}) with an interaction port to construct the Dirac structure 
$\diracKinM \subset \FKinM\oplus \FdlKinM$, with $\FKinM:= T\spXkinM \times \spvecFrmPhi{0}$ and its dual space $\FdlKinM:= T^*\spXkinM \times \spcovFrmPhi{n}$, whose fiber at any $\chiM\in\spXkinM$ is defined as
\begin{equation}\label{eq:Dirac_kin_M}
	\begin{split}
		\diracKinM(\chiM) = &\left \{ ((\fM{\text{k}}, \fM{\text{d}}),(\eM{\text{k}}, \eM{\text{d}})) \in \FKinM(\chiM)\times \FdlKinM(\chiM) | \right.\\
		& \TwoVec{\fM{\varphi}}{\fM{\momM}} = \TwoTwoMat{0}{1}{-1}{0} \TwoVec{\eM{\varphi}}{\eM{\momM}} + \TwoVec{0}{1} \eM{\text{d}} ,\\
		& \qquad\fM{\text{d}}= \begin{pmatrix} 0 & 1 \end{pmatrix} \TwoVec{\eM{\varphi}}{\eM{\momM}}, \\
		& \quad \left. \fM{\varphi}^\cup = \eM{\momM}|_{\Bbound}, \qquad \qquad 0 = \eM{\varphi}^\cup \right\},
	\end{split}
\end{equation}
with $\fM{\text{k}} = (\fM{\varphi},\fM{\momM},\fM{\varphi}^\cup) \in T_{\chiM}\spXkinM$ and $\eM{\text{k}} = (\eM{\varphi},\eM{\momM},\eM{\varphi}^\cup)\in T^*_{\chiM}\spXkinM$.
The Dirac structure (\ref{eq:Dirac_kin_M}) characterizes the power balance
\begin{equation}
	\duPairB{\eM{\varphi}}{\fM{\varphi}} + \duPairB{\eM{\momM}}{\fM{\momM}} + \duPair{\eM{\varphi}^\cup}{\fM{\varphi}^\cup}{\Bbound} = \duPairB{\eM{\text{d}}}{\fM{\text{d}}}.
\end{equation}

By setting the inputs to be $\eM{\text{k}} = ({\varD{\HkinM}{\varphi}},{\varD{\HkinM}{\momM}},{\varDbnd{\HkinM}{\varphi}})$ and $ \eM{\text{d}} = \forceM,$
and the outputs to be $ \fM{\text{k}}= ({\partial_t \varphi},{D_t \momM},\partial_t \varphi|_{\Bbound})$ and $ \fM{\text{d}}= \vM$, 
one extends (\ref{eq:kin_dynamics_closed_M}) to become a port-Hamiltonian system such that the energy balance (\ref{eq:HkinM_closed}) becomes
\begin{equation}\label{eq:HkinM_open}
	\HkinMdot = \duPairB{\forceM}{\vM},
\end{equation}
which states that the rate of change of kinetic energy of the elastic body is equal to the work done due to external stress forces. Consequently, the balance of momentum in Corollary \ref{cor:varD_H_M} takes the form $D_t \momM = \forceM$.

To summarize, with reference to Fig. \ref{fig:pH_kin_M}, the open \pH system is given by the tuple $(\spXkinM,\HkinM,\diracKinM,\intKinM)$ where $(\spXkinM,\HkinM)$ represent the storage of kinetic energy, $\intKinM$ represents the power supplied due to stress, and $\diracKinM$ represents the balance laws of the system.


\subsubsection*{Spatial Dirac structure}
Now we turn attention to the spatial counterpart of the kinetic-energy \pH system introduced above.
We start by transforming the material variables to spatial ones by the nonlinear diffeomorphism 
\begin{equation}\label{eq:diffeoMtoS}
	\fullmap{{}^s\Phi}{\spXkinM\times\spPkinM}{\spXkinS\times\spPkinS}{((\varphi,\momM),(\gS,\mFM))}{((\mFS,\momS),\gS)}
\end{equation}
with $\mFS := \varphi_*( \mFM )\in \spFrmS{n}$ denoting the extensive mass form and $\momS := \cfgFleg{f}(\momM) \in \spcovFrmS{n}$ denoting the extensive momentum form in the spatial representation.
We denote the spatial state space by $\spXkinS := \spFrmS{n}\times \spcovFrmS{n}$ and the spatial parameter space by $\spPkinS:= \cl{M}(\cl{S})$.

The spatial kinetic energy Hamiltonian $\map{\HkinS}{\spXkinS\times \spPkinS}{\bb{R}}$ is defined by $\HkinS[\mFS,\momS;\gS] = \HkinM[\varphi,\momM;\gS,\mFM]$ and is expressed as
\begin{equation}\label{eq:HkinS}
	\HkinS[\mFS,\momS] := \intS \half \invhodgeS \momS \wedgedot \momS,
\end{equation}
where the spatial Hodge star $\map{\hodgeS}{\spvecFrmS{k}}{\spcovFrmS{n-k}}$ incorporates a state dependency on the mass form $\mFS$ and a parametric dependency on the spatial metric $\gS$.
Note that in contrast to the material and convective cases which were represented by bundle-valued forms, the spatial case is represented by a combination of scalar-valued and bundle-valued forms.

\begin{remark}\label{remark:momS}
	In local coordinates we have that the spatial momentum is computed from its material counterpart (\cf. Remark \ref{remark:momM}) by
	$$\momS = \cfgFleg{f}(\momM)= \varphi_*(\vM_i \mFM) \otimes e^i = \varphi_*(\vM_i) \varphi_*(\mFM) \otimes e^i = \vS_i\mFS\otimes e^i,$$
	and thus $\momS$ is identified with the trivial covector-valued form $\mFS\otimes\vfS$ which is in contrast to $\momM$.
	
	One can easily show the equivalence between $\HkinS$ and the standard kinetic energy expression in (\ref{eq:kinetic_energy}) by
	$$\HkinS = \intS \half \vS \wedgedot \momS = \intS \half \vS \wedgedot (\mFS\otimes\vfS) =\intS \half \vfS(\vS)\mFS = \intS \half \gS(\vS,\vS)\mFS.$$\\
\end{remark}

In general, the spatial counterpart $\map{\scr{F}}{\spXkinS}{\bb{R}}$ of any material functional $\map{\tilde{\scr{F}}}{\spXkinM}{\bb{R}}$ can be defined such that $\scr{F}:= \tilde{\scr{F}} \circ {}^s\Phi^{-1}$, i.e.
\begin{equation}\label{eq:Funct_S_def}
	\scr{F}[\varphi_*( \mFM ),\cfgFleg{f}(\momM)] = \scrTld{F}[\varphi,\momM].
\end{equation}
The variational derivatives of the functional $\scr{F}$ with respect to $(\mFS,\momS)$ are the scalar-valued and bundle-valued forms $\varD{\scr{F}}{\mFS} \in \spFrmS{0}$ and $ \varD{\scr{F}}{\momS} \in \spvecFrmS{0},$ that satisfy for any $\delta\mFS\in \spFrmS{n},\delta\momS \in \spcovFrmS{n}$
\begin{equation}\label{eq:Frech_F_S}
	\extD_\mFS\scr{F}(\mFS,\momS) \cdot \delta\mFS = \intS \delta\mFS \wedge \varD{\scr{F}}{\mFS}  , \qquad 
	\extD_{\momS}\scr{F}(\mFS,\momS) \cdot \delta\momS = \intS \delta\momS \wedgedot \varD{\scr{F}}{\momS}.
\end{equation}
The rate of change of $\scr{F}$ with respect to time is then expressed as
\begin{equation}\label{eq:Funct_dot_S}
	\dot{\scr{F}} = \duPairS{\varD{\scr{F}}{\mFS}}{\partial_t \mFS} + \duPairS{\varD{\scr{F}}{\momS}}{\partial_t \momS},
\end{equation}
with $\partial_t \mFS \in \spFrmS{n}$ and $\partial_t \momS \in \spcovFrmS{n}$.

The pushforward and pullback maps of (\ref{eq:diffeoMtoS}) are given by the following important result which is essential for the derivation of the spatial Dirac structure. We recall first a number of operators for bundle-valued forms from \cite{rashad2023intrinsic}.
For any vector bundle $\bb{E}_\cl{S}$ over $\cl{S}$, we denote by $\map{\cfgPleg{f}}{\spFrm{\cl{S};\bb{E}_\cl{S}}{k}}{\Omega_\varphi^{k}(\cl{B};\bb{E}_\cl{S})}$ the pullback of the form part of a bundle-valued form, transforming it from the spatial representation to the material one, and denote by $\cfgFleg{f}$ its inverse (i.e. the pushforward map).
For any $u \in \spVec{\cl{S}}$, we denote by $\map{\iota_u}{\spFrm{\cl{S};\bb{E}_\cl{S}}{k}}{\spFrm{\cl{S};\bb{E}_\cl{S}}{k-1}}$ the interior product of a covector-valued form defined by inserting $u$ into its form part and we denote by $\Lie{u}{(\cdot)}$ the Lie derivative of any tensor-field along $u$.

\begin{proposition}\label{prop:pf_pl_M_to_S}
	Let $(\varphi,\momM) \in \spXkinM$ and $(\mFS,\momS)\in \spXkinS$ be related by the diffeomorphism ${}^s\Phi$ in (\ref{eq:diffeoMtoS}).
	
	i) The pushforward map of ${}^s\Phi$ is given by
	\begin{equation*}
		\fullmap{{}^s\Phi_*}{T\spXkinM}{T\spXkinS}{(\varphi,\momM,\delphiM,\delta\momM,\delphiM|_{\Bbound})}{(\mFS,\momS,\delta\mFS,\delta\momS)}
	\end{equation*}
	with 
	$$\delta\mFS = -\Lie{\delta\varphi}{\mFS} \in \spFrmS{n} ,\qquad \qquad \delta\momS = -\extcdS (\iota_{\delta\varphi}\momS) + \cfgFleg{f}(\delta\momM) \in \spcovFrmS{n},$$
	where $\delta\varphi = \cfgFleg{f}(\delphiM)\in \spvecFrmS{0} \cong \spVec{\cl{S}}$. 
	
	ii) The pullback map of ${}^s\Phi$ is given by
	\begin{equation*}
		\fullmap{{}^s\Phi^*}{T^*\spXkinS}{T^*\spXkinM}{(\mFS,\momS,\varD{\scr{F}}{\mFS},\varD{\scr{F}}{\momS})}{(\varphi,\momM,\varD{\scrTld{F}}{\varphi}, \varD{\scrTld{F}}{\momM}, \varDbnd{\scrTld{F}}{\varphi})}
	\end{equation*}
	with 
	\begin{align*}
		\varD{\scrTld{F}}{\varphi} =& \cfgPleg{f}(\mFS\otimes(\extd \varD{\scr{F}}{\mFS} + \nabS \varD{\scr{F}}{\momS} \wedgedot \vfS)) &&\in \spcovFrmPhi{n} \\
		\varD{\scrTld{F}}{\momM} =& \cfgPleg{f}(\varD{\scr{F}}{\momS}) &&\in \spvecFrmPhi{0}\\
		\varDbnd{\scrTld{F}}{\varphi} =& -\cfgPleg{f}(\varD{\scr{F}}{\mFS} \mFS + \iota_{\varD{\scr{F}}{\momS}}\momS)|_{\Bbound} &&\in \spcovFrmPhibnd{n-1}.
	\end{align*}
\end{proposition}
\begin{proof}
	See Appendix \ref{app:proof_1}.	
	\qed
\end{proof}

From the Hamiltonian reduction theory, the diffeomorphism (\ref{eq:diffeoMtoS}) induces a Poisson bracket on $\spXkinS$ using the canonical Poisson bracket (\ref{eq:can_poisson}) on $\spXkinM$ such that 
\begin{equation}\label{eq:Pois_brac_S}
	{}^s\{\scr{F},\scr{G}\} \circ {}^s \Phi := {}^m\{\scr{F}\circ {}^s \Phi,\scr{G}\circ {}^s \Phi\}.
\end{equation}
Similarly, ${}^s\Phi$ induces from (\ref{eq:Dirac_kin_M}) a spatial Dirac structure that extends the Poisson structure (\ref{eq:Pois_brac_S}) with an interaction port.
This spatial Dirac structure will be denoted by $\diracKinS$ and is derived in the following result.

\begin{theorem}\label{th:Dirac_kin_S}
	The spatial Dirac structure corresponding to (\ref{eq:Dirac_kin_M}) under the diffeomorphism (\ref{eq:diffeoMtoS}) is the sub-bundle
	$\diracKinS \subset \FKinS\oplus \FdlKinS$ with 
	\begin{align*}
		\FKinS:=& {}^s \Phi_*(T\spXkinM) \times \cfgFleg{f}(\spvecFrmPhi{0}) = T\spXkinS \times \spvecFrmS{0}\\
		\FdlKinS :=& ({}^s \Phi^*)^{-1}(T^*\spXkinM) \times \cfgFleg{f}(\spcovFrmPhi{n}) = T^*\spXkinS \times \spcovFrmS{n},
	\end{align*}
	whose fiber at any $\chiS:= (\mFS,\momS)\in \spXkinS$ is defined by
	\begin{equation}\label{eq:Dirac_kin_S}
		\begin{split}
			\diracKinS(\chiS) := &\left\{ ((\subTxt{f}{k},\subTxt{f}{d}),(\subTxt{e}{k},\subTxt{e}{d})) \in \FKinS(\chiS)\times \FdlKinS(\chiS) | \right.\\
			& \TwoVec{\fS{\mFS}}{\fS{\momS}} = \TwoTwoMat{0}{ - \extd \iota_{(\cdot)}\mFS}{-\mFS\otimes \extd(\cdot)}{-\Lie{(\cdot)}{\momS}} \TwoVec{\eS{\mFS}}{\eS{\momS}} + \TwoVec{0}{1} \subTxt{e}{d} ,\\
			& \qquad \subTxt{f}{d}= \begin{pmatrix} 0 & 1 \end{pmatrix} \TwoVec{\eS{\mFS}}{\eS{\momS}}, \\
			& \quad \left.	0 = (\eS{\mFS} \mFS + \iota_{\eS{\momS}} \momS)|_{\Sbound} \right\},
		\end{split}
	\end{equation}
	with $\subTxt{f}{k} = (\fS{\mFS},\fS{\momS}) \in T_\chiS\spXkinS$ and $\subTxt{e}{k} = (\eS{\mFS},\eS{\momS})\in T^*_\chiS\spXkinS$.
	The Dirac structure (\ref{eq:Dirac_kin_S}) characterizes the power balance
	\begin{equation}\label{eq:Dirac_Balance_S}
		\duPairS{\eS{\mFS}}{\fS{\mFS}} + \duPairS{\eS{\momS}}{\fS{\momS}} = \duPairS{\subTxt{e}{d}}{\subTxt{f}{d}}.
	\end{equation}
\end{theorem}

\begin{proof}
	The proof follows by deriving the expression of (\ref{eq:Pois_brac_S}) from (\ref{eq:can_poisson}) using the results of Prop.\ref{prop:pf_pl_M_to_S}.
	Using the change of variables formula and the expressions of $\varD{\scrTld{G}}{\varphi}$ and $\varDbnd{\scrTld{G}}{\varphi}$ in Prop.\ref{prop:pf_pl_M_to_S} (ii), one has that
	\begin{align}
		\intB \varD{\scrTld{F}}{\momM} \wedgedot \varD{\scrTld{G}}{\varphi} =& \intS \varD{\scr{F}}{\momS} \wedgedot \mFS\otimes(\extd \varD{\scr{G}}{\mFS} + \nabS \varD{\scr{G}}{\momS} \wedgedot \vfS),\label{eq:th_proof_1}\\
		\int_{\bndB} \varD{\scrTld{F}}{\momM}|_{\bndB} \wedgedot \varDbnd{\scrTld{G}}{\varphi} =& -\int_{\Sbound} \varD{\scr{F}}{\momS}|_{\Sbound} \wedgedot (\varD{\scr{G}}{\mFS} \mFS + \iota_{\varD{\scr{G}}{\momS}}\momS)|_{\Sbound}.\label{eq:th_proof_2}
	\end{align}
	Furthermore, using the change of variables formula and the steps in the proof of Prop.\ref{prop:pf_pl_M_to_S} (ii), one can show that
	\begin{equation}
		\intB \varD{\scrTld{G}}{\momM} \wedgedot \varD{\scrTld{F}}{\varphi} + \int_{\bndB} \varD{\scrTld{G}}{\momM}|_{\bndB} \wedgedot \varDbnd{\scrTld{F}}{\varphi} = - \intS \varD{\scr{F}}{\mFS} \wedge \Lie{\varD{\scr{G}}{\momS}}{\mFS} + \varD{\scr{F}}{\momS} \wedgedot \extcdS \iota_{\varD{\scr{G}}{\momS}}\momS,\label{eq:th_proof_3}
	\end{equation}
	where (\ref{eq:prop_proof_3}) and (\ref{eq:prop_proof_4}) were used in reverse with $\varD{\scr{G}}{\momS}$ instead of $\delta\varphi$.
	Substituting (\ref{eq:th_proof_1}-\ref{eq:th_proof_3}) in (\ref{eq:can_poisson}), and using the Lie-derivative identity \cite{Gilbert2023}
	\begin{equation}\label{eq:Lie_deriv_id}
			\Lie{\eta}{\momS} = \extcdS(\iota_\eta \momS) + \mFS\otimes (\nabS \eta \wedgedot \vfS), \qquad \forall \eta \in \spVec{\cl{S}},
	\end{equation}
	allows us to express the spatial Poisson bracket (\ref{eq:Pois_brac_S}) as
	\begin{align*}
		{}^s\{\scr{F},\scr{G}\} =& -\intS  \varD{\scr{F}}{\mFS} \wedge \Lie{\varD{\scr{G}}{\momS}}{\mFS} + \varD{\scr{F}}{\momS} \wedgedot \mFS\otimes\extd \varD{\scr{G}}{\mFS} \\
		&\qquad+ \varD{\scr{F}}{\momS} \wedgedot (\extcdS \iota_{\varD{\scr{G}}{\momS}}\momS +  \mFS\otimes( \nabS \varD{\scr{G}}{\momS} \wedgedot \vfS)) \\
		&\qquad + \int_{\Sbound} \varD{\scr{F}}{\momS}|_{\Sbound} \wedgedot (\varD{\scr{G}}{\mFS} \mFS + \iota_{\varD{\scr{G}}{\momS}}\momS)|_{\Sbound}\\
		=&-\intS  \varD{\scr{F}}{\mFS} \wedge \extd \iota_{\varD{\scr{G}}{\momS}}{\mFS} + \varD{\scr{F}}{\momS} \wedgedot \mFS\otimes\extd \varD{\scr{G}}{\mFS}  + \varD{\scr{F}}{\momS} \wedgedot\Lie{\varD{\scr{G}}{\momS}}{\momS}\\
		&\qquad + \int_{\Sbound} \varD{\scr{F}}{\momS}|_{\Sbound} \wedgedot (\varD{\scr{G}}{\mFS} \mFS + \iota_{\varD{\scr{G}}{\momS}}\momS)|_{\Sbound}.
	\end{align*}
	
	Now by letting $\scr{F}=\scr{G}$ one has that ${}^s\{\scr{G},\scr{G}\} =0$ and by introducing $, \eS{\mFS}:= \varD{\scr{G}}{\mFS}, \eS{\momS}:= \varD{\scr{G}}{\momS}$, and 
	$$\fS{\mFS}:=-\extd \iota_{\eS{\momS}}\mFS,\qquad  \fS{\momS}:= - \mFS\otimes \extd \eS{\mFS} - \Lie{\eS{\momS}}{\momS},\qquad (\eS{\mFS} \mFS + \iota_{\eS{\momS}} \momS)|_{\Sbound} = 0,$$
	then the above Poisson bracket expression becomes
	$\duPairS{\eS{\mFS}}{\fS{\mFS}} + \duPairS{\eS{\momS}}{\fS{\momS}} = 0$.
	Finally, by introducing $\subTxt{f}{d}:= \cfgFleg{f}(\fM{\text{d}}),\subTxt{e}{d}:= \cfgFleg{f}(\subTxt{\tilde{e}}{d})$, the interaction port $(\fM{\text{d}},\eM{\text{d}})$ in (\ref{eq:Dirac_kin_M}) can be written using the change of variables formula as
	$$\intB \fM{\text{d}}\wedgedot \eM{\text{d}} = \intS \cfgFleg{f}(\fM{\text{d}})\wedgedot\cfgFleg{f}(\subTxt{\tilde{e}}{d}) = \intS \subTxt{f}{d}\wedgedot \subTxt{e}{d},$$
	which concludes the transformation of (\ref{eq:Dirac_kin_M}) into (\ref{eq:Dirac_kin_S}).
	 \qed
\end{proof}

\begin{proposition}\label{prop:pH_kin_S}
	For the kinetic energy as a Hamiltonian functional with state variables $ \chiS := (\mFS,\momS) \in \spXkinS$, the equations of motion in the spatial representation are given by:
	\begin{align}
		\TwoVec{\partial_t \mFS}{\partial_t \momS} &= \TwoTwoMat{0}{ - \extd \iota_{(\cdot)} \mFS}{- \mFS \otimes \extd (\cdot)}{- \Lie{(\cdot)}{\momS}} \TwoVec{\varD{\HkinS}{\mFS}}{\varD{\HkinS}{\momS}} + \TwoVec{0}{1} {\forceS}, \label{eq:pH_sys_kin_S_1}\\
		\vS &= \left( 0 \quad 1\right) \TwoVec{\varD{\HkinS}{\mFS}}{\varD{\HkinS}{\momS}},\label{eq:pH_sys_kin_S_2}
	\end{align}
	with $\forceS \in \spcovFrmS{n}$ denoting the external body force field in the spatial representation.
	The variational derivatives of the functional $\HkinS$ in (\ref{eq:HkinS}) with respect to $\mFS \in \spFrmS{n}$ and $\momS \in \spcovFrmS{n}$ are given respectively by
	$$\varD{\HkinS}{\mFS} = - \half \iota_{\vS}\vfS \in \spFrmS{0}, \qquad \qquad \varD{\HkinS}{\momS} = \vS \in \spvecFrmS{0},$$
	where $\vS = \invhodgeS\momS$ is the spatial velocity field.
	Furthermore, the state variables are constrained on the boundary $\Sbound$ such that the extensive momentum flux vanishes, i.e. $\iota_\vS \momS|_{\Sbound} = 0 \in \spcovFrmSbnd{n-1}$.
	
	The Hamiltonian functional (\ref{eq:HkinS}) admits its rate of change such that along trajectories $(\mFS(t),\momS(t))$ of (\ref{eq:pH_sys_kin_S_1}-\ref{eq:pH_sys_kin_S_2}), it holds that
	\begin{equation}\label{eq:HkinS_open}
		\HkinSdot = \duPairS{\forceS}{\vS}.
	\end{equation}
\end{proposition}
\begin{proof}
	
	(i) The proof of (\ref{eq:pH_sys_kin_S_1}-\ref{eq:pH_sys_kin_S_2}) follows from the expression of $\diracKinS(\chiS)$ in (\ref{eq:Dirac_kin_S}) by setting the inputs $(\eS{\mFS},\eS{\momS}, \subTxt{e}{d}) = ({\varD{\HkinS}{\mFS}},{\varD{\HkinS}{\momS}},\forceS)$ which leads to the outputs $(\fS{\mFS},\fS{\momS}, \subTxt{f}{d}) = ({\partial_t \mFS},{\partial_t \momS},\varD{\HkinS}{\momS})$.
	Consequently, using (\ref{eq:Funct_dot_S}) for $\HkinS$, the energy balance (\ref{eq:Dirac_Balance_S}) becomes $\HkinSdot = \duPairS{\forceS}{\varD{\HkinS}{\momS}}$.
	
	(ii) The proof of the variational derivatives goes as follows.
	First, the Hamiltonian (\ref{eq:HkinS}) is rewritten to show explicitly its dependence on $\mFS$ and $\momS$ as
	$$\HkinS[\mFS,\momS] := \intS \frac{1}{2 \mFS} \gS^{ij} \momS_i\momS_j,$$
	where $\gS^{ij}\in \spFrmS{0}$ are the components of the inverse metric $\gS^{-1}$ and $\momS_i\in\spFrmS{n}$ denotes the top-form components of $\momS$.
	Now consider the tangent vector to the curve $s \mapsto \mFS_s \in\spFrmS{n}$ denoted by $\delta\mFS := \dds \mFS_s \in \spFrmS{n},$
	with $\mFS_s|_{s=0} = \mFS$.
	The variational derivative $\varD{\HkinS}{\mFS}\in \spFrmS{0}$ is defined such that
	$$\dds \HkinS[\mFS_s,\momS] = \duPairS{\varD{\HkinS}{\mFS}}{\delta\mFS},$$
	where the LHS can be expressed as
	$$\dds \HkinS[\mFS_s,\momS] = \intS \frac{1}{2} \dds(\frac{1}{\mFS}) \gS^{ij} \momS_i\momS_j.$$
	Using the identity $\dds(\frac{1}{\mFS}) = - \frac{1}{\mFS^2}\dds(\mFS_s)$, the definition of $\delta\mFS$ and the fact that $\momS_i = \gS_{ij} \vS^j \mFS$ leads to
	$$\dds \HkinS[\mFS_s,\momS] = - \intS \frac{1}{2} \gS_{km} \vS^k \vS^m \delta\mFS = - \intS \frac{1}{2} \gS(\vS,\vS) \delta\mFS.$$
	Using exterior calculus, one has that $\gS(\vS,\vS) = \iota_{\vS}\vfS \in \spFrmS{0}$, which concludes the derivation of $\varD{\HkinS}{\mFS}$.
	As for $\varD{\HkinS}{\momS} = \vS$, one can also follow similar steps in deriving it as above or alternatively use the relation $\varD{\HkinM}{\momM} = \cfgPleg{f}(\varD{\HkinS}{\momS})$ derived in Prop. \ref{prop:pf_pl_M_to_S}. Consequently, using the expression of $\varD{\HkinM}{\momM}$ it follows that $\varD{\HkinS}{\momS} = \vS$.
	
	(iii) Finally the boundary constraint in (\ref{eq:Dirac_kin_S}) can be rewritten using the expressions of the variational derivatives as
	\begin{align*}
		0 &= (\varD{\HkinS}{\mFS} \mFS + \iota_{\varD{\HkinS}{\momS}} \momS)|_{\Sbound} = (-\half \iota_\vS \vfS \mFS +  \iota_\vS\mFS\otimes \vfS)|_{\Sbound}\\
		&= (-\half \iota_\vS\mFS\otimes \vfS +  \iota_\vS\mFS\otimes \vfS)|_{\Sbound} = (\half \iota_\vS\mFS\otimes \vfS )|_{\Sbound} = \half \iota_\vS\momS|_{\Sbound},
	\end{align*}
	 which concludes the proof.
	
	
		\qed
\end{proof}

\begin{corollary}\label{cor:balance_laws_S}
	The \pH dynamic equations (\ref{eq:pH_sys_kin_S_1}) incorporate the conservation of mass and momentum laws which can be expressed, respectively, in:
	
	(i) Advection formulation:
		$$\partial_t \mFS = -\Lie{\vS}{\mFS}, \qquad \partial_t \momS = -\Lie{\vS}{\momS} + \half \mFS\otimes \extd |\vS|_{\gS}^2 +  \forceS,$$
	with $|\vS|_{\gS} := \sqrt{\gS(\vS,\vS)}$ denoting the norm with respect to $\gS$.
		
	(ii) Conservation formulation:
		$$\partial_t \mFS = -\extd \iota_{\vS}{\mFS}, \qquad \partial_t \momS = -\extcdS \iota_{\vS}\momS + \forceS.$$
\end{corollary}
\begin{proof}
	The proof of the mass balance in (ii) and the momentum balance in (i) follows immediately from substituting the expressions of the variational derivatives. Whereas, the mass balance in (i) is derived using Cartan's homotopy formula $\Lie{\vS}{} = \extd \iota_v + \iota_v\extd$, while the momentum balance in (ii) follows from using the identity $\nabS\vS \wedgedot\vfS = \half \extd |\vS|_{\gS}^2$ and the Lie derivative identity (\ref{eq:Lie_deriv_id}). 
	\qed
	
\end{proof}

\subsubsection*{Convective Dirac structure}
Finally we conclude by presenting the convective representation of the kinetic energy subsystem which follows exactly the same line of thought presented above for the reduction to the spatial representation.
For compactness purposes, the proofs of this section will be shorter versions of their counterparts in the spatial representation.

The transformation of the material variables to the convective ones is carried out using the nonlinear diffeomorphism 
\begin{equation}\label{eq:diffeoMtoC}
	\fullmap{{}^c\Phi}{\spXkinM\times\spPkinM}{\spXkinC\times\spPkinC}{((\varphi,\momM),(\gS,\mFM))}{((\gC,\momC),\mFC)}
\end{equation}
with $\gC := \varphi^* (\gS) \in \spMetB$ denoting the convective metric and $\momC := \cfgPleg{v}(\momM) \in \spcovFrmB{n}$ denoting the extensive momentum in the convective representation, while $\mFC = \mFM \in \spFrmB{n}$.
We denote the convective state space by $\spXkinC := \spMetB\times \spcovFrmB{n}$ and the convective parameter space by $\spPkinC:=\spFrmB{n}$.

The convective kinetic energy Hamiltonian $\map{\HkinC}{\spXkinC\times \spPkinC}{\bb{R}}$ is defined by $\HkinC[\gC,\momC;\mFC] = \HkinM[\varphi,\momM;\gS,\mFM]$ and is expressed as
\begin{equation}\label{eq:HkinC}
	\HkinC[\gC,\momC] := \intB \half \invhodgeC \momC \wedgedot \momC,
\end{equation}
where the convective Hodge star $\map{\hodgeC}{\spvecFrmB{k}}{\spcovFrmB{n-k}}$ incorporates a state dependency on the convective metric $\gC$ and a parametric dependency on the mass form $\mFC$.

\begin{remark}
	In local coordinates we have that the convective metric expressed as $\gC = F^i_I F^j_J g_{ij}\circ \varphi E^I \otimes E^J$, with $F^i_I\in \spFrmB{0}$ denoting the local components of the deformation gradient $F:= T\varphi$.
	The convective momentum is computed from its material counterpart (\cf. Remark \ref{remark:momM}) by
	$$\momC = \cfgPleg{v}(\momM)= \vM_i \mFM \otimes \varphi^*{(e^i)} = \vM_i \mFM \otimes F^i_I E^I = \vC_I \mFC \otimes E^I,$$
	and thus $\momC$ is identified with the trivial covector-valued form $\mFC\otimes\vfC$ similar to the spatial momentum $\momS$.
	
	
\end{remark}

In general, we denote by $\map{\scrHat{F}:= \tilde{\scr{F}} \circ {}^c\Phi^{-1}}{\spXkinC}{\bb{R}}$ the convective counterpart of any material functional $\map{\tilde{\scr{F}}}{\spXkinM}{\bb{R}}$ defined such that:
$$\scrHat{F}[\varphi^* (\gS),\cfgPleg{v}(\momM)] = \scrTld{F}[\varphi,\momM].$$
In our work we identify tangent and cotangent spaces of $\spMetB$ by \cite{rashad2023intrinsic}
$T_{\gC}\spMetB \cong \spcovFrmBsub{1}{sym}\subset\spcovFrmB{1}$ and $T_{\gC}^*\spMetB \cong\spvecFrmBsub{n-1}{sym}\subset \spvecFrmB{n-1}$.
The restriction to symmetric subspaces comes from the symmetric nature of the Riemannian metric $\gC$.
More details on the technical constructions of these spaces will be discussed later in Sec. \ref{sec:decomposition}.
The variational derivatives of the functional $\scrHat{F}$ with respect to $(\gC,\momC)$ are the bundle-valued forms $\varD{\scrHat{F}}{\gC} \in \spvecFrmBsub{n-1}{sym}$ and $ \varD{\scrHat{F}}{\momC} \in \spvecFrmB{0},$ that satisfy for any $\delta\gC\in \spcovFrmBsub{1}{sym},\delta\momC \in \spcovFrmB{n}$
$$	\extD_{\gC}\scrHat{F}(\gC,\momC) \cdot \delta\gC = \intB \delta\gC \wedgedot \varD{\scrHat{F}}{\gC}  , \qquad \qquad 
\extD_{\momC}\scrHat{F}(\gC,\momC) \cdot \delta\momC = \intB \delta\momC \wedgedot \varD{\scrHat{F}}{\momC}  .$$
The rate of change of $\scrHat{F}$ with respect to time is then expressed as
\begin{equation}\label{eq:Funct_dot_C}
	\dot{\scrHat{F}} = \duPairB{\varD{\scrHat{F}}{\gC}}{\partial_t \gC} + \duPairB{\varD{\scrHat{F}}{\momC}}{\partial_t \momC},
\end{equation}
with $\partial_t \gC \in \spcovFrmBsub{1}{sym}$ and $\partial_t \momC \in \spcovFrmB{n}$.

Let $\map{\cfgPleg{v}}{\spvecFrmPhi{k}}{\spvecFrmB{k}}$ denote the pullback of the value part of a vector-valued form, transforming it from the material representation to the convective one, and $\cfgFleg{v}$ denote its inverse. We shall also use the same notation for covector-valued forms. Then the convective counterpart of Prop. \ref{prop:pf_pl_M_to_S} can be stated as follows.

\begin{proposition}\label{prop:pf_pl_M_to_C}
	Let $(\varphi,\momM) \in \spXkinM$ and $(\gC,\momC)\in \spXkinC$ be related by the diffeomorphism ${}^c\Phi$ in (\ref{eq:diffeoMtoC}).
	
	i) The pushforward map of ${}^c\Phi$ is given by
	\begin{equation*}
		\fullmap{{}^c\Phi_*}{T\spXkinM}{T\spXkinC}{(\varphi,\momM,\delphiM,\delta\momM,\delphiM|_{\Bbound})}{(\gC,\momC,\delta\gC,\delta\momC)}
	\end{equation*}
	
	$$\delta\gC = \Lie{\delphiC}{\gC} \in\spvecFrmBsub{1}{sym} ,\qquad \qquad \delta\momC = \mFC\otimes (\nabC\delphiC \wedgedot \vfC) + \cfgPleg{v}(\delta\momM) \in \spcovFrmB{n},$$
	with $\delphiC := \cfgPleg{v}(\delphiM) = T \varphi^{-1} \circ \delphiM \in \spvecFrmB{0} \cong \spVec{\cl{B}}$.

	ii) The pullback map of ${}^c\Phi$ is given by
	\begin{equation*}
		\fullmap{{}^c\Phi^*}{T^*\spXkinC}{T^*\spXkinM}{(\gC,\momC,\varD{\scrHat{F}}{\gC},\varD{\scrHat{F}}{\momC})}{(\varphi,\momM,\varD{\scrTld{F}}{\varphi}, \varD{\scrTld{F}}{\momM}, \varDbnd{\scrTld{F}}{\varphi})}
	\end{equation*}
		with 
	\begin{align*}
		\varD{\scrTld{F}}{\varphi} =& - \cfgFleg{v} (\extcdC(2 (\varD{\scrHat{F}}{\gC}) ^\flat + \iota_{\varD{\scrHat{F}}{\momC}}\momC))  &&\in \spcovFrmPhi{n} \\
		\varD{\scrTld{F}}{\momM} =& \cfgFleg{v}(\varD{\scrHat{F}}{\momC}) &&\in \spvecFrmPhi{0}\\
		\varDbnd{\scrTld{F}}{\varphi} =& \cfgFleg{v}(2 (\varD{\scrHat{F}}{\gC}) ^\flat + \iota_{\varD{\scrHat{F}}{\momC}}\momC)|_{\Bbound} &&\in \spcovFrmPhibnd{n-1}.
	\end{align*}

\end{proposition}
\begin{proof}
	See Appendix \ref{app:proof_2}.
	 \qed
\end{proof}

\begin{theorem}
	The convective Dirac structure corresponding to (\ref{eq:Dirac_kin_M}) under the diffeomorphism (\ref{eq:diffeoMtoC}) is the sub-bundle
	$\diracKinC \subset \FKinC\oplus \FdlKinC$ with 
	\begin{align*}
		\FKinC:=& {}^c \Phi_*(T\spXkinM) \times \cfgPleg{v}(\spvecFrmPhi{0}) = T\spXkinC \times \spvecFrmB{0}\\
		\FdlKinC :=& ({}^c \Phi^*)^{-1}(T^*\spXkinM) \times \cfgPleg{v}(\spcovFrmPhi{n}) = T^*\spXkinC \times \spcovFrmB{n},
	\end{align*}
	whose fiber at any $\chiC:= (\gC,\momC)\in \spXkinC$ is defined by
	\begin{equation}\label{eq:Dirac_kin_C}
		\begin{split}
			\diracKinC(\chiC) := \{ ((\subTxt{\hat{f}}{k}, &\subTxt{\hat{f}}{d}),(\subTxt{\hat{e}}{k},\subTxt{\hat{e}}{d} )) \in \FKinC(\chiC)\times \FdlKinC(\chiC) | \\
			\TwoVec{\fC{\gC}}{\fC{\momC}} &=\TwoTwoMat{0}{2\ \sym\circ \gC \circ \extcdC}{2\ \extcdC \circ \gC}{\Lie{(\cdot)}{\momC}} \TwoVec{\eC{\gC}}{\eC{\momC}} + \TwoVec{0}{1} \subTxt{\hat{e}}{d} ,\\
			\subTxt{\hat{f}}{d}&= \begin{pmatrix} 0 & 1 \end{pmatrix} \TwoVec{\eC{\gC}}{\eC{\momC}}, \\
			0 &= (2 \gC \cdot \eC{\gC} + \iota_{\eC{\momC}}\momC)|_{\Bbound} \},
		\end{split}
	\end{equation}
	with $\subTxt{\hat{f}}{k} = (\fC{\gC},\fC{\momC}) \in T_{\chiC}\spXkinC$ and $\subTxt{\hat{e}}{k} = (\eC{\gC},\eC{\momC})\in T^*_{\chiC}\spXkinC$.
	The Dirac structure (\ref{eq:Dirac_kin_C}) characterizes the power balance
	\begin{equation}\label{eq:diracKinC_balance}
		\duPairB{\eC{\gC}}{\fC{\gC}} + \duPairB{\eC{\momC}}{\fC{\momC}} = \duPairB{\subTxt{\hat{e}}{d}}{\subTxt{\hat{f}}{d}}.
	\end{equation}
\end{theorem}

\begin{proof}
	
	Using the expressions of $\varD{\scrTld{G}}{\varphi}$ and $\varDbnd{\scrTld{G}}{\varphi}$ in Prop.\ref{prop:pf_pl_M_to_C} (ii) in addition to the duality of $\cfgFleg{v}$ and $\cfgPleg{v}$, one can rewrite the different integrals in (\ref{eq:can_poisson}) as
	\begin{equation}
		\intB \varD{\scrTld{F}}{\momM} \wedgedot \varD{\scrTld{G}}{\varphi} + \int_{\bndB} \varD{\scrTld{F}}{\momM}|_{\bndB} \wedgedot \varDbnd{\scrTld{G}}{\varphi} = -\intB \varD{\scrHat{F}}{\momC} \wedgedot \extcdC \cl{E}_{\scrHat{G}} + \int_{\bndB} \varD{\scrHat{F}}{\momC}|_{\Bbound} \wedgedot \cl{E}_{\scrHat{G}}|_{\Bbound},\label{eq:th_proof_1_C}
	\end{equation}
	where $\cl{E}_{\scrHat{G}}:= 2 (\varD{\scrHat{G}}{\gC}) ^\flat + \iota_{\varD{\scrHat{G}}{\momC}}\momC \in \spcovFrmB{n-1}$.
	Furthermore, using the steps in the proof of Prop.\ref{prop:pf_pl_M_to_C} (ii) in reverse, one can show that
	\begin{align}
		\intB \varD{\scrTld{G}}{\momM} \wedgedot \varD{\scrTld{F}}{\varphi} +& \int_{\bndB} \varD{\scrTld{G}}{\momM}|_{\bndB} \wedgedot \varDbnd{\scrTld{F}}{\varphi} = \intB \nabC \varD{\scrHat{G}}{\momC} \wedgedot (2 \gC \varD{\scrHat{F}}{\gC} + \iota_{\varD{\scrHat{F}}{\momC}}\momC)\nonumber\\
		=& \intB \varD{\scrHat{F}}{\gC} \wedgedot 2 \sym(\gC\cdot \nabC \varD{\scrHat{G}}{\momC}) + \varD{\scrHat{F}}{\momC} \wedgedot \mFC \otimes (\nabC\varD{\scrHat{G}}{\momC} \wedgedot \vfC).\label{eq:th_proof_2_C}
	\end{align}
	
	Let ${}^c\{\scrHat{F},\scrHat{G}\}$ denote the convective representation of the canonical Poisson bracket defined similar to (\ref{eq:Pois_brac_S}).
	By substituting (\ref{eq:th_proof_1_C},\ref{eq:th_proof_2_C}) in (\ref{eq:can_poisson}) and using the convective version of (\ref{eq:Lie_deriv_id}), one can express this convective Poisson bracket as
	\begin{align*}
		{}^c\{\scrHat{F},\scrHat{G}\} =& \intB  \varD{\scrHat{F}}{\gC} \wedgedot 2 \sym(\gC\cdot \nabC \varD{\scrHat{G}}{\momC}) + \varD{\scrHat{F}}{\momC} \wedgedot \mFC \otimes (\nabC\varD{\scrHat{G}}{\momC} \wedgedot \vfC) \\
		&+ \varD{\scrHat{F}}{\momC} \wedgedot \extcdC ((\varD{\scrHat{G}}{\gC}) ^\flat + \iota_{\varD{\scrHat{G}}{\momC}}\momC) - \int_{\bndB} \varD{\scrHat{F}}{\momC}|_{\Bbound} \wedgedot \cl{E}_{\scrHat{G}}|_{\Bbound}\\
		=& \intB  \varD{\scrHat{F}}{\gC} \wedgedot 2 \sym(\gC\cdot \nabC \varD{\scrHat{G}}{\momC}) +\varD{\scrHat{F}}{\momC} \wedgedot (\extcdC (\varD{\scrHat{G}}{\gC}) ^\flat + \Lie{\varD{\scrHat{G}}{\momC}}{\momC}) \\
		&- \int_{\bndB} \varD{\scrHat{F}}{\momC}|_{\Bbound} \wedgedot (2 (\varD{\scrHat{G}}{\gC}) ^\flat + \iota_{\varD{\scrHat{G}}{\momC}}\momC)|_{\Bbound}.
	\end{align*}
	Finally, following the same line of thought of the proof of Th. \ref{th:Dirac_kin_S}, one can transform the Poisson bracket ${}^c\{\scrHat{F},\scrHat{G}\}$ into (\ref{eq:Dirac_kin_C}).
	\qed
\end{proof}

\begin{proposition}\label{prop:pH_kinetic_C}
	For the kinetic energy as a Hamiltonian functional with state variables $ \chiC := (\gC,\momC) \in \spXkinC$, the equations of motion in the convective representation are given by:
\begin{align}
	\TwoVec{\partial_t \gC}{\partial_t \momC}  &= \TwoTwoMat{0}{2\ \sym\circ \gC \circ \extcdC}{2\ \extcdC \circ \gC}{\Lie{(\cdot)}{\momC}} \TwoVec{\varD{\HkinC}{\gC}}{\varD{\HkinC}{\momC}}+ \TwoVec{0}{1} {\forceC}, \label{eq:pH_sys_kin_C_1}\\
	\vC &= \begin{pmatrix} 0 & 1 \end{pmatrix} \TwoVec{\varD{\HkinC}{\gC}}{\varD{\HkinC}{\momC}}.\label{eq:pH_sys_kin_C_2}
\end{align}
	with $\forceC \in \spcovFrmB{n}$ denoting the external body force field in the convective representation.	
	The variational derivatives of the functional $\HkinC$ in (\ref{eq:HkinC}) with respect to $\gC \in \spMetB$ and $\momC \in \spcovFrmB{n}$ are given respectively by
	$$\varD{\HkinC}{\gC} = - \half\iota_{\vC}\mFC \otimes \vC \in \spvecFrmB{n-1}, \qquad \qquad \varD{\HkinC}{\momC} = \vC \in \spvecFrmB{0},$$
	where $\vC = \invhodgeC\momC$ is the convective velocity field.
	
	The Hamiltonian functional (\ref{eq:HkinC}) admits its rate of change such that along trajectories $(\gC(t),\momC(t))$ of (\ref{eq:pH_sys_kin_C_1}-\ref{eq:pH_sys_kin_C_2}), it holds that
	\begin{equation}\label{eq:HkinC_open}
		\HkinCdot = \duPairB{\forceC}{\vC}.
	\end{equation}
\end{proposition}
\begin{proof}
		i) The proof of (\ref{eq:pH_sys_kin_C_1}-\ref{eq:pH_sys_kin_C_2}) follows from (\ref{eq:Dirac_kin_C}) by setting the inputs $(\eC{\gC},\eC{\momC}, \subTxt{\hat{e}}{d}) = ({\varD{\HkinC}{\gC}},{\varD{\HkinC}{\momC}},\forceC)$ which leads to the outputs $(\fC{\gC},\fC{\momC}, \subTxt{\hat{f}}{d}) = ({\partial_t \gC},{\partial_t \momC},\varD{\HkinC}{\momC})$.
		Consequently, using (\ref{eq:Funct_dot_C}), $\HkinCdot = \duPairB{\forceC}{\varD{\HkinC}{\momC}}$ follows from (\ref{eq:diracKinC_balance}).
	
		ii) The Hamiltonian (\ref{eq:HkinC}) can be rewritten to show explicitly its dependence on $\gC$ and $\momC$ as
	$$\HkinC[\gC,\momC] := \intB \frac{1}{2 \mFC} \gC^{IJ} \momC_I\momC_J,$$
	where $\gC^{IJ}\in \spFrmB{0}$ are the components of the inverse metric $\gC^{-1}$ such that $\gC^{IJ}\gC_{JK} = \delta^I_K$, with $\gC = \gC_{JK} E^J\otimes E^K$. Furthermore, $\momC_J\in\spFrmB{n}$ denotes the top-form components of $\momC$ such that $\momC = \momC_J\otimes E^J$.

	
	Consider the tangent vector to the curve $s \mapsto \gC_s \in\spMetB$ denoted by
	$$\delgC := \dds \gC_s \in T_{\gC}\spMetB \cong \spcovFrmB{1},$$
	with $\gC_s|_{s=0} = \gC$.
	The variational derivative $\varD{\HkinC}{\gC}\in \spvecFrmB{n-1}$ is then defined implicitly such that
	$$\dds \HkinC[\gC_s,\momC] = \duPairB{\varD{\HkinC}{\gC}}{\delgC}.$$
	Using the expression of $\HkinC$ in (i), one can express the LHS as
	$$\dds \HkinC[\gC_s,\momC] = \intB \frac{1}{2 \mFC} \dds(\gC_s^{IJ}) \momC_I\momC_J.$$
	Using the identity $\dds(\gC_s^{IJ}) = - \gC^{IK}\gC^{JL}\dds((\gC_s)_{KL})$, the definition of $\delgC$ and the fact that $\momC_I = \gC_{IJ} \vC^J \mFC$ leads to
	$$\dds \HkinC[\gC_s,\momC] = - \intB \frac{1}{2 \mFC} \gC^{IK}\gC^{JL} \momC_I \momC_J \delgC_{KL} = - \intB \half \vC^K \vC^L \delgC_{KL} \mFC.$$
	The above expression can be written in tensor and exterior, respectively, as
	$$\dds \HkinC[\gC_s,\momC] = - \intB  (\half\vC\otimes\vC):\delgC\ \mFC = - \intB  \delgC \wedgedot \hodgeC(\half \vC\otimes\vC).$$
	Using the identity $\hodgeC(\vC\otimes\vC) = \iota_{\vC}\mFC \otimes \vC $ concludes the derivation of $\varD{\HkinC}{\gC}$.
	As for $\varD{\HkinC}{\momC} = \vC$, it is straightforward following the steps in proof of Prop. \ref{prop:pH_kin_S}.
	

	(iii) Finally the boundary constraint in (\ref{eq:Dirac_kin_C}) can be rewritten using the expressions of the variational derivatives as
\begin{align*}
	0 = (2 (\varD{\HkinC}{\gC})^\flat + \iota_{\vC} \momC)|_{\Bbound} = (\iota_{\vC}\mFC \otimes \vfC - \iota_\vC\momC)|_{\Bbound},
\end{align*}
which is naturally satisfied by the state variables since $\iota_{\vC}\mFC \otimes \vfC = \iota_\vC\momC$.
		\qed
\end{proof}

\begin{corollary}\label{cor:balance_laws_C}
	The \pH dynamic equations (\ref{eq:pH_sys_kin_C_1}) incorporate the convective metric advection and conservation of momentum laws which can be expressed, respectively, in:
	
	(i) Advection formulation:
	$$\partial_t \gC = \Lie{\vC}{\gC}, \qquad \partial_t \momC = \Lie{\vC}{\momC} - \extcdC \iota_{\vC}\momC + \forceC.$$

	(i) Conservation formulation:
	$$\partial_t \gC = 2\ \sym(\nabC \vfC), \qquad \partial_t \momC = \half \mFC\otimes \extd |\vC|_{\gC}^2  + \forceC,$$
	with $|\vC|_{\gC} := \sqrt{\gC(\vC,\vC)}$ denoting the norm with respect to $\gC$.

\end{corollary}
\begin{proof}
		The proof of the advection equation for $\gC$ in (ii) and the momentum balance in (i) follows immediately from substituting the expressions of the variational derivatives and using the fact that $\iota_\vC\momC = \iota_{\vC}\mFC \otimes \vfC$. 
		Whereas, the advection equation in (i) follows from the identity $\half \Lie{\vC}{\gC} = \sym(\nabC \vfC)$, while the momentum balance in (ii) follows from the identity $\nabC\vC \wedgedot\vfC = \half \extd |\vC|_{\gC}^2$ and the convective counterpart of (\ref{eq:Lie_deriv_id}). 
		\qed
\end{proof}

\begin{remark}
	We conclude this section by a number of remarks:\\
	
	(i) It is important to note that the expression of $\diracKinM(\chiM)$ in (\ref{eq:Dirac_kin_M}) is constant and independent from the base point $\chiM$ which is not the case for $\diracKinS(\chiS)$ in (\ref{eq:Dirac_kin_S}) and $\diracKinC(\chiC)$ in (\ref{eq:Dirac_kin_C}). Thus, in contrast to its material counterpart $\diracKinM$, the spatial and convective Dirac structures are said to be \textit{modulated} by the states $\chiS = (\mFS,\momS)$ and $\chiC = (\gC,\momC)$, respectively, which is a consequence of the Hamiltonian reduction procedure.
	
	(ii) Another interesting observation is that in the spatial dynamics of nonlinear elasticity in Prop. \ref{prop:pH_kin_S}, the momentum flux $\iota_{\vS}\momS \in \spcovFrmS{n-1}$ is constrained to vanish on the boundary $\Sbound:= \varphi(\Bbound)$ of the embedded body in the ambient space, which is unlike the convective dynamics in Prop.\ref{prop:pH_kinetic_C}.
	This is a consequence of the fact that the boundary constraint in (\ref{eq:Dirac_kin_C}) is naturally satisfied by the kinetic energy Hamiltonian (\ref{eq:HkinC}) because it naturally encloses the matter particles for all time. It would be interesting to investigate further the implications of such difference between the convective and spatial representation on the analysis of these dynamic equations.
	
\end{remark}

\subsection{Stress power subsystem}\label{sec:stress_power}
\begin{figure}
	\centering
	\includegraphics[width=1\columnwidth]{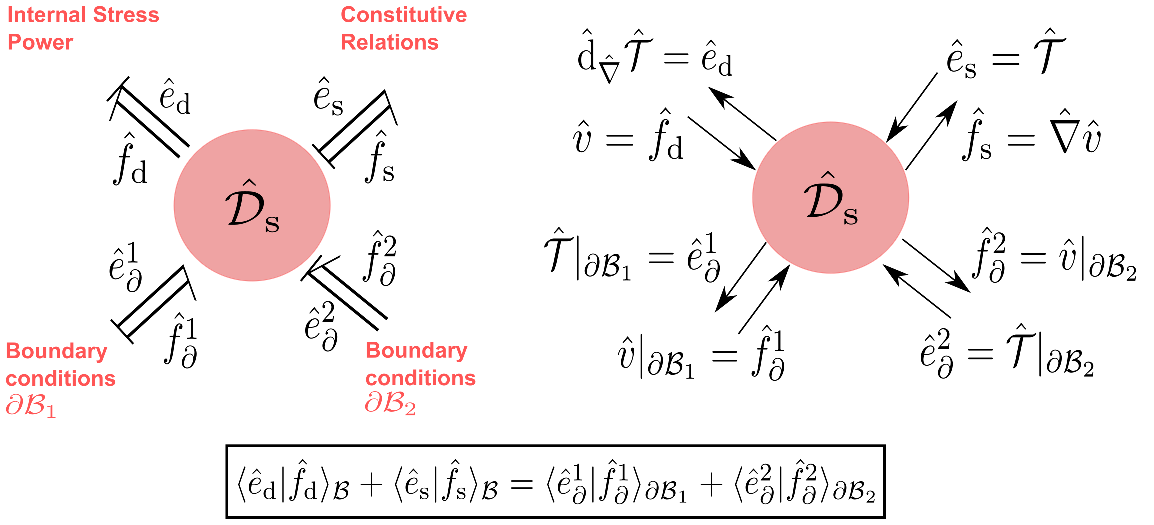}
	\caption{Convective \pH model of the stress power Stokes-Dirac structure expressed in terms of bond graphs (left) and block diagrams (right)}
	\label{fig:pH_str_C}
\end{figure}

The second subsystem of the \pH model for continuum mechanics characterizes the internal stress acting on the elastic body.
We shall only present the convective representation of this \pH model depicted in Fig. \ref{fig:pH_str_C}.
Its spatial and material counterparts are easily deducible.

In general, one has on part of the boundary, which we denote by $\Bbound_1$, the velocity is an input (i.e. boundary data) and traction is an output. On the other part of the boundary, which we denote by $\Bbound_2$, the traction is an input and the velocity in an output.
One has that $\Bbound = \Bbound_1 \cup\Bbound_2$, $\Bbound_1 \cap \Bbound_2 = \emptyset$.

The pairing of the convective velocity $\vC\in \spvecFrmB{0}$ and stress $\stC\in \spcovFrmB{n-1}$ on the boundary of the body is defined as
\begin{equation*}\label{eq:Pst_C}
	\mathscr{P}_{st} := \int_{\bndB} i^*(\vC\wedgedot\stC) = \int_{\bndB} \vC |_{\bndB}\wedgedot\stC|_{\bndB} = \int_{\bndB_1} \vC |_{\bndB_1}\wedgedot\stC|_{\bndB_1} + \int_{\bndB_2} \vC |_{\bndB_2}\wedgedot\stC|_{\bndB_2},
\end{equation*}
where $\vC|_\Bbound:= \ptr(\vC) \in \spvecFrmBbnd{0}$ and $\stC|_\Bbound:= \ptr(\stC) \in \spcovFrmBbnd{n-1}$
denote, respectively, the (partial) pullback of the convective velocity and stress on the boundary under the body inclusion map $\map{i}{\Bbound}{\cl{B}}$.
One has that $\vC|_\Bbound$ represent the boundary's velocity and $\stC|_\Bbound$ the traction on the boundary.


Furthermore, from the integration by parts formula for bundle-valued forms \cite{rashad2023intrinsic}, one has that
$$\intB \nabC\vC\wedgedot\stC + \vC\wedgedot \extcdC\stC  = \intB \extd(\vC\wedgedot\stC) = \int_{\bndB} i^*(\vC\wedgedot\stC).$$
Combining the above expressions together yields the stress power balance
\begin{equation}\label{eq:stress_power_bal}
	\intB \nabC\vC\wedgedot\stC + \vC\wedgedot \extcdC\stC  = \int_{\bndB_1} \vC |_{\bndB_1}\wedgedot\stC|_{\bndB_1} + \int_{\bndB_2} \vC |_{\bndB_2}\wedgedot\stC|_{\bndB_2}.
\end{equation}
The above power balance (\ref{eq:stress_power_bal}) can be encompassed into a Stokes-Dirac\footnote{The term Stokes-Dirac structure has been introduced in the seminal work of van der Schaft and Mashke \cite{van2002hamiltonian} to refer to special Dirac structures that utilize Stokes' theorem to couple energy exchange in the domain and through the boundary.} structure denoted by
$\diracStrC \subset \FStrC\times \FdlStrC$, with 
$\FStrC:= \spvecFrmB{0}\times \spvecFrmB{1}\times \spvecFrmBbnd{0}$ and $\FdlStrC$ denoting its dual.
We define this (constant) Stokes-Dirac structure $\diracStrC$ as
\begin{equation}\label{eq:Dirac_str_C}
	\begin{split}
		\diracStrC := &\left \{ ((\subTxt{\hat{f}}{d}, \subTxt{\hat{f}}{s}, \hat{f}_\partial),(\subTxt{\hat{e}}{d}, \subTxt{\hat{e}}{s}, \hat{e}_\partial)) \in \FStrC\times \FdlStrC | \right.\\
		& \quad \subTxt{\hat{f}}{s}= \nabC \subTxt{\hat{f}}{d} ,\qquad\qquad\ \ \subTxt{\hat{e}}{d}= \extcdC \subTxt{\hat{e}}{s}\\
		& \quad \subTxt{\hat{f}}{s}|_{\Bbound_1}=\hat{f}_\partial^1 ,\qquad\qquad \subTxt{\hat{e}}{s}|_{\Bbound_2}=\hat{e}_\partial^2\\
		& \quad \left. \hat{f}_\partial^2=\subTxt{\hat{f}}{s}|_{\Bbound_2},\qquad\qquad \hat{e}_\partial^1= \subTxt{\hat{e}}{s}|_{\Bbound_1}  \right\},
	\end{split}
\end{equation}
with $\hat{f}_\partial = (\hat{f}_\partial^1,\hat{f}_\partial^2) \in \spvecFrmBbndi{0}{1}\times\spvecFrmBbndi{0}{2}$ 
and $\hat{e}_\partial = (\hat{e}_\partial^1,\hat{e}_\partial^2) \in \spcovFrmBbndi{n-1}{1}\times\spcovFrmBbndi{n-1}{2}$.
The Dirac structure (\ref{eq:Dirac_str_C}) characterizes the power balance
\begin{equation}\label{eq:diracStrC_balance}
	\duPairB{\subTxt{\hat{e}}{d}}{\subTxt{\hat{f}}{d}}+ \duPairB{\subTxt{\hat{e}}{s}}{\subTxt{\hat{f}}{s}}= \duPair{\hat{e}_\partial^1}{\hat{f}_\partial^1}{\Bbound_1} + \duPair{\hat{e}_\partial^2}{\hat{f}_\partial^2}{\Bbound_2},
\end{equation}
which is equivalent to (\ref{eq:stress_power_bal}) by setting the inputs $(\subTxt{\hat{f}}{d},\subTxt{\hat{e}}{s},\hat{f}_\partial^1,\hat{e}_\partial^2) = (\vC,\stC,\vC |_{\bndB_1},\stC |_{\bndB_2})$ which yields the outputs
$(\subTxt{\hat{e}}{d}, \subTxt{\hat{f}}{s}, \hat{e}_\partial^1, \hat{f}_\partial^2)=(\extcdC\stC, \nabC\vC, \stC |_{\bndB_1}, \vC |_{\bndB_2})$.

Finally, it is straightforward to transform $\diracStrC$ into a material one $\diracStrM$ (or spatial one $\diracStrS$) which will have identical structure to (\ref{eq:Dirac_str_C}) with $\extcdC$ replaced by its material counterpart $\extcdM$ (or spatial counterpart $\extcdS$).

\subsection{Decomposition of power ports}\label{sec:decomposition}
So far we constructed the \pH subsystems that describe storage of kinetic energy and incorporate traction and velocity boundary conditions into the dynamics. The last step is to add the elastic body's constitutive model to the power port $(\nabC\vC,\stC)$. 

\begin{figure}
	\centering
	\includegraphics[width=0.8\columnwidth]{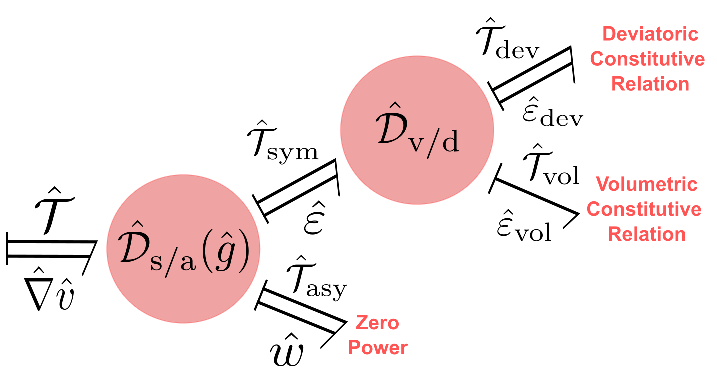}
	\caption{Symmetric-asymmetric decomposition of $(\nabC\vC,\stC)$ and the volumetric-deviatoric decomposition of $(\epC,\subTxt{\stC}{sym})$}
	\label{fig:power_decompositions}
\end{figure}

Before doing so, we will discuss an essential decomposition of the power port $(\nabC\vC,\stC)$ that will split it into a symmetric part $(\epC,\subTxt{\stC}{sym})$ and an asymmetric part $(\hat{w},\subTxt{\stC}{asy})$. By doing so, we will factor out the rigid body motion information included in the flow variable $\nabC\vC$ and the constitutive relation will be added to the symmetric port $(\epC,\subTxt{\stC}{sym})$ only.
Furthermore, we will also discuss another decomposition of the port $(\epC,\subTxt{\stC}{sym})$ which will split it into a volumetric part $(\epCvol,\subTxt{\stC}{vol})$ and a deviatoric part $(\epCdev,\subTxt{\stC}{dev})$. Such decomposition is optional but very useful to distinguish the material's response due to volumetric deformation from its response to isochoric processes (e.g. shear deformation). The two aforementioned decompositions will be implemented in a power consistent manner using Dirac structures, as illustrated in Fig. \ref{fig:power_decompositions}.
The technical details of the aforementioned decompositions are discussed in Appendix \ref{app:decompo}.

\subsubsection{Symmetric-asymmetric decomposition}


Based on the fact that $\spvecFrmB{1} \cong \spVec{^1_1\cl{B}}$ and utilizing the Riemannian metric structure of $(\cl{B},\gC)$, one can define the projection map
\begin{equation}\label{eq:piCsym}
	\map{\piCsym}{\spvecFrmB{1}}{\spvecFrmB{1}}, \qquad \qquad \piCsym:=\gC^{-1}\circ \sym \circ \gC,
\end{equation}
where $\sym$ denotes the symmetrization operation of 2-covariant tensor-fields. Consequently, one can decompose the space of vector-valued one-forms such that
$$\spvecFrmB{1} = \spvecFrmBsub{1}{sym} \oplus\spvecFrmBsub{1}{asy},$$
where $\spvecFrmBsub{1}{sym} := \mathrm{im}(\piCsym)$ and $\spvecFrmBsub{1}{asy} := \mathrm{ker}(\piCsym)$.
Applying the above decomposition on the velocity gradient $\nabC\vC$ allows us to express it as
$$\nabC\vC = \epC + \hat{w}, \qquad \qquad \epC:= \piCsym(\nabC\vC), \qquad  \hat{w} := \piCasy(\nabC\vC) := \nabC\vC- \piCsym(\nabC\vC),$$
where $\epC\in \spvecFrmBsub{1}{sym}$ represents the convective rate of strain field while $\hat{w}\in \spvecFrmBsub{1}{asy}$ represents the convective vorticity field
\footnote{In local Cartesian coordinates, $\epC$ and $\hat{w}$ are given by the familiar expressions $\epC^I_J = \half\left(\frac{\partial \vC^I}{\partial X^J} + \frac{\partial \vC^J}{\partial X^I}\right)$ and $\hat{w}^I_J = \half\left(\frac{\partial \vC^I}{\partial X^J} - \frac{\partial \vC^J}{\partial X^I}\right)$.}.

While the velocity field $\vC$ includes information on how the elastic body deforms as well as the translation and rotation (i.e. rigid body motion) it performs in the ambient space, the deformation is described only by the symmetric component $\epC$ and not $\hat{w}$. This can be seen by inspecting the covariant velocity gradient $\nabC\vfC := \gC \cdot \nabC\vC \in \spcovFrmB{1}$ which can be written as \cite[Prop. 1]{rashad2023intrinsic}
$$\nabC \vfC= \sym(\nabC \vfC) + \asym(\nabC \vfC) = \half \Lie{\vC}{\gC} + \half \extd \vfC,\label{eq:vel_grad_identity_C_}.$$
Since $\Lie{\vC}{\gC} = \varphi^*(\Lie{\vS}{\gS}) = 0$ is the equation characterizing rigid body motion, one can see that  only the symmetric part describes the deformation of the body.
Thus, the operation $\sym\circ \nabC$ factors out the rigid body motion information included in $\vfC$.
The rate of strain tensor $\epC$ plays an important role in finite-strain elasticity theory as will be shown in the coming section. 

By duality, one can decompose the stress tensor field $\stC \in \spcovFrmB{n-1}$ as the sum of $ \subTxt{\stC}{sym} \in \spcovFrmBsub{n-1}{sym}$ and $\subTxt{\stC}{asy} \in \spcovFrmBsub{n-1}{asy}$. With reference to Fig. \ref{fig:power_decompositions}, the above symmetric-asymmetric decomposition of the power port $(\nabC\vC,\stC)$ can be encompassed in the Dirac structure $\hat{\cl{D}}_\text{s/a}(\gC)$,  which is modulated by $\gC$ and defined as the relation corresponding to the following map:
\begin{equation}\label{eq:Dsa}
	\ThrVec{\epC}{\hat{w}}{\stC} = 
	\begin{pmatrix}
		0 & 0 & \piCsym\\
		0 & 0 & \piCasy\\
		\mathds{1} & \mathds{1} & 0
	\end{pmatrix} 
	\ThrVec{\subTxt{\stC}{sym}}{\subTxt{\stC}{asy}}{\nabC\vC},
\end{equation}
which characterizes the power balance
\begin{equation}\label{eq:Dsa_balance}
	\duPairB{\stC}{\nabC\vC} = \duPairB{\subTxt{\stC}{sym}}{\epC} + \duPairB{\subTxt{\stC}{asy}}{\hat{w}}.
\end{equation}

\begin{remark}\label{remark:stress_symmetry}
	Note that if one assumes that $\stC \in \spcovFrmBsub{n-1}{sym}$ then the power balance (\ref{eq:Dsa_balance}) simplifies to
	\begin{equation}\label{eq:Dsa_balance_2}
	\duPairB{\stC}{\nabC\vC} = \duPairB{\stC}{\epC},
	\end{equation}
	which leads to the condition stated in \cite{Kanso2007OnMechanics,Gilbert2023}:
	\begin{equation*}
		(\hat{\alpha} \otimes \hat{\beta}^\sharp)\wedgedot \stC = (\hat{\beta} \otimes \hat{\alpha}^\sharp)\wedgedot \stC, \qquad \forall \hat{\alpha},\hat{\beta} \in \spFrmB{1}.
	\end{equation*} 
	In the formulation above, we have shown that it follows naturally from duality as a consequence of factoring out rigid body motions.
	In numerical methods, assuming $\stC \in \spcovFrmBsub{n-1}{sym}$ corresponds to the strong imposition of symmetry while using the formulation in (\ref{eq:Dsa}) with the vorticity $\hat{w}$ as a Lagrange multiplier corresponds to the weak imposition of symmetry. The interested reader is referred to \cite{arnold2018finite}.
\end{remark}

\begin{remark}[\textbf{Spatial and Material counterparts of $\hat{\cl{D}}_{s/a}$}]
	The above construction can be identically repeated for the spatial representation by defining the projection map $\subTxt{\pi}{sym}:= \gS^{-1}\circ \sym\circ \gS$ to factor out rigid body motions. This will be used later in Sec. \ref{sec:pH_fluid} when we discuss fluids.
	On the other hand, the above construction does not hold for the material representation. An immediate reason is the fact that the symmetrization operation of two-point tensors is not defined simply because its two indices belong to different spaces. This has a strong relation to the axiom of material frame independence of constitutive laws. The interested reader can refer to \cite[Sec. 6.5]{rashad2023intrinsic} and \cite[Ch.3]{Marsden1994}.
\end{remark}

\subsubsection{Volumetric-deviatoric decomposition}
Now we turn attention to the volumetric-deviatoric decomposition of the power port $(\epC,\stC)$ which can be constructed along the same line of thought as above and detailed in Appendix \ref{app:decompo}. For notational simplicity, we assume the symmetry $\stC$ to be enforced strongly (\cf Remark \ref{remark:stress_symmetry}).

Let $\map{\piCvol}{\spvecFrmB{1}}{\spvecFrmBsub{1}{vol}}$ denote the volumetric projection map defined by $\piCvol(\alpha): = \frac{1}{n}\tr(\alpha)I_n,$ for any $\alpha \in \spvecFrmB{1}$, with $\map{\tr}{\spvecFrmB{1}}{\spFrmB{0}}$ denoting the trace map. Furthermore, let $\map{\piCdev}{\spvecFrmB{1}}{\spvecFrmBsub{1}{dev}}$ denote the deviatoric projection map defined by
$\piCdev(\alpha) := \alpha - \piCvol(\alpha)$.
The volumetric-deviatoric decomposition is implemented by the Dirac structure $\hat{\cl{D}}_{v/d}$ defined as the relation corresponding to the following map:
\begin{equation}\label{eq:Dvd}
	\ThrVec{\epCvol}{\epCdev}{\stC} = 
	\begin{pmatrix}
		0 & 0 & \tr\circ \piCvol\\
		0 & 0 & \piCdev\\
		\tr^* & \mathds{1} & 0
	\end{pmatrix} 
	\ThrVec{\subTxt{\stC}{vol}}{\subTxt{\stC}{dev}}{\epC},
\end{equation}
where $\epCvol \in \spFrmB{0}$ and $\subTxt{\stC}{vol} \in \spFrmB{n}$ are the volumetric components and $\epCdev \in \spvecFrmBsub{1}{dev}$ and $\subTxt{\stC}{dev} \in \spcovFrmBsub{n-1}{dev}$ are the deviatoric components of the rate of strain and stress tensors, respectively, while
$\map{\tr^*}{\spFrmB{n}}{\spcovFrmB{n-1}}$ denotes the dual trace map.

The volumetric component of the rate of strain $\epCvol$ is an important quantity that describes the volumetric deformation of the elastic body. In fact, one can show that is equal to the divergence of the convective velocity:
\begin{equation}\label{eq:tr_epC}
	\divrC(\vC) = \tr(\nabC \vfC) = \tr(\epC) + \tr(\hat{w}) = \epCvol,
\end{equation}
which follows since the trace of $\hat{w}$ is equal to zero due to its skew-symmetric nature.

Finally, the power balance characterized by $\hat{\cl{D}}_{v/d}$ is given by
\begin{equation}\label{eq:Dvd_balance}
	\duPairB{\stC}{\epC} = \duPairB{\subTxt{\stC}{vol}}{\epCvol} + \duPairB{\subTxt{\stC}{dev}}{\epCdev},
\end{equation}
where the first pairing in the RHS is in terms of scalar-valued forms.

\subsection{Intensive Rougee stress tensor field}

The last point to consider is the relation between the (extensive) stress variables we introduced so far as covector-valued forms in $\spcovFrmB{n-1}$ and their usual representation as (intensive) second rank tensor fields.
The identification between the two is achieved using the metric structure represented by the Hodge star $\hodgeC$.
Let $\stTauC := \invhodgeC \stC \in \spvecFrmB{1} \cong \spVec{^1_1\cl{B}}$ denote the intensive version of $\stC\in \spcovFrmB{n-1}$.
In local coordinates for $n=3$, one can see that \cite{rashad2023intrinsic}
\begin{equation}\label{eq:stC_local}
	\stC_{AIJ} = \stTauC_A^B \mFC_{BIJ} \in \spFrmB{n-1},
\end{equation}
which indicates that $\stC$ contains two types of information; one related to the intrinsic mass form $\mFC$ and another related to the intensive (i.e. mass and volume independent) stress which should be provided by the elastic properties of the body.
Indeed this combination in (\ref{eq:stC_local}) is what makes the form-part and value-part explicit in the exterior calculus formulation. However, for the purposes of defining the stress-strain constitutive relation, it suffices to provide an expression relating $\stTauC$ and $\epC$.

The condition that $\stC \in \spcovFrmBsub{n-1}{sym}$ is equivalent to the symmetrization of the 2-contravariant tensor $\stTauC^\sharp\in \spVec{^2_0\cl{B}}$, i.e. $\stTauC^{IJ} = \stTauC^{JI} $  which is the standard stress-symmetry requirement in tensor calculus.
Furthermore, the expression of $\stC$ in (\ref{eq:Dvd}):
$$\stC = \tr^*(\subTxt{\stC}{vol}) + \subTxt{\stC}{dev}$$
is equivalent to 
\begin{equation}\label{eq:stress_v_d_decom}
	\stTauC = \subTxt{\stTauC}{vol} I_n + \subTxt{\stTauC}{dev},
\end{equation}
with $\subTxt{\stTauC}{vol} = \star^{-1} \subTxt{\stC}{vol} \in \spFrmB{0}$ and $\subTxt{\stTauC}{dev} = \invhodgeC \subTxt{\stC}{dev} \in \spvecFrmB{1}$ denoting the volumetric and deviatoric components of $\stTauC$, respectively.
The power balance (\ref{eq:Dvd_balance}) can then be rewritten in exterior and tensor calculus, respectively, as  
\begin{equation}\label{eq:Dvd_balance_int}
	\intB\epC \wedgedot \hodgeC \stTauC = \intB \epCvol \wedge \star\stTauCvol  + \epCdev \wedgedot \hodgeC \stTauCdev = \intB (\epCvol\stTauCvol  + \epCdev : \stTauCdev) \mFC.
\end{equation}
Note that in the above power balance the inner product of $\epCvol$ and $\stTauCvol$ is in terms of scalar-valued forms with the Hodge star $\hat{\star}$ including mass information such that $\hat{\star} \mFC = \mathds{1}$ where $\mathds{1}$ denotes the identity function on $\cl{B}$.
In Sec. \ref{sec:const_relations_elast}, we will discuss a number of constitutive models to close the relation between $\epC$ and $\stTauC$.

\subsection{Summary of the convective \pH dynamics}

\begin{figure}
	\centering
	\includegraphics[width=0.8\columnwidth]{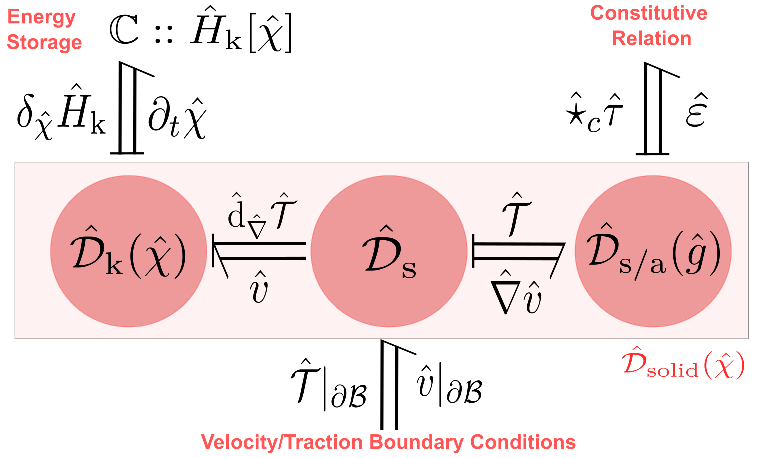}
	\caption{Convective \pH structure of nonlinear elasticity showing its constituting subsystems in bond graph notation. The boundary ports are depicted without causality information for simplicity.}
	\label{fig:overall_k_plus_s}
\end{figure}
In conclusion, the \pH dynamics we constructed so far is characterized by the storage of kinetic energy, one port for traction boundary input, one port for velocity boundary input, and the open distributed port $(\epC,\hodgeC \stTauC)$ which will be closed by the stress constitutive law. At the heart of the model are the three Dirac structures $\diracKinC, \diracStrC, \hat{\cl{D}}_{s/a}$ which we combine into $\subTxt{\hat{\cl{D}}}{solid}$, and will characterize the combined power balances (\ref{eq:diracKinC_balance},\ref{eq:diracStrC_balance},\ref{eq:Dsa_balance_2}).
With reference to Fig. \ref{fig:overall_k_plus_s}, the resulting \pH model in the convective representation is summarized for convenience by the following theorem.

\begin{theorem}
	The equations of motion governing the convective metric and extensive momentum variables $\hat{\chi} := (\gC,\momC) \in \spMetB\times\spcovFrmB{n} =: \spXkinC$ is represented in the \pH framework as
	\begin{align}
		\TwoVec{\partial_t \gC}{\partial_t \momC} &= \TwoTwoMat{0}{2\ \sym\circ \gC \circ \extcdC}{2\ \extcdC \circ \gC}{\Lie{(\cdot)}{\momC}} \TwoVec{\varD{\HkinC}{\gC}}{\varD{\HkinC}{\momC}} + \TwoVec{0}{\extcdC} \hodgeC \stTauC, \label{eq:pH_conv_summary_1}\\
		\epC &= \left( 0 \quad \subTxt{\hat{\pi}}{sym} \circ  \nabC \right) \TwoVec{\varD{\HkinC}{\gC}}{\varD{\HkinC}{\momC}}\label{eq:pH_conv_summary_2}
	\end{align}
	where the Hamiltonian functional $\map{\HkinC}{\spXkinC}{\bb{R}}$ is given by (\ref{eq:HkinC}), the variational derivatives of $\HkinC$ are given by Prop. \ref{prop:pH_kinetic_C}, and the projection map $\subTxt{\hat{\pi}}{sym}$ is given by (\ref{eq:piCsym})..
	The symmetry of the stress $\stTauC\in\spvecFrmBsub{1}{sym}$ is assumed, while the boundary data are given by $\vC |_{\bndB}$ and $\stC|_{\bndB}$, with $\stC = \hodgeC \stTauC \in \spcovFrmB{n-1}$.
	
	The rate of change of $\HkinC$ along trajectories of (\ref{eq:pH_conv_summary_1}-\ref{eq:pH_conv_summary_2}) is given by
	\begin{equation}
		\HkinCdot = \int_{\bndB} \vC |_{\bndB}\wedgedot\stC|_{\bndB} - \intB \epC \wedgedot\hodgeC \stTauC.
	\end{equation}
\end{theorem}

%
%

We finally conclude by a number of insights that the \pH model above reveals, in contrast to the tensor calculus PDEs in Sec. \ref{sec:geom_mech_tens_calc}.
\begin{enumerate}
	\item The formulation of nonlinear elasticity above explicates i) the geometric structure using coordinate free expressions, ii) the topological structures using exterior calculus, and iii) the energetic structure using the Dirac structures and power ports.
	\item There is a clear role of duality between effort and flow variables that make up the power ports.
	\item There is a clear identification of the skew-symmetric operator $\map{\hat{J}(\hat{\chi})}{T^* \spXkinC}{T \spXkinC}$ which represents the Poisson structure, associated to the convective Poisson bracket ${}^c\{\cdot,\cdot\}$ which can be defined similar to (\ref{eq:Pois_brac_S}).
	The skew-symmetry of $\hat{J}(\hat{\chi})$ is directly related to the conservation of energy characterized by the Poisson bracket.
	\item There is a clear separation between the constitutive relations and the Dirac structures. One key consequence is that one distinguishes between the nonlinearities of each, for example as in the state dependency of $\diracKinC(\hat{\chi})$.
	\item It shows that the evolution of the dynamic equation (in the convective and spatial case) is independent of the configuration $\map{\varphi}{\cl{B}}{\cl{S}}$. In fact, one case reconstruct it by solving the ordinary differential equation
		\begin{equation}\label{eq:reconst_phi_C}
			\partial_t \varphi^{-1}(x,t) = - \vC (\varphi^{-1}(x,t),t).
		\end{equation}
	\item The boundary conditions and how they affect the power balance can be easily identified. In fact, it shows that one does not need to impose displacement boundary conditions but rather velocity boundary conditions. This is also a crucial point when interconnecting flexible and rigid bodies that leads either to displacement constraints or velocity constraints. Indeed both types of boundary conditions are one-to-one as shown in (\ref{eq:reconst_phi_C}). However, it has several consequences on numerical methods. We shall come back to this point later in Sec. \ref{sec:incomp_hyper_elast} when discussing incompressibility.
	\item The equivalence between the balance laws corresponding to the \pH model (\ref{eq:pH_conv_summary_1}-\ref{eq:pH_conv_summary_2}), \cf Corollary \ref{cor:balance_laws_C}, and their tensor calculus counterparts (\ref{eq:mDC_dot}-\ref{eq:DE_C_std}) can be found in \cite{rashad2023intrinsic}.
\end{enumerate}

\section{Port-Hamiltonian model of fluid flow}\label{sec:pH_fluid}

Nonlinear elasticity has a lot of similarity with fluid mechanics. A compressible fluid can be considered a special case of visco-elastic materials \cite{Marsden1994}.
Even though the governing equations of motion and the Hamiltonian reduction process of both dynamical systems are identical, there a few fundamental differences between fluids and solids from the geometric point of view.
In this small section, we shall highlight these similarities and differences by presenting a \pH formulation of the dynamics of fluid flow.

\subsection{Geometric formulation}

In principle, the geometric setting of fluids can also be described by a matter space $\cl{B}$ that is embedded in an ambient space $\mathscr{A}$ along the same line of Sec. \ref{sec:geom_mech_tens_calc}. In this setting, one should consider an additional control volume subdomain of $\mathscr{A}$ where the dynamics will be described as in \cite{califano2022energetic}. However, it is more common in the literature to premise the geometric formulation of fluids by assuming the fluid particles to form a continuum filling the whole spatial domain, denoted by $\cl{S}$, which we assume to have a Riemannian structure \cite{arnold1965topologie,marsden1982hamiltonian}. The configuration of the fluid is given by the diffeomorphism $\map{\varphi}{\cl{S}}{\cl{S}}$ with the configuration space given by $\spC:= \text{Diff}(\cl{S})$, the space of diffeomorphisms on $\cl{S}$. The configuration space has the structure of a Frechet-Lie group \cite{modin2011euler}.

Similar to solids, the material velocity of the fluid is represented as a vector on the infinite-dimensional manifold $\spC$ and one can define the spatial and convective velocity fields $\vS, \vC \in\spVec{\cl{S}}$ similar to (\ref{eq:vel_relations}).
The Lie group structure of $\spC$ allows us to interpret  (\ref{eq:vel_relations}) as a pushforward of the material velocity $\vM \in T_\varphi\spC$ to the left (convective) and right (spatial) Lie algebras, respectively.

The group structure of $\spC$ and the diffeomorphic property of $\varphi$ imply physically that the fluid particles remain within $\cl{S}$ for all time. This is only the case if $\cl{S}$ is a compact manifold without a boundary or that it has a boundary $\Sbound$ that is impermeable, i.e. the fluid velocity is always parallel to the boundary \cite{Rashad2021Port-HamiltonianEnergy}.
This boundary impermeabiltiy condition is expressed in exterior calculus as $\iota_{\vS}\vFS|_{\Sbound} = 0 \in \spFrm{\Sbound}{n-1}$, where $\vFS\in\spFrm{\cl{S}}{n}$ denotes the volume form of the spatial domain associated to the metric $g$. In vector calculus, this condition corresponds to the vanishing of the normal component of the velocity at the boundary.
All related works on the Hamiltonian formulation of fluids (e.g. \cite{marsden1983coadjoint,marsden1984semidirect,holm1986hamiltonian}) are limited to this case of impermeable or no boundary.
However, we shall show next how this assumption can be relaxed using Dirac structures during the Hamiltonian reduction procedure.

\subsection{Port-Hamiltonian model}
The \pH modeling process proceeds in the same manner of Sec. \ref{sec:pH_modeling_solid}:
\begin{enumerate}
	\item The material equations of motion are given by the canonical Poisson structure (\ref{eq:can_poisson}) of the cotangent bundle $T^*\spC$ such that the kinetic energy functional (\ref{eq:HkinM}) is conserved.
	\item Following the Hamiltonian reduction procedure in Theorem \ref{th:Dirac_kin_S}, one can transform the canonical Poisson structure to a Lie-Poisson structure in the spatial representation.
	\item By extending the Poisson structure to a Dirac structure, one reaches the \pH dynamical system (\ref{eq:pH_sys_kin_S_1}-\ref{eq:pH_sys_kin_S_2}) representing the kinetic energy subsystem.
	\item Then one adds the stress and velocity boundary conditions via the stress subsystem, as in Sec. \ref{sec:stress_power}, and factorizes the rigid body motions, as in Sec. \ref{sec:decomposition}.
	\item Finally, one adds the constitutive relation between the rate of strain and stress variables $\epS,\stTauS \in \spvecFrmS{1}$ based on the class of fluids to be modeled.
\end{enumerate}

One prominent difference between fluids and solids is that the former are allowed to have momentum flux through the spatial image $\Sbound$ of the boundary, i.e. $\iota_\vS \momS|_{\Sbound} \neq 0$. In contrast to the Hamiltonian formalism, one can account for such energy exchange through the boundary in Step 3 by extending the Dirac structure (\ref{eq:Dirac_kin_S}) with the boundary interaction port $(f_\partial,e_\partial)\in \spvecFrmSbnd{0}\times \spcovFrmSbnd{n-1}$ with $f_\partial:= \eS{\momS}|_{\Sbound}$ and $e_\partial:= -(\eS{\mFS}\mFS + \iota_{\eS{\momS}}\momS)|_{\Sbound}$. Consequently, the power balance (\ref{eq:Dirac_Balance_S}) then becomes
\begin{equation}\label{eq:Dirac_Balance_S_fluid}
	\duPairS{\eS{\mFS}}{\fS{\mFS}} + \duPairS{\eS{\momS}}{\fS{\momS}} = \duPairS{\subTxt{e}{d}}{\subTxt{f}{d}} + \duPair{e_\partial}{f_\partial}{\Sbound},
\end{equation}
and as a result the energy exchange due to momentum flux via $\Sbound$ appears in (\ref{eq:HkinS_open}).

Interesting, but not surprisingly, the vanishing condition of $\iota_\vS \momS|_{\Sbound}$ in Prop. \ref{prop:pH_kin_S} is equivalent to the boundary impermeability condition since $\iota_\vS \momS|_{\Sbound} = \iota_{\vS}\mFS\otimes\vfS$. This in fact shows that the group structure of $\spC$ and the diffeomorphic property of $\varphi$ are pertinent to the Hamiltonian formulation which necessitates such zero-exchange of energy through the boundary \cite{Rashad2021Port-HamiltonianEnergy}.

\begin{figure}
	\centering
	\includegraphics[width=0.95\columnwidth]{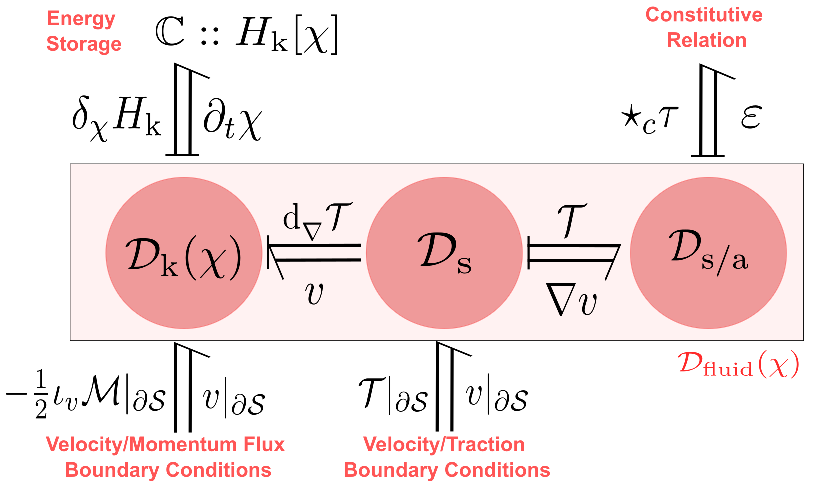}
	\caption{Spatial \pH structure of fluid flow showing its constituting subsystems in bond graph notation. The boundary ports are depicted without causality information for simplicity.}
	\label{fig:pH_fluid_flow}
\end{figure}

In conclusion, the \pH representation of fluid dynamics, summarized in Fig. \ref{fig:pH_fluid_flow}, is characterized by the storage of kinetic energy, boundary ports for traction, momentum flux, and velocity on $\Sbound$, and the open distributed port $(\epS,\hodgeS \stTauS)$ which will be closed by the stress constitutive law. At the heart of the model are the three Dirac structures $\diracKinS, \diracStrS, {\cl{D}}_{s/a}$ which we combine into $\subTxt{{\cl{D}}}{fluid}$.
The resulting \pH model in the spatial representation is summarized by the following theorem.

\begin{theorem}
	The equations of motion governing the extensive mass and momentum variables $\chi := (\mFS,\momS) \in \spFrmS{n}\times\spcovFrmS{n} =: \spXkinS$ is represented in the \pH framework as
	\begin{align}
		\TwoVec{\partial_t \mFS}{\partial_t \momS} &= \TwoTwoMat{0}{ - \extd \iota_{(\cdot)} \mFS}{- \mFS \otimes \extd (\cdot)}{- \Lie{(\cdot)}{\momS}} \TwoVec{\varD{\HkinS}{\mFS}}{\varD{\HkinS}{\momS}} + \TwoVec{0}{\extcdS} {\hodgeS \stTauS}, \label{eq:EoM_fluid_1}\\
		\epS &= \left( 0 \quad \subTxt{\pi}{sym}\circ \nabS\right) \TwoVec{\varD{\HkinS}{\mFS}}{\varD{\HkinS}{\momS}},\label{eq:EoM_fluid_2}
	\end{align}
	where the Hamiltonian functional $\map{\HkinS}{\spXkinS}{\bb{R}}$ is given by (\ref{eq:HkinS}), the variational derivatives of $\HkinS$ are given by Prop. \ref{prop:pH_kin_S}, and the projection map $\map{\subTxt{\pi}{sym}}{\spvecFrmS{1}}{\spvecFrmS{1}}$ is given by $\subTxt{\pi}{sym}:= \gS^{-1}\circ \mathrm{sym}\circ \gS$.
	The symmetry of the stress $\stTauS\in\spvecFrmSsub{1}{sym}$ is assumed, while the boundary data are given by $\vS |_{\Sbound}$ and $(\stS-\half \iota_{\vS}\momS)|_{\Sbound}$, with $\stC = \hodgeC \stTauC \in \spcovFrmB{n-1}$ and $\vS = \invhodgeS \momS \in \spvecFrmS{0}$.
	
	The rate of change of $\HkinS$ along trajectories of (\ref{eq:EoM_fluid_1}-\ref{eq:EoM_fluid_2}) is given by
	$$\HkinSdot = \int_{\Sbound} \vS|_{\Sbound} \wedgedot (\stS - \half\iota_\vS\momS)|_{\Sbound} - \intS \epS \wedgedot\hodgeS \stTauS.$$
\end{theorem}
%

Finally, the equivalence between the balance laws corresponding to the \pH model (\ref{eq:EoM_fluid_1}-\ref{eq:EoM_fluid_2}), \cf Corollary \ref{cor:balance_laws_S}, and their tensor calculus counterparts (\ref{eq:EoM_S_mDS}-\ref{eq:EoM_S_Sig}) can be found in \cite{rashad2023intrinsic}.

\section{Constitutive relations}\label{sec:const_relations_elast}

Now we turn attention to the constitutive relation between the rate of strain and stress variables, which are both symmetric tensor-fields.
We aim in this section to highlight how different classes of constitutive relations are incorporated in the \pH framework in addition to the importance of the convective representation for constitutive modeling. In particular, we aim to emphasize the role of $\spMetB$ in defining constitutive equations that are geometrically consistent, attributed to the work of Rougee \cite{Rougee2006}.

In general, neglecting thermodynamic effects, the constitutive law is abstractly a relation between the stress and rate-of-strain variables $\stTauC, \EpsC\in \spvecFrmBsub{1}{sym}$ and some time-integral variable $\StrC\in \mathfrak{D}$ of $\EpsC$ that characterizes deformation.
The different choices for the construction of $\StrC$ and its corresponding space $\mathfrak{D}$ will be discussed in detail in this section.
Using $\EpsC$, one can rewrite the evolution equation for the convective metric (\ref{eq:gC_dot}) as
\begin{equation}\label{eq:dt_gC}
	\partial_t \gC = 2 \gC\cdot \epC \in T_{\gC}\spMetB \cong \spcovFrmB{1},
\end{equation}
which identifies vectors on $\spMetB$ with the covariant version of $\epC$. Therefore, it is expected that $\StrC$ to be constructed from the metric state $\gC$.
The constitutive law will be denoted by 
$$\mathscr{R}(\stTauC, \EpsC, \StrC) = 0.$$
In general, $\mathscr{R}$ depends also on $X\in \cl{B}$ to incorporate non-homogeneous effects in the elastic material. Without loss of generality, we shall only focus on the homogeneous case in this work.

From an energetic perspective, the power port $(\EpsC,\hodgeC \stTauC)$ has to be closed by either 1) an energy-storage unit characterized by a relation between $\stTauC$ and $\StrC$, 2) an energy dissipation unit characterized by a relation between $\stTauC$ and $\EpsC$, or 3) a combination of both.
The three cases above classify elastic, viscous, and visco-elastic materials, respectively.
The relations could be either linear or nonlinear, as well as static or dynamic, as in hyper-elasticity or hypo-elasticity, respectively.

In what follows, we shall only focus on the case of (homogeneous) hyper-elastic materials in Secs. \ref{sec:gen_hyper_elast}-\ref{sec:incomp_hyper_elast} and viscous fluid flow in Sec. \ref{sec:visc_fluid}.

\subsection{General hyper-elasticity}\label{sec:gen_hyper_elast}
%

The power port $(\EpsC,\hodgeC \stTauC)$ is closed for hyper-elastic materials by an energy-storage unit characterizing the functional
$$\hat{\Psi}[\StrC] = \intB \hat{\psi}(\StrC) \mFC,$$
where $\map{\hat{\psi}}{\mathfrak{D}}{\spFn{\cl{B}}}$ is its specific storage energy function (which is independent of mass and volume).

\begin{proposition}\label{prop:DE_formula}
	Let the variable $\StrC \in \mathfrak{D}$ be constructed from the metric state $\gC$ by some function $\map{f}{\spMetB}{\mathfrak{D}}$ and let $\map{Tf}{T_{\gC} \spMetB}{T_{\StrC}\mathfrak{D}}$ denote its tangent map. Then, the stress constitutive law takes the following generalized form of the Doyle-Ericskon formula
	\begin{equation}\label{eq:DE_formula}
		\stTauC = 2 \invhodgeC \hat{f}^*\left( \hodgeC\frac{\partial \hat{\psi}}{\partial\StrC}(\StrC)\right),
	\end{equation}
	where $\map{\hat{f}^*}{T^*_{\StrC}\mathfrak{D}}{\spcovFrmB{n-1}}$ denotes the dual of $\map{\hat{f} := Tf \circ \gC}{\spvecFrmB{1}}{T_{\StrC}\mathfrak{D}}$ with respect to the duality pairing $\duPairB{\cdot}{\cdot}$.
\end{proposition}
\begin{proof}
	Let $\delta_\StrC \hat{\Psi} \in T^*_{\StrC}\mathfrak{D}$ denote the variational derivative of $\hat{\Psi}$ with respect to $\StrC$.
	The rate of change of $\hat{\Psi}$ is then given by
	$$\dot{\hat{\Psi}} 
	= \intB \partial_t \StrC \wedgedot \delta_\StrC \hat{\Psi} 
	= \intB Tf(\partial_t \gC)\wedgedot \delta_\StrC \hat{\Psi}
	= \intB 2 Tf(\gC \EpsC)\wedgedot \delta_\StrC \hat{\Psi} 
	= \intB 2 \EpsC\wedgedot \hat{f}^*(\delta_\StrC \hat{\Psi} ).$$
	A necessary condition for the energy balance 
	$\dot{\hat{\Psi}} = \intB\EpsC\wedgedot \hodgeC \stTauC,$
	to hold and be covariant\footnote{i.e. invariant to spatial transformations of the ambient space} is that $\hat{\Psi}$ should be independent of derivatives of $\StrC$ \cite[Ch.3]{Marsden1994}.
	Consequently, the variational derivative is identified with the partial derivative of the top form $\hat{\psi}(\StrC) \mFC$ leading to the expression $\delta_\StrC \hat{\Psi} = \hodgeC \frac{\partial \hat{\psi}}{\partial\StrC}$.
	\qed
\end{proof}

Using the definition of $\dot{\hat{\Psi}} = \duPairB{\delta_\StrC \hat{\Psi} }{\partial_t \StrC}$, the total energy balance of the \pH system of nonlinear elasticity (\ref{eq:pH_conv_summary_1}-\ref{eq:pH_conv_summary_2}) after adding the constitutive relation (\ref{eq:DE_formula}) can be expressed as
$$\frac{\extd}{\extd t} (\HkinC + \hat{\Psi}) = \int_{\Bbound} \vC|_{\Bbound} \wedgedot \stC|_{\Bbound}.$$
The above power balance states that the rate of change of total (kinetic plus strain) energy within the elastic body's domain $\cl{B}$ is equal to the power flow through the boundary $\Bbound$, due to stress.

The most natural choice for the quantity $\StrC$ is $\gC\in\spMetB$ itself since the space of Riemannian metrics serves as a natural space of deformation \cite{rashad2023intrinsic,stramigioli2024principal}. In this case, $\StrC$ represents a \textit{state-of-deformation} and one has $f = \text{id}$ and $\hat{f} = \hat{f}^* = \gC$ which simplifies the stress law (\ref{eq:DE_formula}) to be
\begin{equation}\label{eq:DE_formula_standard}
	\stTauC = 2 \gC \frac{\partial \hat{\psi}}{\partial\gC}(\gC),
\end{equation}
which is the standard Doyle-Erickson formula.

The alternative choice which we shall adopt, is to define  $\StrC$ as a \textit{strain-variable} represented by a vector-valued 1-form, i.e. $\mathfrak{D} = \spvecFrmB{1}$.
In the vast literature of nonlinear hyper-elasticity, there have been several propositions for defining the strain $\StrC$ and several propositions for defining the storage energy $\hat{\psi}$ leading to numerous expressions for the stress $\stTauC$ which very quickly ramifies when we consider their spatial and material counterparts along with tensorial variants.
Thanks to the work of Rougee \cite{Rougee2006} and Fiala \cite{Fiala2011GeometricalMechanics}, taking the (non-Euclidean) geometric nature of $\spMetB$ into account leads to the fact that there is \textit{only one natural way} to define the strain as the relative deformation between two states in $\spMetB$ as discussed next.


\subsection{Finite-strain hyper-elasticity}

\begin{figure}
	\centering
	\includegraphics[width=\columnwidth]{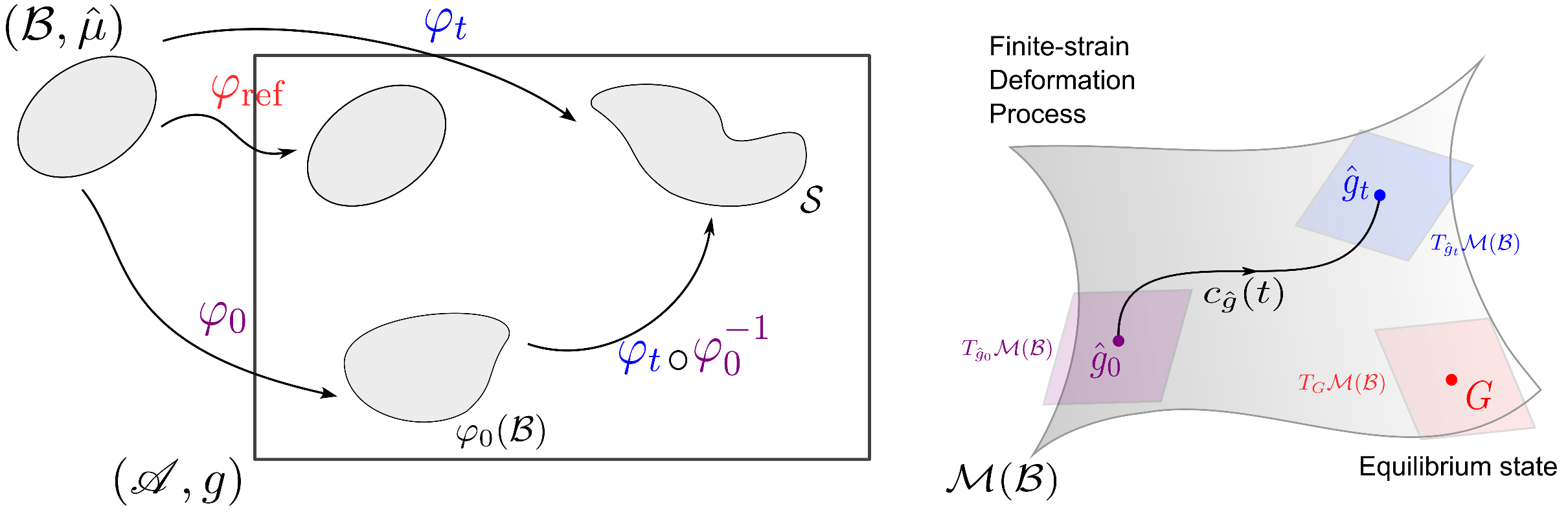}
	\caption{Deformation process in finite-strain theory.}
	\label{fig:deformation_process_schem}
\end{figure}

At this point  it is essential to introduce the concept of the reference metric $G\in \spMetB$ which will characterize the equilibrium reference state of deformation of the body. This equilibrium state also corresponds to the minimum of the storage elastic energy $\hat{\psi}$ considered as a function on $\spMetB$.
Given $G$, one can associate to it the reference configuration $\map{\cfgref}{\cl{B}}{\mathscr{A}}$ such that $G = \cfgPref(\gS)$.
With reference to Fig. \ref{fig:deformation_process_schem}, the deformation process will be characterized by the curve $c_{\gC}(t)$ in $\spMetB$ starting at the initial state $\gC_0:= c_{\gC}(0) \in \spMetB$. If the deformation process starts from the reference configuration then $\cfgref = \varphi_0$ and $\gC_0 = G$, but in general the elastic body could be in a stressed state at $t=0$.

The strain variable $\StrC_t$ at any time $t$ is defined as the geodesic connecting $\gC_t$ to the reference state $G$. It has been shown in \cite{Fiala2011GeometricalMechanics} that this geodesic is characterized by $\partial_t \epC = 0$, i.e. $\epC$ is constant along the geodesic, which leads to the following exponential and logarithm maps of $\spMetB$ \cite{Fiala2011GeometricalMechanics}
\begin{equation}\label{eq:def_Exp_Log}
	\fullmap{\mathrm{Exp}_G}{T_{G}\spMetB}{\spMetB}{\delta G}{G \Exp(G^{-1} \delta G)}
	\qquad \qquad
	\fullmap{\mathrm{Log}_G}{\spMetB}{T_G\spMetB}{\gC}{ G\Ln(G^{-1}\gC)}
\end{equation}
where $\Exp$ and $\Ln$ denote the matrix exponential and natural logarithm maps, respectively, applied to mixed second-rank tensors.
By introducing $\hat{C}_t := G^{-1} \gC_t$ as shorthand notation, we define the strain variable to the vector-valued 1-form given by
\begin{equation}\label{eq:def_StrC}
	\StrC_t := \half \ln (\hat{C}_t) \in \spvecFrmB{1},
\end{equation}
which leads to the nonlinear map $f(\gC) : = \half \ln (G^{-1}\gC)$.

\begin{remark}
	It is important to note that the expressions of (\ref{eq:def_Exp_Log}) and (\ref{eq:def_StrC}) are mere notation for expressing the geodesic flow on $\spMetB$. Thus, one should not interpret $G^{-1} \gC$ as an index raising of $\gC$ by $G$ since both metrics are two different points on the manifold $\spMetB$. In \cite{Fiala2011GeometricalMechanics} this distinction between thee geometric objects and their algebraic representation was made by denoting the latter with bold.
\end{remark}

From the definition of the strain (\ref{eq:def_StrC}), one has that 
$$\StrC_t = 0 \implies \hat{C}_t = I \implies \gC_t = G,$$
stating that $G$ is the zero-stress undeformed state of the body. However, it should warned that $\spMetB$ does not have this privileged state intrinsically.
In fact, it has been shown by Fiala \cite{Fiala2011GeometricalMechanics} that this is a consequence of the geodesic completeness of $\spMetB$, i.e. any two deformation states can be connected through a geodesic. From the construction we presented here, it can be seen that $G$ is actually a property of the elastic body's material. Some composite materials might even have multiple zero-stress states, such as bielastic structures used in soft-robotics.

As reviewed by \cite{Neff2015TheConvexity}, the logarithmic strain measure (\ref{eq:def_StrC}) possesses a number of remarkable properties. Perhaps the most important one is that it additively separates dilation from pure distortion, as shown in the following important result.

\begin{theorem}\label{th:vol-dev_decomp}
	Consider the volumetric-deviatoric decomposition of the strain tensor
	\begin{equation}\label{eq:strC_decomp}
		\StrC = \frac{1}{n} \StrCvol I_n + \StrCdev,
	\end{equation}
	where $\StrCvol := \tr(\StrC) \in \spFrmB{0}$ and $\StrCdev := \StrC - \frac{1}{n} \StrCvol I_n \in \spvecFrmBsub{1}{dev}$.
	Then, 
	$$\partial_t \StrCvol = \epCvol,$$
	where $\epCvol:=\tr(\epC) \in\spFrmB{0}$ is the volumetric component of the rate of strain tensor.
\end{theorem}
\begin{proof}
	First, we rewrite (\ref{eq:def_StrC}) as $\alpha = \Ln(\hat{C})$ with $\alpha := 2\StrC$. Using the power series definition of the matrix exponential we have that
	$$\hat{C} = \Exp(\alpha) = I + \alpha + \frac{1}{2!} \alpha ^2 + \frac{1}{3!} \alpha ^3 + \cdots, $$
	and similarly using the properties of the matrix exponential
	\begin{equation}\label{eq_proof:Cinv}
		\hat{C}^{-1} = \Exp(-\alpha) = I - \alpha + \frac{1}{2!} \alpha ^2 - \frac{1}{3!} \alpha ^3 + \cdots. 
	\end{equation}
	From the definition of $\hat{C} = G^{-1}\gC$, it follows that that
	$\dot{\hat{C}} = 2 G^{-1}\gC \epC = 2 \hat{C} \epC,$
	and consequently $2 \tr(\epC) = \tr(\hat{C}^{-1}\dot{\hat{C}})$.
	
	The rate of change of $\hat{C}$ can be expressed using the product rule as
	\begin{equation}\label{eq_proof:Cdot}
		\dot{\hat{C}} = \partial_t \Exp(\alpha) = \dot{\alpha} + \frac{1}{2!} (\alpha \dot{\alpha} + \dot{\alpha}\alpha) + \frac{1}{3!}(\dot{\alpha}\alpha^2  + \alpha \dot{\alpha}\alpha + \alpha^2 \dot{\alpha}) + \cdots .
	\end{equation}
	By multiplying the expressions of $\hat{C}^{-1}$ and $\dot{\hat{C}}$ in (\ref{eq_proof:Cinv}) and (\ref{eq_proof:Cdot}), then expanding the product of sums, keeping only the fourth order terms and lower, one can show that
	$$\hat{C}^{-1} \dot{\hat{C}} = \dot{\alpha} + \frac{1}{2!}( \dot{\alpha}\alpha - \alpha \dot{\alpha})+ \frac{1}{3!}( \dot{\alpha}\alpha^2 - 2\alpha \dot{\alpha}\alpha + \alpha^2 \dot{\alpha}) +\frac{1}{4!}( \dot{\alpha}\alpha^3 - 3\alpha \dot{\alpha}\alpha^2 + 3\alpha^2\dot{\alpha} \alpha- \alpha^3 \dot{\alpha}).$$
	Taking the trace of the above expression and using the cyclic property of the trace will lead to the vanishing of all terms in the RHS except $\dot{\alpha}$ leading to
	$$2 \tr(\epC) = \tr(\hat{C}^{-1}\dot{\hat{C}}) = \tr(\dot{\alpha}) = 2 \tr(\dot{\StrC}).$$
	Due to the commutativity of the trace with time-differentiation, then $\tr(\dot{\StrC}) = \partial_t\tr(\StrC) = \partial_t \StrCvol $ which concludes the proof.
	\qed
\end{proof}

\begin{corollary}\label{cor:vol-dev_decomp_1}
	The rate of change of the deviatoric part of $\StrC$ is not equal to $\epCdev$ in general but can be computed by
	$$\partial_t \StrCdev = \partial_t \StrC - \frac{1}{n} \epCvol.$$
\end{corollary}
\begin{proof}
	For $\partial_t \StrCdev = \epCdev$ to hold then one must have that $\partial_t \StrC = \epC$ to hold.  As shown earlier in the proof of Prop. \ref{prop:DE_formula}, one has that 
	$$\partial_t \StrC = 2 Tf \circ \gC \cdot \epC, $$
	which for the logarithmic strain definition $f(\gC) = \half \ln (G^{-1}\gC)$ one does not have that $Tf = \half \gC^{-1}$.
	\qed
\end{proof}

The above decomposition of $\StrC$ implies that the constitutive closure problem can be done separately for the volumetric deformation $\StrCvol$ and the isochoric deformation $\StrCdev$. Thus, one can express $\hat{\Psi}$ as
$$\hat{\Psi}[\StrCvol,\StrCdev] = \intB (\psiCvol(\StrCvol) + \psiCdev(\StrCdev))\mFC,$$
such that 
$$\dot{\hat{\Psi}} 
= \intB \partial_t\StrCvol \wedge 
\hat{\star} \frac{\partial \psiCvol}{\partial\StrCvol} ( \StrCvol) + 
\partial_t\StrCdev \wedgedot\hodgeC \frac{\partial \psiCdev}{\partial\StrCdev} ( \StrCdev),$$
provided the nonlinearity of $\hat{\psi}$ does not couple $\StrCvol$ and $\StrCdev$.

The decomposition above is quite favorable in practice since the volumetric part $\StrCvol$ distinguishes compressible and incompressible materials from each other, as will be discussed later. The remarkable point that Theorem \ref{th:vol-dev_decomp} highlights is that $\StrCvol$ is an intrinsic quantity that is independent of the equilibrium configuration $G$ since it is the time integral of another intrinsic quantity. This result is consistent from a geometric point of view since it follows from (\ref{eq:tr_epC}) that
\begin{equation}\label{eq:strvol_dot}
	\partial_t \StrCvol = \epCvol = \tr(\epC) = \divrC(\vC) = \varphi^*(\divrS(v)),
\end{equation}
which shows that the volumetric deformation $\StrCvol \in \spFrmB{0}$ is simply the integral of the vector field's divergence, in the convective representation, which is indeed an intrinsic quantity independent of $G$.

If we consider the volumetric-deviatoric decomposition of the stress in (\ref{eq:stress_v_d_decom}) then $\stTauC$ is computed as the combination of the scalar function $\stTauCvol\in \spFrmB{0}$ and the vector-valued 1 form $\stTauCdev \in \spvecFrmBsub{1}{dev}$.
However, due to Corollary \ref{cor:vol-dev_decomp_1}, one cannot in general compute 
$\stTauCvol$ from the gradient of $\psiCvol$ and $\stTauCdev$ from the gradient of $\psiCdev$ separately.
Although this might seem counter-intuitive, it is in face a natural consequence of the (intrinsic) nonlinearity of $\hat{f}^*$ in the Doyle-Erickson formula (\ref{eq:DE_formula}). In other words, even though one can decompose the power ports  $(\epC,\hodgeC\stTauC)$ and $(\partial_t \StrC,\varD{\hat{\Psi}}{\StrC})$ each into two, the nonlinearity of the transformation by $\hat{f}$ and $\hat{f}^*$ mixes these two components.

\begin{figure}
	\centering
	\includegraphics[width=0.9\columnwidth]{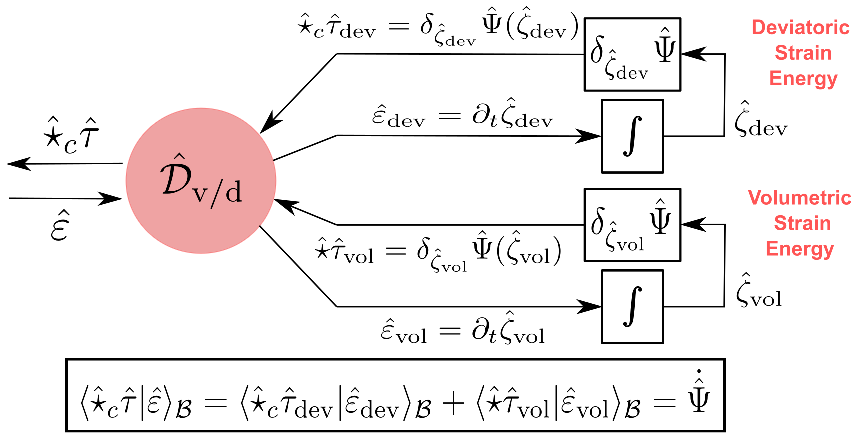}
	\caption{Convective constitutive relations of the for finite-strain isotropic hyper-elasticity}
	\label{fig:finite_hyperelast}
\end{figure}

In the special case of isotropic hyper-elasticity, luckily one can have the aforementioned separation \cite{Sansour2001OnIssues}. Consider for example the classic Hencky quadratic strain energy \cite{Neff2016GeometryMechanics} which is expressed in exterior calculus as
\begin{equation}\label{eq:Hencky_strain_energy}
	\hat{\Psi}[\StrCvol,\StrCdev] = \intB \half \left(\kappa \StrCvol \wedge \hat{\star} \StrCvol + 2 \theta \StrCdev \wedgedot \hodgeC \StrCdev \right),
\end{equation}
where $\kappa,\theta \in \spFn{\cl{B}}$ denote the bulk and shear moduli respectively. The stress-constitutive law, summarized in Fig. \ref{fig:finite_hyperelast}, takes then the form
\begin{equation}\label{eq:Hencky_const_model}
	\stTauCvol = \frac{\partial \psiCvol}{\partial\StrCvol} = \kappa \StrCvol \qquad\qquad \stTauCdev = \frac{\partial \psiCdev}{\partial\StrCdev} =  2 \theta \StrCdev.
\end{equation}

It has been shown in \cite{anand1979h,anand1986moderate} that the isotropic Hencky constitutive model above agrees with experiments for moderate principal stretch values.
One can extend the application range to high principal stretches by using a nonlinear strain energy $\hat{\psi}$ combined with the nonlinearity of $\StrC$, e.g. the exponentiated Hencky-strain energy \cite{Neff2015TheConvexity,Neff2016GeometryMechanics}. The extension of the isotropic Hencky model to the general anisotropic case and to isotropic plasticity can be found in \cite{Sansour2001OnIssues}.

\subsection{Infinitesimal-strain hyper-elasticity}

There are numerous hyper-elasticity models in the literature that rely on linear definitions of the strain measure $\StrC$ to represent finite-strains. However, as shown by Fiala \cite{Fiala2011GeometricalMechanics}, linear definitions of $\StrC$ is not consistent with the (non-Euclidean) geometry of $\spMetB$ unless one assumes the deformations to be infinitesimally small.

The theory of small (infinitesimal) strains represents the deformation of the elastic body $\cl{B}$ linearized about the reference metric $G\in\spMetB$.
Then, the deformation process characterized by the curve $c_{\gC}(t)$ in $\spMetB$ is approximated by a trajectory in $T_G\spMetB$. Consequently, the dynamics takes place in the vector space $T_G\spMetB$ which we identify with $\spvecFrmB{1}$.

At a time instant $t$, the strain tensor is then defined as 
\begin{equation}\label{eq:StrC_linear}
	\StrC_t := \half(\hat{C}_t - I), \qquad \qquad \hat{C}_t:= G^{-1} \gC_t\in \spvecFrmB{1},
\end{equation}
where $I \in \spvecFrmB{1}\cong \spVec{^1_1\cl{B}}$ denotes the identity tensor.
Consequently, using (\ref{eq:dt_gC}), one has that 
$$\partial_t \StrC_t := \half\partial_t\hat{C}_t = \half G^{-1} \partial_t\gC_t =  G^{-1} \gC_t \EpsC = \hat{C}_t \EpsC.$$
Thus, for the case of infinitesimal strain, $\hat{f}$ in Prop. \ref{prop:DE_formula} is given by the linear map $\hat{f} = \hat{C}$.
One can in fact show that linearizing the logarithmic strain measure (\ref{eq:def_StrC}) at $\hat{C} = I$ using Taylor series leads to the above linear strain definition.

Using the linear strain measure above, one can then define different classes of hyper-elastic models by choosing either linear or nonlinear strain energy functions.
For instance, the St. Venant-Kirchhoff model has a linear elastic strain energy of the form
$$\hat{\Psi}[\StrC] = \intB \half \StrC \wedgedot \hodgeC \bb{E} \StrC,$$
where $\map{\bb{E}}{\spvecFrmB{1}}{\spvecFrmB{1}}$ denotes the fourth rank elasticity tensor, which is independent of $\StrC$ but could vary at different points in $\cl{B}$ in the non-homogeneous case.
The resulting linear stress-strain constitutive law takes the form $\stTauC = \gC G^{-1} \bb{E} \StrC,$
which can be reformulated as the 2-contravariant tensor $\stTauC^\sharp:= \gC^{-1}\stTauC$ with local components $\stTauC^{AB}$:
$$\stTauC^\sharp = \hat{\bb{E}}(\gC - G) \in \spVec{^2_0\cl{B}}, \qquad \qquad \stTauC^{AB} = \hat{\bb{E}}^{ABIJ}(\gC_{IJ} - G_{IJ}) \in \spFn{\cl{B}},$$
with $\hat{\bb{E}}:= \half G^{-1} \bb{E} G^{-1}$.
%

The constitutive law above represents a generalized Hooke's law with the elasticity tensor $\hat{\bb{E}}$ having 21 independent components due to its (major and minor) symmetric properties.
The presence of material symmetries reduces the independent components of $\hat{\bb{E}}$ with the simplest case being homogeneous isotropic hyper-elasticity with only 2 independent components $\kappa,\theta$ denoting the bulk and shear modulus.
In this simple case, the strain energy is also given by (\ref{eq:Hencky_strain_energy}) with the stress volumetric and deviatoric components given by (\ref{eq:Hencky_const_model}).
While this linear stress-strain constitutive model is physically valid only in the (infinitesimally) small-strain regime, it accounts for arbitrary rigid body motions represented in the nonlinearity of the \pH model in (\ref{eq:pH_conv_summary_1}-\ref{eq:pH_conv_summary_2}).

Nonlinear models for isotropic hyper-elasticity, such as the Mooney-Rivlin and Neo-Hookean models, ar usually expressed in terms of rotational invariants of (\ref{eq:StrC_linear}):
$$I_1(\StrC): = \tr(\StrC), \qquad I_2(\StrC): = \half(\tr(\StrC)^2 - \tr(\StrC^2)), \qquad I_3(\StrC): = \det(\StrC).$$
The strain functional for the (compressible) Mooney Rivlin model takes the form
\begin{equation}\label{eq:Moon_Riv}
	\hat{\Psi}[\StrC] = \intB (c_1 I_1(\StrC) + c_2 I_2(\StrC) + c_3 I_3(\StrC))\mFC,
\end{equation}
where the material parameters $c_1,c_2\in\spFn{\cl{B}}$ characterize distortion response while $c_3\in\spFn{\cl{B}}$ characterizes dilatation response.
For the special case $c_2 = 0$, one gets the Neo-Hookean model.

It is interesting to note that both the Mooney-Rivlin model (\ref{eq:Moon_Riv}) and the quadratic Hencky model (\ref{eq:Hencky_strain_energy}) constitute a nonlinear closure relation for the port $(\epC,\hodgeC\stTauC)$.
While the Mooney Rivlin relies on a linear strain definition and a nonlinear strain energy, the Hencky model has a linear strain energy and a (geometrically) nonlinear strain definition.
Even though both models have a comparable number of material parameters to fit, the Hencky model has been shown to outperform the Mooney-Rivlin and Neo-Hookean models for moderate strains \cite{Neff2015TheConvexity}.
This again highlights the importance of taking the geometry of $\spMetB$ into account for defining finite-strain.

\begin{remark}
	We emphasize again that in principle, one does need to define strain at all (\cf \cite[Sec. 43]{Truesdell1966}) by defining the strain energy in terms of the state of deformation $\gC$ as in (\ref{eq:DE_formula_standard}). Consequently, the stress law can be perceived as a covector field on $\spMetB$ \cite{Rougee2006}. However, the concept of strain and its geometric nature as a mixed tensor comes with numerous useful properties (such as the trace, volumetric-deviatoric decomposition, rotational invariants, and eigenvalues) that are not available to the deformation state $\gC$ intrinsically due to its fully-covariant nature.
\end{remark}

\begin{remark}
	The pushforward of the mixed tensor-field $\hat{C}_t$ onto the reference configuration $\cfgref(\cl{B})$ corresponds to the \textit{Cauchy-Green deformation} tensor. Whereas, the pushforward of the strain measures (\ref{eq:StrC_linear}) and (\ref{eq:def_StrC}) onto $\cfgref(\cl{B})$ correspond, respectively, to standard \textit{Cauchy-Green strain} tensor and the \textit{logarithmic strain} tensor introduced by Hencky extending the work of Becker, \cf \cite{Martin2018AElasticity} for a historic review.
	Furthermore, let the reference mass density function be denoted by $\subTxt{\rho}{ref} \in \spFn{\cl{B}}$ such that $\mFC = \subTxt{\rho}{ref} \omega_G$ with $\omega_G \in \spFrmB{n}$ denoting the volume form corresponding to the metric $G$. Then, the pushforward of the stress $\subTxt{\rho}{ref} \stTauC$ onto $\cfgref(\cl{B})$ corresponds to the \textit{second Piola-Kirchhoff stress} tensor represented as a mixed tensor-field.
\end{remark}

%

\subsection{Incompressible hyper-elasticity}\label{sec:incomp_hyper_elast}

\begin{figure}
	\centering
	\includegraphics[width=0.9\columnwidth]{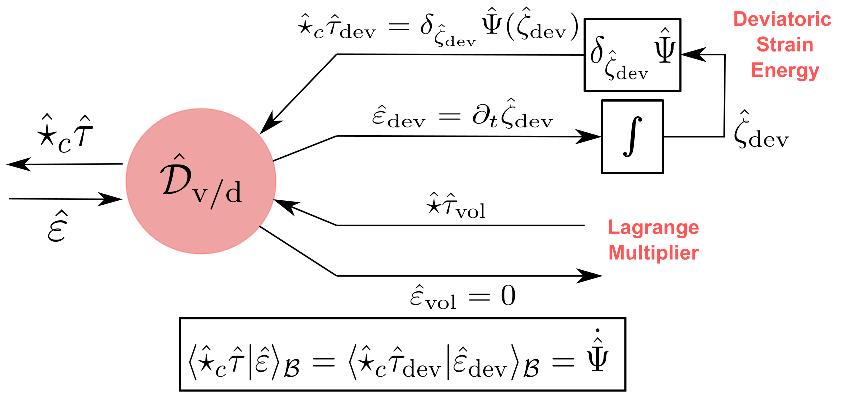}
	\caption{Convective constitutive relations of incompressible isotropic hyper-elasticity}
	\label{fig:incom_hyperelast}
\end{figure}

Certain elastic materials can be characterized, in certain strain-regimes, as being incompressible. A prototypical example is rubber. Such incompressibility is characterized intrinsically, in the spatial and convective representations, respectively, by the geometric conditions
$$\Lie{\vS}{\vFS} = 0 \implies \divrS(\vS) = 0, \qquad \qquad \Lie{\vC}{\vFC} = 0 \implies \divrC(\vC) = 0,$$
where $\vFS \in \spFrmS{n}, \vFC \in \spFrmB{n} $ denote the volume forms associated to the Riemannian metrics $\gS$ and $\gC$, respectively.

From the volumetric-deviatoric decomposition (\ref{eq:strC_decomp}) of the logarithmic strain measure discussed earlier, (\ref{eq:strvol_dot}) implies that the incompressibility constraint is equivalently expressed as
$$\epCvol = 0 \implies \StrCvol = \text{constant},$$
for any finite-strain deformation. 
Consequently, the volumetric stress $\stTauCvol$ is interpreted as a Lagrange multiplier enforcing this constraint and no longer is computed from a constitutive relation.
The stress-strain constitutive equations for incompressible hyper-elasticity, summarized in Fig. \ref{fig:incom_hyperelast}, can be expressed in general as
\begin{align*}
	\stTauC =& \stTauCvol I_n + \frac{\partial \psiCdev}{\partial \StrCdev},\\
	\partial_t \StrCdev =& \epCdev ,\\
	0 =& \epCvol.
\end{align*}
One can also model near-incompressibility effects trivially by choosing $\stTauCvol = \frac{\partial \subTxt{f}{pen}}{\partial \StrCvol}(\StrCvol)$ where $\map{\subTxt{f}{pen}}{\spFn{\cl{B}}}{\spFn{\cl{B}}}$ denotes the penalty function of choice.

The aforementioned constitutive equation for incompressible hyper-elasticity can be recovered from the compressible models discussed earlier as the volumetric parameters of the model (given by the bulk modulus $\kappa$ in the isotropic case) tend to zero.
In addition, from the trace property $\tr(\Ln(\hat{C})) = \Ln(\det(\hat{C})),$ one can  see that the incompressibility constraint expressed in terms of $\hat{C}$ is given by
$$\Ln(\det(\hat{C})) = 0 \implies \det(\hat{C}) = 1.$$
Using the linear strain measure (\ref{eq:StrC_linear}) then implies that the rotational invariant $I_3=\det(\StrC) = 0.$
Thus, incompressible versions of isotropic nonlinear models, such as the Mooney-Rivlin model, that use (\ref{eq:StrC_linear}), the invariant $\det(\StrC)$ is the variable that vanishes and not $\tr(\StrC)$.

\subsection{Viscous fluid flow}\label{sec:visc_fluid}

\begin{figure}
	\centering
	\includegraphics[width=0.9\columnwidth]{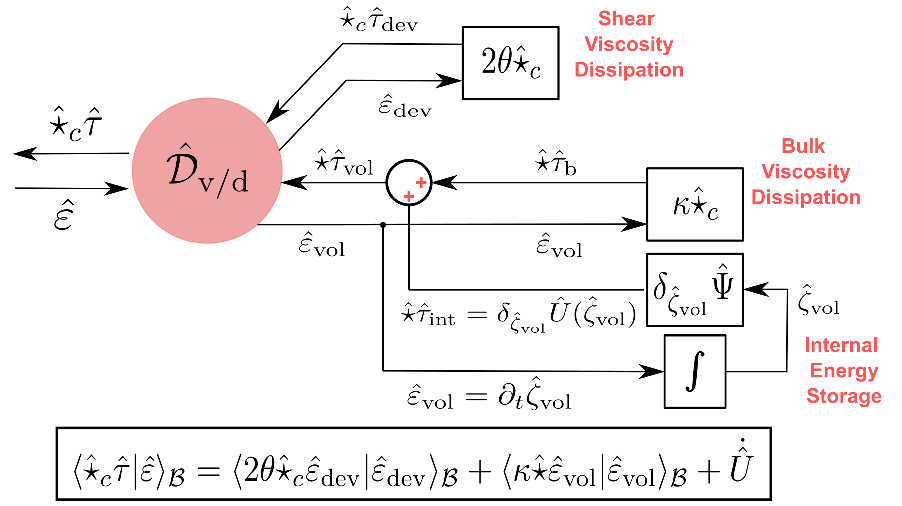}
	\caption{Convective constitutive relations of Newtonian fluids}
	\label{fig:const_Newton_fluid}
\end{figure}

The last type of constitutive relations that we discuss is viscous fluid flow, in particular Newtonian fluids.
Newtonian fluids have the simplest mathematical form as they are characterized by an isotropic linear constitutive relation similar to (\ref{eq:Hencky_const_model}).
As illustrated in Fig. \ref{fig:const_Newton_fluid}, the constitutive model for Newtonian fluids, neglecting thermal effects, is characterized by 
\begin{enumerate}
	\item a linear resistive relation between $(\epCdev,\stTauCdev)$ representing the viscous response to distortion,
	\item a combined resistive and storage relation between $(\epCvol,\stTauCvol)$ representing a visco-elastic response to volumetric deformation.
\end{enumerate}

Let the convective internal energy functional be given by
\begin{equation}\label{eq:int_energy_func_1}
	\hat{U}[\StrCvol;\mFC] = \intB \hat{\mathcal{U}}(\StrCvol) \mFC,
\end{equation}
with $\map{\hat{\mathcal{U}}}{\spFrmB{0}}{\spFrmB{0}}$ denoting the specific internal energy density function that describes the equation of state of the fluid. The internal energy's state is given by the volumetric strain $\StrCvol$ and depends parametrically on $\mFC$.
The constitutive stress law for Newtonian fluids is then expressed in the convective representation as
\begin{equation}\label{eq:fluid_const_rel}
	\stTauCdev = 2 \theta \epCdev, \qquad\qquad \stTauCvol =  \underbrace{\kappa  \epCvol}_{\subTxt{\stTauC}{b}} + \underbrace{\frac{\partial \hat{\mathcal{U}}}{\partial \StrCvol}(\StrCvol)}_{\subTxt{\stTauC}{int}}, \qquad \qquad \partial \StrCvol = \epCvol,
\end{equation}
with $\kappa,\theta \in \spFn{\cl{B}}$ denoting the bulk and shear viscosity coefficients, respectively, which are in general state dependent.
The variational derivative of the internal energy functional is given by 
$\varD{\hat{U}}{\StrCvol}(\StrCvol) = \hat{\star} \frac{\partial \hat{\mathcal{U}}}{\partial \StrCvol}(\StrCvol)= \frac{\partial \hat{\mathcal{U}}}{\partial \StrCvol}(\StrCvol) \mFC \in \spFrmS{n}.$

%

Using $\dot{\hat{U}} = \duPairB{\varD{\hat{U}}{\StrCvol}}{\partial_t \StrCvol} = \duPairB{\hat{\star}\subTxt{\stTauC}{int}}{\epCvol}$, the above constitutive law has a power balance expressed as
\begin{align*}
	\duPairB{\hodgeC \stTauC}{\epC} =& \duPairB{\hat{\star} \stTauCvol}{\epCvol} + \duPairB{\hodgeC \stTauCdev}{\epCdev} \\
									=& \dot{\hat{U}} + \intB \kappa \epCvol \wedge \hat{\star} \epCvol+ 2 \theta \epCdev \wedgedot \hodgeC \epCdev,
\end{align*}
which characterizes the storage of internal energy $\hat{U}$ and two Rayleigh dissipation functions.

Finally, we conclude this section by deriving the spatial counterpart of the constitutive law (\ref{eq:fluid_const_rel}) for $(\epSdev,\stTauSdev)$ and $(\epSvol,\stTauSvol)$ which is more common for fluids as presented in Sec. \ref{sec:pH_fluid}. We start with re-expressing the internal energy (\ref{eq:int_energy_func_1}) in terms of other state variables that are more common; namely the mass density and specific volume.

\begin{proposition}
	Let $\mDC,\hat{\nu} \in \spFrmB{0}$ denote the convective mass density and specific volume functions, defined such that the mass form $\mFC\in\spFrmB{n}$ can be expressed as 
	$$\mFC = \mDC \vFC, \qquad  \text{or} \qquad \vFC = \hat{\nu} \mFC,$$
	where $\vFC\in \spFrmB{n}$ is the volume form associated to the convective metric $\gC\in \spMetB$. The volumetric strain is related to these two variables by
	$$\StrCvol = \ln(\hat{\nu}) = - \ln(\mDC).$$
\end{proposition}
\begin{proof}
	Recall that $\vFC = \varphi^*(\vFS)$ where $\vFS\in \spFrmS{n}$ is the spatial volume form associated to $\gS$. From the Lie derivative definition, one has that
	$$\partial_t \vFC = \varphi^*(\Lie{\vS}{\vFS}) = \Lie{\varphi^*(\vS)}{\varphi^*(\vFS)} = \Lie{\vC}{\vFC} = \divrC(\vC) \vFC.$$
	Since $\mFC$ is constant, then $\partial_t \vFC = \partial_t(\vC)\mFC = \frac{\partial_t\vC}{\vC}$ which implies that $\partial_t\vC = \vC\divrC(\vC)$. From Th. \ref{th:vol-dev_decomp} and the fact that $\partial_t \StrCvol = \frac{\partial_t\vC}{\vC}$, then one has that $\StrCvol =\ln(\hat{\nu})$. Using the properties of the natural logarithm, it is straightforward that $\StrCvol =\text{ln}(\frac{1}{\mDC}) = -\ln(\mDC)$.
	\qed
\end{proof}

With the convective mass density as the state variable, one can rewrite the internal energy (\ref{eq:int_energy_func_1}) as
\begin{equation}\label{eq:int_energy_func_2}
	\hat{U}[\mDC;\gC] = \intB \hat{\mathcal{U}}(\mDC) \mDC \vFC,
\end{equation}
with $\hat{\mathcal{U}}$ now being a function of $\mDC$ with a slight abuse of notation.
Using $\partial_t \mDC = - \mDC \partial_t \StrCvol  = - \mDC \epCvol$, the rate of change of $\hat{U}$ now takes the form
$$\dot{\hat{U}} = \intB \partial_t \hat{\cl{U}}(\mDC) \mDC \vFC =\intB \partial_t \mDC \frac{\partial \hat{\mathcal{U}}}{\partial \mDC}(\mDC) \mFC = \intB-\epCvol  \mDC \frac{\partial \hat{\mathcal{U}}}{\partial \mDC}(\mDC) \mFC.$$
Since $\dot{\hat{U}}= \duPairB{\hat{\star}\subTxt{\stTauC}{int}}{\epCvol}$, then $\subTxt{\stTauC}{int}$ in (\ref{eq:fluid_const_rel}) can be expressed as
$\subTxt{\stTauC}{int} = - \mDC \frac{\partial \hat{\mathcal{U}}}{\partial \mDC}(\mDC).$

Now let $U:= \hat{U}\circ\varphi$ denote the spatial representation of the internal energy functional defined such that $U[\mDS;\gS] = \hat{U}[\varphi^*(\mDS);\varphi^*(\gS)]$ and expressed as
\begin{equation}\label{eq:int_energy_func_3}
	U[\mDS;\gS] = \intS \mathcal{U}(\mDS) \mDS \vFS,
\end{equation}
with the spatial mass density $\mDS\in \spFrmS{0}$ as the state variable and $\map{\mathcal{U}}{\spFrmS{0}}{\spFrmS{0}}$ denoting the spatial specific internal energy.
Using Reynold's transport theorem, the rate of change of $U$ is given by
$$\dot{U} = \frac{\extd}{\extd t} \intS \mathcal{U}(\mDS) \mDS \vFS =\intS \extD_t (\mathcal{U}(\mDS) \mDS \vFS)=\intS \extD_t (\mathcal{U}(\mDS)) \mDS \vFS ,$$
which follows from the mass balance $\extD_t(\mDS \vFS) = \extD_t \mFS = 0$ with $\extD_t:= \partial_t + \Lie{\vS}{}$ being a shorthand notation.
From the non-equilibrium Thermodynamics principle $\extD_t (\mathcal{U}(\mDS)) = \frac{\partial\mathcal{U}}{\partial\mDS}(\mDS)\extD_t\mDS$ and mass balance $\extD_t\mDS = -\mDS \divrS(\vS)$ \cite{Malek2018DerivationFluids}, one can express $\dot{U} $ as
$$\dot{U} = \intS \mDS  \frac{\partial\mathcal{U}}{\partial\mDS}(\mDS) \extD_t\mDS \vFS = - \intS p \divrS(\vS)\vFS,$$
where $p:= \mDS^2 \frac{\partial\mathcal{U}}{\partial\mDS}(\mDS) \in \spFn{\cl{S}}$ is the thermodynamic pressure function.
Given that $\epSvol =\divrS(\vS)$, then for $\dot{U} = \duPairS{\star \subTxt{\tau}{int}}{\epSvol}$ to hold then one has that $\subTxt{\tau}{int} = - \frac{p}{\mDS}$.

The spatial counterpart of the constitutive stress law for Newtonian fluids (\ref{eq:fluid_const_rel}) is then expressed as
\begin{equation}\label{eq:fluid_const_rel_S}
	\stTauSdev = 2 \theta \epSdev, \qquad\qquad \stTauSvol =  \kappa  \epSvol\subTxt{\stTauS}{b} - p , \qquad \qquad \extD_t\mDS = -\mDS \epSvol,
\end{equation}
with $\kappa,\theta \in \spFn{\cl{S}}$ denoting the bulk and shear viscosity coefficients considered now functions on $\cl{S}$.
The above constitutive law has a power balance expressed as
\begin{align*}
	\duPairS{\hodgeS \stTauS}{\epS} =& \duPairS{\star\stTauSvol}{\epSvol} + \duPairS{\hodgeS \stTauSdev}{\epSdev} \\
	=& \dot{U} + \intS \kappa \epSvol \wedge \star \epSvol+ 2 \theta \epSdev \wedgedot \hodgeS \epSdev.
\end{align*}

By adding the constitutive relation (\ref{eq:fluid_const_rel}) to the \pH system (\ref{eq:EoM_fluid_1}-\ref{eq:EoM_fluid_2}), one has a \pH representation of the \textbf{Navier-Stokes equations} for Newtonian fluids.
The total energy balance after adding the stress constitutive law can be expressed as
\begin{equation*}
	\frac{\extd}{\extd t} (\HkinS + U) = \int_{\Sbound} \vS|_{\Sbound} \wedgedot (\stS-\half\iota_\vS\momS)|_{\Sbound} - \intS \kappa \epSvol \wedge \star \epSvol+ 2 \theta \epSdev \wedgedot \hodgeS \epSdev.
\end{equation*}
The above power balance states that the rate of change of total energy within the domain $\cl{S}$ is equal to the difference between the power flow through the boundary $\Sbound$, due to stress and momentum flux, and the power dissipated in the domain, due to bulk and shear viscosity.

\begin{remark}
	
	i) The stress $\stTauS\in \spvecFrmS{1} \cong \spVec{^1_1\cl{S}}$ is the spatial representation of the intensive Rougee stress discussed in Sec. \ref{sec:const_relations_elast} and is related to the standard Cauchy stress tensor $\stSigS\in \spVec{^1_1\cl{S}}$ by $\stSigS = \mDS \stTauS$.
	The final constitutive law of compressible Newtonian fluids in terms of $\stSigS$ can then be summarized as
	$$\stSigS = \mDS \stTauS = \mDS(\stTauSvol I_n + \stTauSdev) = (-p + \subTxt{\kappa}{dyn} \epSvol) I_n + \subTxt{\theta}{dyn} \epSdev,$$
	where $\subTxt{\kappa}{dyn}:= \mDS \kappa \in \spFn{\cl{S}}$ denotes the dynamic bulk viscosity and $\subTxt{\theta}{dyn}:= \mDS \theta \in \spFn{\cl{S}}$ denotes the dynamic shear viscosity.
	Furthermore, by introducing the mechanical pressure function $p_m := -\frac{1}{n} \tr(\stSigS)\in \spFn{\cl{S}}$, then one recovers the standard relation between mechanical and thermodynamic functions
	$p_m = p - \subTxt{\kappa}{dyn} \divrS(\vS).$
	
	ii) Incorporating incompressibility in the \pH model is done identically to the procedure explained in Sec. \ref{sec:incomp_hyper_elast}. The volumetric component of the stress $\stTauSvol$ is then treated as a Lagrange multiplier enforcing the incompressibility constraint $\epSvol = \divrS(\vS) = 0$.
	Furthermore, the constitutive model (\ref{eq:fluid_const_rel}) can be easily extended to model more complex fluid behavior. For instance, representing the shear viscosity by a fourth-order tensor instead of a function allows modeling anisotropic Newtonian fluids, while representing the bulk and shear viscosity as nonlinear functions of $\epSvol$ and $\epSdev$, respectively, allows modeling non-Newtonian fluids.
	As for modeling the transfer of the dissipated energy to the thermal domain, the interested reader is referred to \cite{califano2022differential,Malek2018DerivationFluids} and references therein for a discussion of the topic.
	
\end{remark}

\section{Conclusion}

In this paper we presented decomposed \pH models that represent the governing equations of continuum mechanics, complementing our recent work \cite{rashad2023intrinsic}.
These models provide a geometric coordinate-free insight by using bundle-valued differential forms to mathematically represent physical variables.
Thanks to the elegant machinery of exterior calculus, the models were derived using Hamiltonian reduction theory from first principles.
A distinguishing feature of our work is the port-based philosophy of \textit{tearing} the overall system into its constituting energetic units.
An immediate benefit of this approach is that one does not have to consider all physical variables of the whole system at once. Consequently, this simplifies greatly the application of Hamiltonian reduction techniques.

The derived \pH models expose the rich geometric, topological and energetic structure underlying the theory of continuum mechanics.
Our work lays the foundations for exploiting this rich structure in analysis and control of distributed parameter system, in addition to the development of efficient numerical algorithms for discretization and model-order reduction of these dynamic equations that preserve this structure at the discrete level.

\setcounter{section}{0} 

\renewcommand\thesection{\Alph{section}}

\section{Appendix}

\subsection{Proof of Prop. \ref{prop:pf_pl_M_to_S}} \label{app:proof_1}
(i) Let $\delphiM \in T_\varphi\spC$ be the tangent vector to the curve $s\mapsto \varphi_s$ at $\varphi\in\spC$ at $s=0$.
Then $\delta\mFS\in \spFrmS{n}$ is the tangent vector to the induced curve $s\mapsto \mFS_s:= \varphi_{s,*}(\mFM)$ at $\mFS\in\spFrmS{n}$ at $s=0$.
One has from the Lie derivative definition that 
$$\delta\mFS:= \dds \varphi_{s,*}(\mFM) = - \Lie{\delta\varphi}{\mFS},$$
with $\delta\varphi:= \cfgFleg{f}(\delphiM) = \delphiM\circ \varphi^{-1}\in\spvecFrmS{0}$.

The variation $\delta\momS \in \spcovFrmS{n}$ consists of two components; one due to the variation of $\varphi$ and another due to the variation of $\momM$, such that: $\delta\momS:= \delta\momS_1 + \delta\momS_2$.
For the later one due to varying $\momM$, it is simply given by 
$\delta\momS_2:= \dds \cfgFleg{f}(\momM+s\delta\momM) =\cfgFleg{f}(\delta\momM).$
As for $\delta\momS_1$, it is more involved due to the intuition mentioned preceding (\ref{eq:Frech_F_phi}).
From Remark \ref{remark:momS}, one can identify $\momS$ with $\mFS\otimes \vfS$ or equivalently $\vFS\otimes m^\flat$, where $\vFS\in\spFrmS{n}$ and $m^\flat\in\spFrmS{1}$ are the volume top form and intensive momentum one-form, respectively.
Since $\vFS$ is constant, one can express $\delta\momS_1 = {}^{\gS}\hodgeS (\delta m) =  \vFS\otimes \delta m^\flat$, with $\delta m^\flat\in\spFrmS{1}$ and ${}^{\gS}\hodgeS$ denotes the (constant) Hodge-star operator constructed from the volume form $\vFS$.
It has been shown in \cite{Mazer1989HamiltonianFlows} that the vector field $\delta m := \gS^{-1}\cdot\delta m^\flat \in \spVec{\cl{S}}$ can be related to $\delta\varphi$ by
$$\delta m = - \divrS(\delta\varphi m) - \nabS_{\delta\varphi} m = - \divrS(m\otimes \delta\varphi),$$
where $m\otimes \delta\varphi$ represents a momentum flux.
Furthermore, it has been shown in \cite[Th. 1]{rashad2023intrinsic} that the divergence of a two-covariant tensor is equivalent in exterior calculus to the exterior covariant derivative of a related covector-valued $n-1$ form.
For the case of $m\otimes \delta\varphi$, one can show that $\delta\momS_1 = - \extcdS ({}^{\gS}\hodgeS(m^\flat\otimes \delta\varphi))$.
Finally, one can show that the covector-valued $n-1$ form ${}^{\gS}\hodgeS(m^\flat\otimes \delta\varphi)$ is equivalent to the extensive momentum flux expressed as $\iota_{\delta\varphi} \momS := \iota_{\delta\varphi}\vFS\otimes m^\flat \in \spcovFrmS{n-1}$, which concludes the first part of the proof.

(ii) From the chain rule of functionals and (\ref{eq:Frech_F_M},\ref{eq:Funct_S_def}), one has that 
\begin{align*}
	\extD_{\momM} \scrTld{F}(\varphi,\momM) \cdot \delta\momM=& \dds \scrTld{F}[\varphi,\momM + s \delta\momM]\\
	=& \dds \scr{F}[\varphi_*( \mFM ),\cfgFleg{f}(\momM + s \delta\momM)] = \extD_{\momS}\scr{F}(\mFS,\momS) \cdot \delta\momS_2.
\end{align*}
Using the change of variables formula and the expression for $\delta\momS_2$ in (i) above, one has that
\begin{align*}
	\extD_{\momS}\scr{F}(\mFS,\momS) \cdot \delta\momS_2 &= \intS  \delta\momS_2  \wedgedot \varD{\scr{F}}{\momS}  = \intB \varphi^*(\delta\momS_2 \wedgedot \varD{\scr{F}}{\momS}  ) \\
	&= \intB \cfgPleg{f}(\delta\momS_2) \wedgedot \cfgPleg{f}(\varD{\scr{F}}{\momS})   = \intB \delta\momM \wedgedot \cfgPleg{f}(\varD{\scr{F}}{\momS}) .
\end{align*}
Comparing the above two expressions and using (\ref{eq:FrechD_2_M}) yields that $\varD{\scrTld{F}}{\momM} = \cfgPleg{f}(\varD{\scr{F}}{\momS})$.

Similarly, from the chain rule of functionals and (\ref{eq:Frech_F_phi},\ref{eq:Funct_S_def}), one has that
\begin{align}
	\extD_{\varphi} \scrTld{F}(\varphi,\momM) \cdot \delphiM =& \dds \scrTld{F}[\varphi_s,\momM_s] = \dds \scr{F}[\varphi_{s,*}( \mFM ),(\varphi_s)_{\mathrm{f},*}(\momM_s)] \nonumber\\
	=& \extD_{\mFS}\scr{F}(\mFS,\momS) \cdot \delta\mFS + \extD_{\momS}\scr{F}(\mFS,\momS) \cdot \delta\momS_1.\label{eq:prop_proof_1}
\end{align}
Using (\ref{eq:Frech_F_S}) and the expressions for $\delta\mFS,\delta\momS_1$ in (i) above, one can rewrite (\ref{eq:prop_proof_1}) as 
\begin{equation}\label{eq:prop_proof_2}
	\extD_{\mFS}\scr{F}(\mFS,\momS) \cdot \delta\mFS + \extD_{\momS}\scr{F}(\mFS,\momS) \cdot \delta\momS_1 = - \intS \varD{\scr{F}}{\mFS} \wedge \Lie{\delta\varphi}{\mFS} + \varD{\scr{F}}{\momS} \wedgedot \extcdS \iota_{\delta\varphi} \momS.
\end{equation}

Consider the following identities between scalar-valued and bundle-valued forms
$$\iota_{\delta\varphi}(\extd f) \wedge\mFS = \delta\varphi \wedgedot (\mFS\otimes \extd f) \in \spFrmS{n} , \qquad \iota_{\delta\varphi} (f \wedge \mFS) = \delta\varphi\wedgedot (f \wedge \mFS) \in \spFrmS{n-1},$$ 
for any $f\in \spFn{\cl{S}}$, with $\mFS$ in the second LHS is considered as a top-form and in the RHS as a covector-valued $n-1$ form.
Using the identities above, the Leibniz rule and Cartan's formula for the Lie derivative, and Stokes theorem, one has that 
\begin{align}
	\intS \varD{\scr{F}}{\mFS} \wedge \Lie{\delta\varphi}{\mFS} &= - \intS \Lie{\delta\varphi}{(\varD{\scr{F}}{\mFS})} \wedge \mFS + \intS \Lie{\delta\varphi}{(\varD{\scr{F}}{\mFS}\wedge\mFS)} \nonumber\\
	&= - \intS \iota_{\delta\varphi} \extd{(\varD{\scr{F}}{\mFS})} \wedge \mFS + \int_{\Sbound} \iota_{\delta\varphi}{(\varD{\scr{F}}{\mFS}\wedge\mFS)}|_{\Sbound} \nonumber\\
	&= - \intS \delta\varphi \wedgedot (\mFS\otimes \extd \varD{\scr{F}}{\mFS}) +
	\int_{\Sbound} \delta\varphi|_{\Sbound} \wedgedot \varD{\scr{F}}{\mFS}\wedge\mFS|_{\Sbound}.\label{eq:prop_proof_3}
\end{align}

Now consider the following identities for any vector field $\eta \in \spVec{\cl{S}}$
$$\eta \wedgedot \iota_{\delta\varphi}\momS = \delta\varphi \wedgedot \iota_{\eta}\momS\in \spFrmS{n}, \qquad \nabS \eta \wedgedot \iota_{\delta\varphi} \momS = \delta\varphi \wedgedot \mFS \otimes (\nabS \eta\wedgedot \vfS) \in \spFrmS{n}, $$
with $\momS = \mFS \otimes \vfS$.
Using the above identities, the Leibniz rule for the exterior covariant derivative, and Stokes theorem, one has that
\begin{align}
	\intS \varD{\scr{F}}{\momS} \wedgedot \extcdS \iota_{\delta\varphi} \momS =& - \intS \nabS(\varD{\scr{F}}{\momS}) \wedgedot \iota_{\delta\varphi} \momS + \intS \extd( \varD{\scr{F}}{\momS}\wedgedot \iota_{\delta\varphi} \momS) \nonumber\\
	=& - \intS \delta\varphi \wedgedot \mFS \otimes (\nabS(\varD{\scr{F}}{\momS})\wedgedot \vfS) + \int_{\Sbound} \delta\varphi|_{\Sbound} \wedgedot \iota_{\varD{\scr{F}}{\momS}} \momS|_{\Sbound}.\label{eq:prop_proof_4}
\end{align}

Finally, by substituting (\ref{eq:prop_proof_3},\ref{eq:prop_proof_4}) in (\ref{eq:prop_proof_2}), combining (\ref{eq:prop_proof_2}) with (\ref{eq:prop_proof_1}) and using the change of variables formula, the expressions of $\varD{\scrTld{F}}{\varphi}$ and $\varDbnd{\scrTld{F}}{\varphi}$ are concluded by comparing the result with (\ref{eq:FrechD_1_M}).

\subsection{Proof of Prop. \ref{prop:pf_pl_M_to_C}} \label{app:proof_2}
(i) Recall the curve $s\mapsto\varphi_s$ in $\spC$ introduced in Prop. \ref{prop:pf_pl_M_to_S}.
One has that $\delta\gC:= \dds \gC_s$, where $\gC_s:= \varphi_s^*(\gS)$ is the induced curve on $\spMetB$. From the Lie derivative properties, it follows that $\delta\gC= \dds \gC_s = \varphi^*(\Lie{\delta\varphi}{\gS}) = \Lie{\delphiC}{\gC}$, where $\delphiC = T \varphi^{-1} \circ \delphiM$.

The variation $\delta\momC = \delta\momC_1 + \delta\momC_2 \in \spcovFrmB{n}$ also consists of two components; one due to the variation of $\varphi$ and another due to $\momM$, where the latter is given by $\delta\momC_2:= \cfgPleg{v}(\delta\momM)$. The first part is defined as $\delta\momC_1:= \dds \momC_s$, where $s\mapsto\momC_s$ is the induced curve in $\spcovFrmB{n}$ defined in local components as $\momC_s:= \mFC {}^sF^i_I \delphiM_i \otimes E^I$, where $\delphiM_i\in \spFrmB{0}$ are the components of $\delphiM$ and ${}^sF^i_I \in \spFrmB{0}$ are the components of ${}^sF := T\varphi_s$.
Using the identities $\dds {}^sF^i_I := (\nabM \delphiM)^i_I $ and $(\nabM \delphiM)^i_I = F^i_J (\nabC \delphiC)^J_I$ with $\nabM$ denoting the material covariant derivative (\cf \cite{Simo1988ThePlates,rashad2023intrinsic}), then one has that
\begin{align*}
	\delta\momC_1:= \dds \momC_s =& \mFC \delphiM_i \dds{}^sF^i_I \otimes E^I = \mFC \delphiM_i(\nabM \delphiM)^i_I\otimes E^I\\
	=& \mFC \delphiM_i F^i_J (\nabC \delphiC)^J_I \otimes E^I = \mFC \delphiC_J (\nabC \delphiC)^J_I \otimes E^I,
\end{align*}
which can be written as $\delta\momC_1 = \mFC\otimes (\nabC\delphiC \wedgedot \vfC)$.

(ii) The derivation of $\varD{\scrTld{F}}{\momM} = \cfgFleg{v}(\varD{\scrHat{F}}{\momC})$ follows identically its spatial counterpart in Prop. \ref{prop:pf_pl_M_to_S} (ii).
For the other variational derivatives, one has from the chain rule similar to (\ref{eq:prop_proof_1}) that
$$\extD_{\varphi} \scrTld{F}(\varphi,\momM) \cdot \delphiM = \extD_{\gC}\scrHat{F}(\gC,\momC) \cdot \delta\gC + \extD_{\momC}\scrHat{F}(\gC,\momC) \cdot \delta\momC_1,$$
where the RHS can be further expanded using identities $\half \Lie{\vC}{\gC} = \sym(\extcdC \vfC)$ and $\nabC \hat{\eta} \wedgedot \iota_{\delphiC} \momC = \delphiC \wedgedot \mFC \otimes (\nabC \hat{\eta}\wedgedot \vfC)$ as
\begin{align*}
	\extD_{\varphi} \scrTld{F}(\varphi,\momM) \cdot \delphiM 
	=& \intB \varD{\gC}{\scrHat{F}} \wedgedot \delta\gC + \varD{\momC}{\scrHat{F}} \wedgedot \delta\momC_1
	= \intB \varD{\gC}{\scrHat{F}} \wedgedot \Lie{\delphiC}{\gC} + \varD{\momC}{\scrHat{F}} \wedgedot \mFC\otimes (\nabC\delphiC \wedgedot \vfC)\\
	=& \intB \varD{\gC}{\scrHat{F}} \wedgedot 2 \sym(\extcdC \vfC) + \nabC\delphiC \wedgedot \iota_{\varD{\momC}{\scrHat{F}}} \momC
	= \intB \nabC\delphiC \wedgedot \cl{E}_{\hat{F}},
\end{align*}
with $\cl{E}_{\scrHat{F}}:= (2 (\varD{\scrHat{F}}{\gC}) ^\flat + \iota_{\varD{\scrHat{F}}{\momC}}\momC \in \spcovFrmB{n-1}$.
Using integration by parts, the definition of $\delphiC$ and the duality of $\cfgFleg{v}$ and $\cfgPleg{v}$, one has that 
\begin{align*}
	\intB \nabC\delphiC \wedgedot \cl{E}_{\hat{F}} =& - \intB \delphiC \wedgedot \extcdC \cl{E}_{\hat{F}} + \int_{\bndB} \delphiC|_{\Bbound} \wedgedot \cl{E}_{\hat{F}}|_{\Bbound}\\ =& - \intB \delphiM \wedgedot \cfgFleg{v}(\extcdC \cl{E}_{\hat{F}}) + \int_{\bndB} \delphiM|_{\Bbound} \wedgedot \cfgFleg{v}(\cl{E}_{\hat{F}})|_{\Bbound}.
\end{align*}
Finally, the results follow by comparing with (\ref{eq:FrechD_1_M}).

\subsection{Vector space decomposition}\label{app:decompo}
\newcommand{\gln}{\mathfrak{gl}_n}
\newcommand{\bln}{\mathfrak{bl}_n}
\newcommand{\glnD}{\mathfrak{gl}_n^*}
\newcommand{\blnD}{\mathfrak{bl}_n^*}
\newcommand{\duPairbln}[2]{\duPair{#1}{#2}{\bln}}
\newcommand{\duPairgln}[2]{\duPair{#1}{#2}{\gln}}
Let $V,W$ be two vector spaces over the field $\bb{R}$. Let $\text{Hom}(V,W)$ denote te set of linear maps from $V$ to $W$. An endomorphism of $V$ is an element of $\text{End}(V) := \text{Hom}(V,V)$. The dual vector space to $V$ is $V^*:= \text{Hom}(V,\bb{R})$ with the duality product between any $v\in V,\tilde{v}\in V^*$ denoted by $\duPair{\tilde{v}}{v}{V} := \tilde{v}(v) \in \bb{R}$. The orthogonal complement of a subspace $W\subset V$ is the subspace $W^\perp \subset V^*$ defined as 
$$W^\perp := \{\tilde{w}\in V^* | \duPair{\tilde{w}}{w}{V} = 0,\ \forall w\in W\}.$$

Let $M$ be a smooth manifold of dimension $n$, and let $T_p M,T_p^*M$ denote the tangent and cotangent spaces of $M$ at a point $p\in M$, respectively.
The space of all $(r,s)$ tensor at $p\in M$ is denoted by $T^r_{s,p}M$ with $T^0_{0,p}M = \bb{R}, T^1_{0,p}M =T_p M, $ and $T^0_{1,p}M = T_p^*M$. Furthermore, $T^0_{2,p}M = \text{Hom}(T_p M,T_p^*M)$ and $T^1_{1,p}M = \text{End}(T_p M)$.

Given a choice of basis for $T_p M$ and its dual basis for $T_p^*M$, one has that 
$$T^0_{2,p}M \cong \bln, \qquad \qquad T^1_{1,p}M \cong \gln,$$ 
where $\bln$ and $\gln$ denote the vector spaces of bilinear forms and linear transformations on $\bb{R}^n$, respectively.
Elements of $\bln$ and $\gln$ are expressed in components as
$$A = A_{ij} e^i\otimes e^j \in \bln , \qquad \qquad X = X^i_j e_i\otimes e^j \in \gln,$$
where $\{e_i\}, \{e^j\}$ denote the standard basis and its dual on $\bb{R}^n$.
Furthermore, elements of the dual spaces $\blnD$ and $\glnD$ are expressed in components as
$$B = B^{ij} e_i\otimes e_j \in \blnD , \qquad \qquad Y = Y^i_j e^i\otimes e_j\in \glnD.$$
The duality product $\map{\duPairbln{\cdot}{\cdot}}{\blnD\times\bln }{\bb{R}}$ is defined by
$\duPairbln{B}{A} := B^{ij}A_{ij},$
whereas the duality product $\map{\duPairgln{\cdot}{\cdot}}{\glnD\times\gln }{\bb{R}}$ is defined by
$\duPairbln{Y}{X} := Y_i^j X^i_j.$
If $\B{A},\B{B} \in \bb{R}^{n\times n}$ denote the matrix representations of $A\in \bln$ and $B\in\blnD$, respectively, then the duality products is more commonly expressed as $\duPairbln{B}{A} = \tr(\B{B}\B{A})$. The same expression holds also for $\gln$.

\subsubsection{Symmetric-asymmetric decomposition of $\bln$}
\newcommand{\pisym}{\subTxt{\pi}{sym}}
\newcommand{\piasy}{\subTxt{\pi}{asy}}
\newcommand{\piTsym}{\subTxt{\tilde{\pi}}{sym}}
\newcommand{\piTasy}{\subTxt{\tilde{\pi}}{asy}}
\newcommand{\im}{\text{im}}
\newcommand{\krn}{\text{ker}}
\newcommand{\Asym}{\subTxt{A}{sym}}
\newcommand{\Aasy}{\subTxt{A}{asy}}
\newcommand{\Bsym}{\subTxt{B}{sym}}
\newcommand{\Basy}{\subTxt{B}{asy}}
\newcommand{\blnS}{\mathfrak{bl}_s}
\newcommand{\blnA}{\mathfrak{bl}_a}
\newcommand{\blnDS}{\mathfrak{bl}_s^*}
\newcommand{\blnDA}{\mathfrak{bl}_a^*}
Consider the following projection map
\begin{equation}\label{eq:pisym}
	\map{\pisym}{\bln}{\bln}, \qquad \qquad A \mapsto \half (A + A^\top).
\end{equation}
It is straightforward to see that $\pisym$ is idempotent (i.e. $\pisym\circ \pisym = \pisym$).
Let $\blnS := \im(\pisym) \subset \bln$ and $\blnA := \krn(\pisym) \subset \bln$ denote the vector subspaces of symmetric and asymmetric bilinear forms, respectively.
Any $A\in \bln$ can be decomposed then as $A = \Asym + \Aasy$ with
$$ \Asym:= \pisym(A) \in \blnS, \qquad \Aasy:= \piasy(A) := A -\pisym(A)  \in \blnA,$$
where $\Asym$ and $\Aasy$ are called the symmetric and asymmetric components of $A$, respectively.
By construction, one has that $\pisym(\piasy(A)) = 0$.
We refer to the decomposition 
$\bln = \blnS \oplus \blnA,$
as the symmetric-asymmetric decomposition of $\bln$.
While $\text{dim}(\bln) = n^2$, one has that $\text{dim}(\blnS) = \half n(n+1)$ and $\text{dim}(\blnA) = \half n(n-1)$.

The dual space of $\bln$ is decomposed as $\blnD = \blnDS \oplus \blnDA$ where 
\begin{align*}
	\blnDS &= \blnA^\perp := \{B \in \blnD | \duPairbln{B}{\Aasy} = 0,\ \forall \Aasy \in \blnA \},\\
	\blnDA &= \blnS^\perp := \{B \in \blnD | \duPairbln{B}{\Asym} = 0,\ \forall \Asym \in \blnS \}.
\end{align*}
If $\map{\pisym}{\bln}{\blnS}$ and $\map{\piasy}{\bln}{\blnA}$ are defined as above, then $\map{\piTsym}{\blnD}{\blnDS}$ and $\map{\piTasy}{\blnD}{\blnDA}$ are defined such that 
$$\duPairbln{\piTsym(B)}{\piasy(A)} = 0, \qquad \qquad \duPairbln{\piTasy(B)}{\pisym(A)} = 0.$$
One can show that 
$\piTsym(B) = \half(B+B^\top)=: \Bsym$ and $ \piTasy(B) = \half(B-B^\top)=: \Basy.$

The above construction of the symmetric-asymmetric decomposition of $\bln$ can be summarized in the following map $\map{J_{s/a}}{\blnDS\times\blnDA\times\bln}{\blnS\times\blnA\times\blnD}$
\begin{equation*}
	\ThrVec{\Asym}{\Aasy}{B} = 
	\begin{pmatrix}
		0 & 0 & \pisym\\
		0 & 0 & \piasy\\
		\mathds{1} & \mathds{1} & 0
	\end{pmatrix} 
	\ThrVec{\Bsym}{\Basy}{A},
\end{equation*}
such that the duality pairing between $A\in \bln$ and $B\in \blnD$ is given by
$$\duPairbln{B}{A} = \duPairbln{\Bsym}{\Asym} + \duPairbln{\Basy}{\Aasy},$$
with $\mathds{1}$ denoting the identity map.

\subsubsection{Volumetric-deviatoric decomposition of $\gln$}
\newcommand{\pivol}{\subTxt{\pi}{vol}}
\newcommand{\pidev}{\subTxt{\pi}{dev}}
\newcommand{\sln}{\mathfrak{sl}_n}
\newcommand{\ksp}{\mathfrak{k}}
\newcommand{\Xvol}{\subTxt{X}{vol}}
\newcommand{\Xdev}{\subTxt{X}{dev}}
\newcommand{\slnD}{\mathfrak{sl}_n^*}
\newcommand{\kspD}{\mathfrak{k}^*}
\newcommand{\piTvol}{\subTxt{\tilde{\pi}}{vol}}
\newcommand{\piTdev}{\subTxt{\tilde{\pi}}{dev}}
\newcommand{\Yvol}{\subTxt{Y}{vol}}
\newcommand{\Ydev}{\subTxt{Y}{dev}}

Consider the following projection map
\begin{equation}\label{eq:pivol}
	\map{\pivol}{\gln}{\gln}, \qquad \qquad {X}\mapsto{\frac{1}{n} \tr(X)I_n,}
\end{equation}
where $\map{\tr}{\gln}{\bb{R}}$ denotes the trace map and $I_n \in \gln$ denotes the identity element (with components $\delta_j^i$).
It is straightforward to see that $\pivol$ is idempotent (which necessitates the division by $n$).
Let $\ksp := \im(\pivol) \subset \gln$ and $\sln := \krn(\pivol) \subset \gln$ denote the vector subspaces of scalar operators and traceless operators, respectively.
Any $X\in \gln$ can be decomposed then as $X = \Xvol + \Xdev$ with
$$ \Xvol:= \pivol(X) \in \ksp, \qquad \Xdev:= \pidev(X) := X -\pivol(X)  \in \sln,$$
where $\Xvol$ and $\Xdev$ will be referred to as the volumetric and deviatoric components of $X$, respectively.
By construction, one has that $\pivol(\pidev(X)) = 0$.
We refer to the decomposition 
$\gln = \ksp \oplus \sln,$
as the volumetric-deviatoric decomposition of $\gln$.
While $\text{dim}(\gln) = n^2$, one has that $\text{dim}(\ksp) = 1$ and $\text{dim}(\sln) = n^2 -1$.

The dual space of $\gln$ is decomposed as $\glnD = \kspD \oplus \slnD$ where 
\begin{align*}
	\kspD &= \sln^\perp := \{Y \in \glnD | \duPairgln{Y}{\Xdev} = 0,\ \forall \Xdev \in \sln \},\\
	\slnD &= \ksp^\perp := \{Y \in \glnD | \duPairgln{Y}{\Xvol} = 0,\ \forall \Xvol \in \ksp \}.
\end{align*}
If $\map{\pivol}{\gln}{\ksp}$ and $\map{\pidev}{\gln}{\sln}$ are defined as above, then $\map{\piTvol}{\glnD}{\kspD}$ and $\map{\piTdev}{\glnD}{\slnD}$ are defined such that 
$\duPairgln{\piTvol(Y)}{\pidev(X)} = 0$ and $ \duPairbln{\piTdev(Y)}{\pivol(X)} = 0.$
One can show that 
$\piTvol(Y) = \frac{1}{n} \tr(X)I_n=: \Yvol$ and $ \piTdev(Y) = Y -\Yvol  =: \Ydev.$
Consequently, one has that the duality pairing between any $X\in \gln$ and $Y\in \glnD$ to be decomposed as
$$\duPairgln{Y}{X} = \duPairgln{\Yvol}{\Xvol} + \duPairgln{\Ydev}{\Xdev}.$$

Since the dimension of the subspace $\ksp$ is 1, it is isomorphic to $\bb{R}$. However, this isomorphism is not canonical.
In this work, we choose to identify $\ksp$ with $\bb{R}$ with the following isomorphism
\begin{equation*}
	\map{\tr}{\ksp}{\bb{R}},\qquad \qquad {\Xvol}\mapsto{\tr(\Xvol)=: \subTxt{x}{vol}.}
\end{equation*}
The dual map $\map{\tr^*}{\bb{R}}{\kspD}$ is given by $\tr^*(\subTxt{y}{vol}) = \subTxt{y}{vol} I_n$.
Consequently, one can express the duality pairing on $\ksp$ as a product of scalars such that
$$\duPairgln{\Yvol}{\Xvol} = \subTxt{y}{vol} \subTxt{x}{vol}.$$
One can see that $\subTxt{x}{vol} = \tr(\Xvol) = \tr(X)$ while $\subTxt{y}{vol} = \frac{1}{n} \tr(Y)$. In principle, one can modify the above isomorphism between $\ksp$ and $\bb{R}$ such that the factor $\frac{1}{n}$ is multiplicatively distributed between $\subTxt{x}{vol}$ and $\subTxt{y}{vol}$ in any way.

The above construction of the volumetric-deviatoric decomposition of $\gln$ can be summarized in the following map $\map{J_{v/d}}{\kspD\times\slnD\times\gln}{\ksp\times\sln\times\glnD}$
\begin{equation*}
	\ThrVec{\subTxt{x}{vol}}{\Xdev}{Y} = 
	\begin{pmatrix}
		0 & 0 & \tr \circ \pivol\\
		0 & 0 & \pidev\\
		\tr^* & \mathds{1} & 0
	\end{pmatrix} 
	\ThrVec{\subTxt{y}{vol}}{\Ydev}{X},
\end{equation*}
such that the duality pairing between $X\in \gln$ and $Y\in \glnD$ is given by
$$\duPairgln{Y}{X} = \subTxt{y}{vol} \subTxt{x}{vol} + \duPairgln{\Ydev}{\Xdev}.$$

\subsubsection{Decomposition of second-rank tensor fields on $M$}
\newcommand{\spVecSub}[2]{\Gamma_{\text{#2}}(T#1)}
\newcommand{\duPairM}[2]{\left\langle\left\langle  #1 | #2 \right\rangle\right\rangle_{M}}

The above constructions allow us to extend the symmetric-asymmetric decomposition of $\bln \cong T^0_{2,p}M$ to the space of (0,2) tensor-fields $\spVec{^0_2M}$ and the volumetric-deviatoric decomposition of $\gln \cong T^1_{1,p}M$ to the space of (1,1) tensor-fields $\spVec{^1_1M}$ such that
$$\spVec{^0_2M} = \spVecSub{^0_2M}{s} \oplus \spVecSub{^0_2M}{a}, \qquad \spVec{^1_1M} = \spVecSub{^1_1M}{v} \oplus \spVecSub{^1_1M}{d}.$$

The duality product between $A\in \spVec{^0_2M}$ and $B\in \spVec{^2_0M}$ can then be written as
$$\duPairM{B}{A} = \duPairM{\Bsym}{\Asym}  + \duPairM{\Basy}{\Aasy},$$
with $\duPairM{B}{A} := \int_M B:A\ \omega$, such that $B:A\in \spFn{M}$ denotes the double contraction of the tensor fields $A$ and $B$ and $\omega$ denotes the volume form of $M$.
Similarly, the duality product between $X\in \spVec{^1_1M}$ and $Y\in \spVec{^1_1M}$ can then be written as
$$\duPairM{X}{Y} = \duPair{\subTxt{y}{vol}}{\subTxt{x}{vol}}{M}  + \duPairM{\Ydev}{\Xdev},$$
while $\map{\duPair{\cdot}{\cdot}{M} }{\spFn{M}\times \spFn{M}}{\bb{R}}$ is defined as $\duPair{\subTxt{y}{vol}}{\subTxt{x}{vol}}{M} := \int_M \subTxt{y}{vol} \subTxt{x}{vol}\ \omega$.

It is important to note that the above decomposition of second rank tensor fields did not require the manifold $M$ to have a Riemannian $g$. In the presence of this extra metric structure, then one could apply the volumetric-deviatoric decomposition to any $A\in \spVec{^0_2M}$ using its tensorial variants by raising and lowering indices. This is achieved by extending the projection map (\ref{eq:pivol}) to
$$\map{\subTxt{\bar{\pi}}{vol} := g \circ \pivol \circ g^{-1}}{\spVec{^0_2M}}{\spVec{^0_2M}}.$$
In the same manner, one could apply the symmetric-asymmetric decomposition to any $X\in \spVec{^1_1M}$ by extending (\ref{eq:pisym}) to
$$\map{\subTxt{\bar{\pi}}{sym} := g^{-1} \circ \pisym \circ g}{\spVec{^1_1M}}{\spVec{^1_1M}}.$$

\begin{acknowledgements}
	This work was supported by the PortWings project funded by the European Research Council
	[Grant Agreement No. 787675]
\end{acknowledgements}

\bibliographystyle{ieeetr}
\bibliography{references}

\end{document}